# Data Coding Means and Event Coding Means Multiplexed Over the 1000BASE-T1 PCS Payload

Alexander Ivanov

*Abstract*—**We make an attempt to apply the linguistic multiplexing approach onto a known coding means featuring a very long binary transport word, we'll call a microframe, and by this extend such a means with an event transfer ability, considering the 1000BASE-T1 physical layer the point we start in.**

## INTRODUCTION

GIGABIT Ethernet, type 1000BASE-T1 [1] is a relatively new IEEE 802.3 physical layer intended to work over a single unshielded twisted pair media, in a harsh environment like of or similar to automotive and industrial.

Abstracting from comprehensive details, we say we employ some coding means that features a very long binary transport word—we refer to as the microframe—consisting of $450 = 405{+}1{+}44$ GF($2^9$) information symbols or $450 \times 9 = 4{,}045$ bits, respectively, and incurs a duty to convey $45 \times 10 = 450$ separate instances of a regular payload element—we refer to as the transfer unit—each corresponding a) either with one of $2^8 = 256$ possible data octet values or b) with one of at least only possible control value, alongside the preserved $1 \times 9 = 9$ service and $44 \times 9 = 396$ check bits, accordingly, per every microframe time period,[1] see Table I.

Operating on the top of that means, we expectedly preserve its mentioned duty in the whole, but then apply the linguistic multiplexing approach, as it was developed up to in [2], to get what that we refer to as the spare bits in the terminology of our search, choice, and design, see Table II.

Substituting the spare bits, we can simply pass an encoded position of an event—sensed and fixed on some precision time synchronization interface[2] provided by the physical layer on its top—occurring during the respective microframe time period, into the payload of that microframe, together with the encoded content of the respective transfer units.

In the rest of this paper, we describe a way to embed such the capability inside the 1000BASE-T1 physical layer, keeping its extended use still compliant with its original duty.

A manuscript of this work was submitted to IEEE Communications Letters December 26, 2022 and rejected for fair reasons during its peer review.

Please sorry for the author has no time to find this work a new home, peer reviewed or not, except of arXiv, and just hopes there it meets its reader, one or maybe various, whom the author beforehand thanks for their regard.

A. Ivanov is with JSC Continuum, Yaroslavl, the Russian Federation.

[1]In the case of 1000BASE-T1, its microframe time is 450 octet times, that gives $8 \times 450 = 3600$ bit times, i.e., 3600 ns or 3.6 $\mu$s.

[2]Note that the service interface of the physical layer is now composite and comprises the GMII and a PTSI for data and event, respectively.

TABLE I
ORIGINAL MICROFRAME STRUCTURE

| Hierarchy and Elements | Contents, Comments | Size, bits |
|---|---|---|
| Microframe (or μframe) entirely | issue rate ≈ 0.28M μframes/s | 450 × 9 = Σ 4,050 |
| ├ Payload, visible at GMII | max 450 data octets → 3,600 bits | 45 × 81 = 3,645 |
| │ ├ 80B/81B coded block #1 | •first 10 data octets or controls | 10×8+1 = 9 × 9 = 81 |
| . . . . . . . . . . . . . . . . . . . . | . . . . . . . . . . . . . . . . . . . . . . . | . . . |
| │ └ 80B/81B coded block #45 | •last 10 data octets or controls | 10×8+1 = 9 × 9 = 81 |
| ├ OAM field, hidden by PHY | purposed for link management | 1 × 9 = 9 |
| └ RS-FEC tail, hidden by PHY | found over all the bits above | (450 − 406) × 9 = 396 |

TABLE II
TERMINOLOGY

| Ref. | Objective | Definition, Description, Designation | Measured |
|---|---|---|---|
| $N$ | given | data modulus of the transfer *unit*, $257 \le N \le 274$, | in items |
| $n_p$ | fixed | scope of the *payload*, $n_p = 45{\times}10 = 450$, | in units/μframe |
| $n_e$ | given | scope of an *echo* multiplexing round, | in units/round |
| $k$ | $k = n_p / n_e$ | run count to cover the whole payload, $k \ge 1$, | in rounds/μframe |
| $v$ | fixed | information *volume* of the payload, $v = 45{\times}81$, | in bits/μframe |
| $s$ | maximize | *spare* volume for event coding, $s + k \bullet t = v$, | in bits/μframe |
| $t$ | $t = b + r$ | *total* amount of information handled per a run, | in bits/round |
| $b$ | maximize | its portion that results simply from *bypass*, $b \ge 0$, | in bits/round |
| $r$ | minimize | its portion that results from *representation*, $r \ge 1$, | in bits/round |
| $C_r$ | $C_r = r \ln 2$ | (log) transport *capacity* of the represented portion, | in nepers |
| $M_r$ | $M_r = n_e \ln N - b \ln 2$ | (log) data *modulus* represented per a round, | in nepers |
| U*w* | minimize | unsigned ALU complexity class = required bit *width*, | in steps |

TABLE III
BEST CHOICE TIPS

| Priority | Focus | Label | Formalized Objective Definition | |
|---|---|---|---|---|
| Highest | $s$ | BEST-*s* | maximize $s$, as much as possible | [then...] |
| Medium | $N$ | BEST-*N* | maximize $N$, as much as possible | [then...] |
| Lowest | $b$, $r$, $w$ | BEST-*w* | maximize $b$, minimize $r$, minimize $w$ | [done. ] |

## THE HUNT FOR THE SPARE BITS

We expect that our search may produce many results, thus, we first need to formulate a set of prioritized tips to make the following choice as best as possible among the various results, see Table III, and then pre-plot the search plan, see Table IV, given a data modulus of the transfer unit.[3] [4]

To plot the search plan completely, we also consider a finite set of appropriate multiplexing processes, see Table V, running over all the divisors of the number 450 to comprise exact this

[3]Because there are exact $8^2 = 256$ possible data octet values, 0x00÷0xFF, and at least one possible control value, idle, we iterate for $N > 256$.

[4]Because there is a limit on $N$ resulting from the constraint of $N^{n_p} \le 2^v$, thus $N \le 2^{v/n_p}$, so in the case of 1000BASE-T1, $256 < N \le 274$.

TABLE IV
TRANSFER UNIT VARIANTS

| $N$ | Decomposition | Bypassed Bits | Items' Possible Purpose {Count} |
|---|---|---|---|
| 256 | $= N_r \times N_b = 1 \times 2^8$ | n/a | data octet values {256} only |
| 257 | $= 257\times 1 = 2^8 + 2^0$ | $b/n_e = \log_2 2^0 = 0$ | data octet values {256} + idle {1} |
| 258 | $= 129\times 2 = 2^8 + 2^1$ | $b/n_e = \log_2 2^1 = 1$ | all above {257} + data error {1} |
| 260 | $= 65\times 4 = 2^8 + 2^2$ | $b/n_e = \log_2 2^2 = 2$ | all above {258} + halt {1} + lpi {1} |
| 264 | $= 33\times 8 = 2^8 + 2^3$ | $b/n_e = \log_2 2^3 = 3$ | all above {260} + more controls {4} |
| 272 | $= 17\times 16 = 2^8 + 2^4$ | $b/n_e = \log_2 2^4 = 4$ | all above {264} + extra controls {8} |
| 274 | $= 137\times 2 = 2^8+2^4+2^1$ | $b/n_e = \log_2 2^1 = 1$ | all above {272} + extra controls {2} |

NOTE – Original 80B/81B encoding provides for $2^3 = 8$ possible control codes.
NOTE – 1000BASE-T1 PCS uses 4 out of 8 codes, assuming the rest deprecated:

| | | | |
|---|---|---|---|
| $0 = 000_{(2)}$ | idle when phy is not ready | $4 = 100_{(2)}$ | [do not transmit, treat as error if receive] |
| $1 = 001_{(2)}$ | data error propagation | $5 = 101_{(2)}$ | assert low power idle |
| $2 = 010_{(2)}$ | idle when phy is ready | $6 = 110_{(2)}$ | [do not transmit, treat as error if receive] |
| $3 = 011_{(2)}$ | [do not transmit, treat as error if receive] | $7 = 111_{(2)}$ | [do not transmit, treat as error if receive] |

TABLE V
PROCESS FASHION VARIANTS

| $n_e$ | 10 | 15 | 18 | 25 | 30 | 45 | 75 | 90 | 150 | 225 | 450 |
|---|---|---|---|---|---|---|---|---|---|---|---|
| $k$ | 45 | 30 | 25 | 18 | 15 | 10 | 6 | 5 | 3 | 2 | 1 |
| lim $s$ | 0 | ≤15 | ≤20 | ≤27 | ≤30 | ≤35 | ≤39 | ≤40 | ≤42 | ≤44 | ≤44 |

TABLE VI
SEARCH BEST RESULTS

| $n_e$ | $N$ | $C_r : M_r$ | $t$ | = | $b$ | + | $r$ | ALU | $k$ | $s$ |
|---|---|---|---|---|---|---|---|---|---|---|
| 10 | 274 | 1.0003 | 81 | | 10 | | 71 | U128 | 45 | 0 |
| 15 | 264 | 1.0044 | 121 | | 45 | | 76 | U128 | 30 | 15 |
| 18 | 264 | 1.0022 | 145 | | 54 | | 91 | U128 | 25 | 20 |
| 25 | 260 | 1.0029 | 201 | | 50 | | 151 | U256 | 18 | 27 |
| 30 | 260 | 1.0018 | 241 | | 60 | | 181 | U256 | 15 | 30 |
| 45 | 258 | 1.0016 | 361 | | 45 | | 316 | U512 | 10 | 35 |
| 75 | 258 | 1.0003 | 601 | | 75 | | 526 | U1024 | 6 | 39 |
| 90 | 257 | 1.0007 | 721 | | — | | 721 | U1024 | 5 | 40 |
| 150 | 257 | 1.0001 | 1,201 | | — | | 1,201 | >U1024 | 3 | 42 |
| 225 | 257 | 1.0004 | 1,802 | | — | | 1,802 | >U1024 | 2 | 41 |
| 450 | 257 | 1.0001 | 3,603 | | — | | 3,603 | >U1024 | 1 | 42 |
| 450 | 274 | 1.0003 | 3,645 | | 450 | | 3,195 | >U1024 | 1 | 0 |

TABLE VII
EVENT FIXATION SCALE

| $s$, bits → | 15 | 20 | 27 | 30 | 35 |
|---|---|---|---|---|---|
| Resolution | ~109.9 ps | ~3.4 ps | ~26.8 fs | ~3.6 fs | ~104.8 as |
| Frequency | ~9.1 GHz | ~291.3 GHz | ~37.3 THz | ~298.3 THz | ~9.5 PHz |

TABLE VIII
SEARCH DETAILS EXCERPT

| $n_e$ | $N$ | = | $N_r \times N_b$ | $b/n_e$ | $r/n_e$ | $C_r : M_r$ | $C_r > M_r$? | $t$ | = | $b$ | + | $r$ | ALU | $k$ | $k \bullet t$ | + | $s$ | = | $v$ | Status |
|---|---|---|---|---|---|---|---|---|---|---|---|---|---|---|---|---|---|---|---|---|
| 10 | 274 | = | 137×2 | 1 | 7.1 | 1.00028 | yes | 81 | = | 10 | + | 71 | U128 | 45 | 3,645 | + | 0 | = | 3,645 | *no spare* |
| 10 | 272 | = | 17×16 | 4 | 4.1 | 1.00307 | yes | 81 | = | 40 | + | 41 | U64 | 45 | 3,645 | + | 0 | = | 3,645 | *no spare* |
| 15 | 274 | = | 137×2 | — | — | 0.99558 | no | | | | | | | | | | | | | [unreachable] |
| 15 | 272 | = | 17×16 | — | — | 0.99491 | no | | | | | | | | | | | | | [unreachable] |
| 15 | 264 | = | 33×8 | 3 | < 5.1 | 1.00442 | yes | 121 | = | 45 | + | 76 | U128 | 30 | 3,630 | + | 15 | = | 3,645 | BEST-*N* |
| 15 | 260 | = | 65×4 | 2 | < 6.1 | 1.00736 | yes | 121 | = | 30 | + | 91 | U128 | 30 | 3,630 | + | 15 | = | 3,645 | *spare* > 0 |
| 15 | 258 | = | 129×2 | 1 | < 7.1 | 1.00791 | yes | 121 | = | 15 | + | 106 | U128 | 30 | 3,630 | + | 15 | = | 3,645 | *spare* > 0 |
| 15 | 257 | = | 257×1 | 0 | < 8.1 | 1.00762 | yes | 121 | = | 0 | + | 121 | U128 | 30 | 3,630 | + | 15 | = | 3,645 | *spare* > 0 |
| 18 | 274 | = | 137×2 | — | — | 0.99402 | no | | | | | | | | | | | | | [unreachable] |
| 18 | 272 | = | 17×16 | — | — | 0.99219 | no | | | | | | | | | | | | | [unreachable] |
| 18 | 264 | = | 33×8 | 3 | < 5.1 | 1.00221 | yes | 145 | = | 54 | + | 91 | U128 | 25 | 3,625 | + | 20 | = | 3,645 | BEST-*N* |
| 18 | 260 | = | 65×4 | 2 | < 6.1 | 1.00551 | yes | 145 | = | 36 | + | 109 | U128 | 25 | 3,625 | + | 20 | = | 3,645 | *spare* > 0 |
| 18 | 258 | = | 129×2 | 1 | < 7.1 | 1.00632 | yes | 145 | = | 18 | + | 127 | U128 | 25 | 3,625 | + | 20 | = | 3,645 | *spare* > 0 |
| 18 | 257 | = | 257×1 | 0 | < 8.1 | 1.00624 | yes | 145 | = | 0 | + | 145 | U256 | 25 | 3,625 | + | 20 | = | 3,645 | *spare* > 0 |
| 25 | 260 | = | 65×4 | 2 | < 6.1 | 1.00293 | yes | 201 | = | 50 | + | 151 | U256 | 18 | 3,618 | + | 27 | = | 3,645 | BEST-*s* |
| 25 | 264 | = | 33×8 | 3 | < 5.1 | 1.00706 | yes | 202 | = | 75 | + | 127 | U128 | 18 | 3,636 | + | 9 | = | 3,645 | BEST-*Nw* |
| 30 | 260 | = | 65×4 | 2 | < 6.1 | 1.00182 | yes | 241 | = | 60 | + | 181 | U256 | 18 | 3,618 | + | 30 | = | 3,645 | BEST-*s* |
| 30 | 264 | = | 33×8 | 3 | < 5.1 | 1.00442 | yes | 242 | = | 90 | + | 157 | U256 | 18 | 3,636 | + | 15 | = | 3,645 | BEST-*N* |
| 45 | 258 | = | 129×2 | 1 | < 7.1 | 1.00157 | yes | 361 | = | 45 | + | 316 | U512 | 10 | 3,610 | + | 35 | = | 3,645 | BEST-*s* |
| 45 | 264 | = | 33×8 | 3 | < 5.1 | 1.00001 | yes | 362 | = | 135 | + | 227 | U256 | 10 | 3,620 | + | 25 | = | 3,645 | *spare* > 0 |
| 45 | 264 | = | 33×8 | 3 | < 5.1 | 1.00442 | yes | 363 | = | 135 | + | 228 | U256 | 10 | 3,630 | + | 15 | = | 3,645 | *spare* > 0 |
| 45 | 272 | = | 17×16 | 4 | < 4.1 | 1.00035 | yes | 364 | = | 180 | + | 184 | U256 | 10 | 3,640 | + | 5 | = | 3,645 | BEST-*N* |
| 75 | 258 | = | 129×2 | 1 | < 7.1 | 1.00030 | yes | 601 | = | 75 | + | 526 | U1024 | 6 | 3,606 | + | 39 | = | 3,645 | BEST-*s* |
| 75 | 260 | = | 65×4 | 2 | < 6.1 | 1.00071 | yes | 602 | = | 150 | + | 552 | U512 | 6 | 3,612 | + | 33 | = | 3,645 | *spare* > 0 |
| 75 | 260 | = | 65×4 | 2 | < 6.1 | 1.00293 | yes | 603 | = | 150 | + | 453 | U512 | 6 | 3,618 | + | 27 | = | 3,645 | *spare* > 0 |
| 75 | 264 | = | 33×8 | 3 | < 5.1 | 1.00177 | yes | 604 | = | 225 | + | 379 | U512 | 6 | 3,624 | + | 21 | = | 3,645 | *spare* > 0 |
| 75 | 264 | = | 33×8 | 3 | < 5.1 | 1.00442 | yes | 605 | = | 225 | + | 380 | U512 | 6 | 3,630 | + | 15 | = | 3,645 | *spare* > 0 |
| 75 | 264 | = | 33×8 | 3 | < 5.1 | 1.00706 | yes | 606 | = | 225 | + | 381 | U512 | 6 | 3,636 | + | 9 | = | 3,645 | *spare* > 0 |
| 75 | 272 | = | 17×16 | 4 | < 4.1 | 1.00144 | yes | 607 | = | 300 | + | 307 | U512 | 6 | 3,642 | + | 3 | = | 3,645 | BEST-*N* |
| 90 | 257 | = | 257×1 | 0 | < 8.1 | 1.00069 | yes | 721 | = | 0 | + | 721 | U1024 | 5 | 3,605 | + | 40 | = | 3,645 | BEST-*s* |
| 90 | 274 | = | 137×2 | 1 | < 7.1 | 1.00028 | yes | 729 | = | 90 | + | 639 | U1024 | 5 | 3,645 | + | 0 | = | 3,645 | *no spare* |
| 150 | 257 | = | 257×1 | 0 | < 8.1 | 1.00013 | yes | 1,201 | = | 0 | + | 1,201 | >U1024 | 3 | 3,603 | + | 42 | = | 3,645 | BEST-*s* |
| 150 | 274 | = | 137×2 | 1 | < 7.1 | 1.00028 | yes | 1,215 | = | 150 | + | 1,065 | >U1024 | 3 | 3,645 | + | 0 | = | 3,645 | *no spare* |
| 225 | 257 | = | 257×1 | 0 | < 8.1 | 1.00041 | yes | 1,802 | = | 0 | + | 1,802 | >U1024 | 2 | 3,604 | + | 41 | = | 3,645 | BEST-*s* |
| 225 | 272 | = | 17×16 | 4 | < 4.1 | 1.00035 | yes | 1,820 | = | 900 | + | 920 | U1024 | 2 | 3,640 | + | 5 | = | 3,645 | BEST-*N* |
| 450 | 257 | = | 257×1 | 0 | < 8.1 | 1.00013 | yes | 3,603 | = | 0 | + | 3,603 | >U1024 | 1 | 3,603 | + | 42 | = | 3,645 | BEST-*s* |
| 450 | 274 | = | 137×2 | 1 | < 7.1 | 1.00028 | yes | 3,645 | = | 450 | + | 3,195 | >U1024 | 1 | 3,645 | + | 0 | = | 3,645 | *no spare* |

TABLE IX
CODING SCHEME

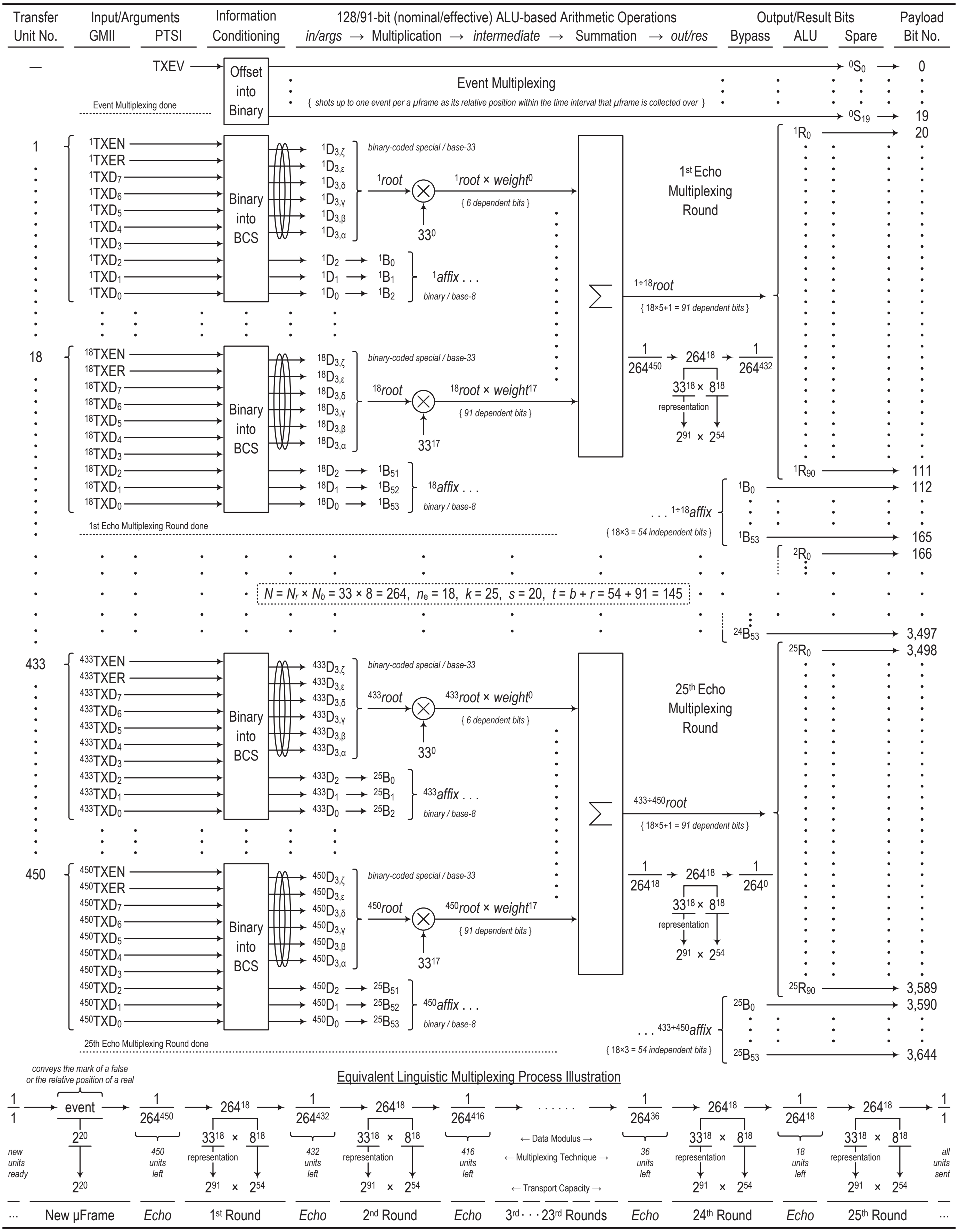

TABLE X
KEY FEATURES OF THE PROPOSED WAY

| Parameter | Description |
|---|---|
| Coding Delay = μFrame Period | 3,600 information bit times = 3,600 ns = 3.6 μs |
| Conditional Coding Complexity | like 128/91-bit nominal/effective unsigned ALU |
| Event Fixation Resolution (Freq.) | about 3.4 ps (291.3 GHz), or 20 bits/ev/μframe |
| Event Transmission Rate | up to one event per a μframe ≈ 277,778 ev/s |

number, i.e., $450 = 10 \times 45 = 15 \times 30 = \cdots = 225 \times 2 = 450 \times 1 = n_\mathrm{e} \times k$, that fully defines the configuration of an appropriate multiplexing process treating on $n_\mathrm{e}$ transfer units per each of its $k$ multiplexing rounds.[5]

Further, following that search plan, we do the search itself, choosing the most suitable results, see Table VI, then briefly estimate their applicability in a precise synchronization task, see Table VII, to make the single choice.

Assuming the need to pass up to one event per a microframe time period, we find out that any such variant provides for an excessive number of the spare bits, $s$, to solve such a task at any desired level, from more than simply particular, e.g., with $s = 15$ or 20, into more than very exotic, e.g., with $s = 27$ or above, that also finally helps us in our choice.

Because the variants with $s = 15$ and $s = 20$ both relate to the same filling of the transfer unit with $N = 264$ as well as both exhibit for the same accepted degree of implementation complexity with $w \leq 128$, we obviously make our choice on the variant with $s = 20$ ($N = 264$, $n_\mathrm{e} = 18$, $k = 25$).

Thus, based on the mentioned above, we proceed to design an appropriate coding means that corresponds with the multiplexing configuration deliberately chosen by us among many possible variants, see Table VIII.

## CODING SCHEME

Applied in the scope of the microframe, every microframe time period, the coding means runs thru its three consecutive stages, see Table IX, operating as the following.

On its multiple service interface stage, leftmost in Table IX, the coding means senses the respective interface signals and conditions the information conveyed by them.

[5] Exact $N^{n_\mathrm{e}}$ out of $2^t$ possible echo samples are used, all considered native, thus, given the constants, $N$, $n_\mathrm{e}$, $k$, iterators, $i = 1 \ldots k$, $j = 1 \ldots n_\mathrm{e}$, and transfer unit values to be sent in the payload during the respective microframe,

$$u_{i,j} \in \{ \underbrace{0 \,, \quad \ldots \quad , 255}_{\text{data octet values}} \;\;, \;\; \underbrace{256 \,, \ldots , N{-}1}_{\text{control values}} \;\} \,,$$

one could find the corresponding native echo samples just as the following:

$$\mathrm{ECHO}_i = \sum_j \underbrace{u_{i,j} \times N^{j-1}}_{value \times weight} < N^{n_\mathrm{e}} \leq 2^t \;,$$

or, assuming $N = N_r \times N_b$ and the latter is a power of two, compositely:

$$\mathrm{root}_i = \sum_j \underbrace{[u_{i,j} \operatorname{div} N_b] \times [N_r]^{j-1}}_{\text{representation}}, \;\; \mathrm{affix}_i = \sum_j \underbrace{[u_{i,j} \bmod N_b] \times [N_b]^{j-1}}_{\text{bypass}},$$

that, compared to the pseudo-code-based normative definition of the 80B/81B encoding in [1], seems better in the terms of clarity, simplicity, unambiguity, of course, related to description, not implementation. Is not it?

On its treating stage, central in Table IX, the coding means consumes and generates all the coupled, dependent bits only, into and from its unsigned arithmetic procedures, respectively, simply bypassing all the rest, independent bits.

On its final stage, rightmost in Table IX, the coding means just maps the spare, generated dependent, and bypassed independent bits onto the designated payload bits.[6]

Altogether, the consecutive stages comprise an appropriate procedure that corresponds with both the payload parameters ($v = 3{,}645$ for $n_\mathrm{p} = 450$) and the multiplexing configuration ($N = 264$, $n_\mathrm{e} = 18$, $k = 25$, $s = 20$), initially predetermined prior and deliberately chosen under us, respectively.

In the terms of static definition, not particular implementation, all those stages are memory-less, thus, the coding means seems to be a sort of memory-less encoding.

In the terms of application and characterization, that means seems like a set of recognizable, scalable, adaptable patterns, many of which are repeated more than once.

In the terms of complexity estimation, it seems to be similar with one after a representation of an $n_\mathrm{e}$-digit base-$N_r$ numeral into an $r$-digit base-2 numeral, both binary-coded.

## CONCLUSION

We described a way to multiplex a data coding means, fully compliant with the original duty of the 1000BASE-T1 physical layer as well as fully interior for its physical coding sublayer, or PCS, and an event coding means, new for that layer, over the 1000BASE-T1 PCS microframe.[7]

The proposed way enables the coding means of interesting properties, see Table X, and is grounded upon the linguistic multiplexing approach, that's involved to release a number of the spare bits, i.e., those are free of the original duty and thus useful the manner we described or another. [8] [9]

The enabled coding means defines—in an abstract form—a scalable multi-stage procedure that operates on dependent and independent bit quantities, doing binary arithmetic and bitwise logic manipulations, all are well-known as well as wide-spread in the field of digital semiconductor devices.

[6] Because 1000BASE-T1 implements a systematic Reed–Solomon forward error correction (RS-FEC) means [1], we may optimize the scheme defined in Table IX by moving the spare bits from the top (first on the wire) into the bottom (last on the wire) bits of the payload, to reduce at least the transmit coding delay, making the multiplexing process inverse [2]. Moreover, there are no restrictions to place the spare bits anyhow (together or apart) and anywhere (sooner or later) in the payload, since we access a microframe.

[7] We call it "micro" comparing to the maximum Ethernet frame length (1526 octets or 12,208 bits, if accounted including the preamble and start-of-frame delimiter elements) and to clearly distinguish between the frames.

[8] One possible use of the spare bits is to consider each of those bits separate and responsible for reflecting an event that occurs during the designated part of the respective microframe time period, forming an event train.

[9] Other possible use of the spare bits is to establish a carrier for a dedicated channel passing exclusively messages (packets) of some time synchronization protocol, like IEEE 1588 PTP or RFC 5904 NTP, across microframes.

# Data Coding Means and Event Coding Means Multiplexed Over the 10GBASE-T PCS Payload

Alexander Ivanov

***Abstract*—We continue to extend known protocols featuring a very long binary transport word, with an event transfer ability, considering now the 10GBASE-T PHY the point we stand in.**

***Index Terms*—Ethernet, linguistic multiplexing, multiplexing, precise synchronization, synchronization, 10GBASE-T.**

## Introduction

TEN-GIGABIT Ethernet, type 10GBASE-T [1] describes a full-duplex 10 Gbps physical layer operating over four twisted pairs of Category 6a or better. It has a duty to convey $50 \times 8 = 400$ separate transfer units, each keeping either one of $2^8 = 256$ possible data octet values or a control code, per every microframe time period, see Table I.

Inside that 10GBASE-T PCS[1] microframe,[2] there are exact $50 \times 65 = 3{,}250$ information bits of the payload—intended to keep the mentioned duty—we consider fully accessible while its user is interior, i.e., related to the PCS, in between of the payload access interface[3] and the PCS service interface,[4] on the bottom and on the top, respectively.

In the rest of this paper, we try to highlight a way to embed an event passing ability into the 1000BASE-T PCS sublayer, replacing the interior user of its payload with an appropriately modified one, that supports for both the event passing and the data passing, also keeping the new task of the physical layer still fully compliant with its original duty.

## Save Box Over Spare Bits

We begin with the same search for the spare bits, as it was described in [2], using the similar conditions, slightly updated to match our new circumstances, see Table II,[5] however, we now assume a distinct purpose for those bits, that is intended to meet the changed behavior of the multiplexing process we study further in this paper, see Table III.

A manuscript of this work was submitted to IEEE Communications Letters December 26, 2022 and rejected for fair reasons during its peer review.

Please sorry for the author has no time to find this work a new home, peer reviewed or not, except of arXiv, and just hopes there it meets its reader, one or maybe various, whom the author beforehand thanks for their regard.

A. Ivanov is with JSC Continuum, Yaroslavl, the Russian Federation.

[1] Physical Coding Sublayer, the upper entity in 10GBASE-T, operates atop the 10GBASE-T Physical Medium Attachment (PMA) sublayer, see [1].

[2] It seems “micro” compared to the maximum Ethernet frame length, 1526 octets or 12,208 bits, named so to distinguish between the frames clearly.

[3] It is an abstract interface we refer only to point out the place of the topic inside the PCS as well as in the whole physical layer, accordingly.

[4] It is a standardized interface, XGMII in the case of 10GBASE-T, provided for the client of the physical layer, by the PCS on its top, see [1], too.

[5] Sensible differences between Table II and [2]’s Table II are underlined.

TABLE I
ORIGINAL MICROFRAME STRUCTURE

| Elements and Hierarchy | | Contents, Comments | Size, bits | |
|---|---|---|---|---|
| Microframe (or μframe) entirely | | issue rate = 3.125M μframes/s | | Σ 3,584 |
| ├ Payload, visible at XGMII | | max 400 data octets ⟶ 3,200 bits | | 50 × 65 = 3,250 |
| │ ├ 64B/65B coded block #1 | | ├ first 8 data octets or controls | | ├ 8×8+1= 65 |
| │ ⋮ . . . . . . . . . . | | ⋮ . . . . . . . . . . | | ⋮ . . . . |
| │ └ 64B/65B coded block #50 | | └ last 8 data octets or controls | | └ 8×8+1= 65 |
| ├ CRC field | hidden by PHY | calculated on the payload bits | Σ334 | 8 |
| ├ LDPC parity | | calculated on some bits above | | 325 |
| └ Auxiliary bit | | has no standardized purpose | | 1 |

TABLE II
PREPARATORY SEARCH TERMINOLOGY

| Ref. | Objective | Definition, Description, Designation | Measured |
|---|---|---|---|
| $N$ | given | data modulus of the transfer *unit*, $257 \le N \le$ <u>279</u>, | in items |
| $n_\mathrm{p}$ | fixed | scope of the *payload*, $n_\mathrm{p}$ = <u>50×8</u> = <u>400</u>, | in units/μframe |
| $n_\mathrm{e}$ | given | scope of <u>each of *k*</u> *echo* multiplexing rounds, | in units/round |
| $k$ | $k = n_\mathrm{p} / n_\mathrm{e}$ | run count to cover the payload <u>w/o events</u>, | in rounds/μframe |
| $v$ | fixed | information *volume* of the payload, $v$ = <u>50×65</u>, | in bits/μframe |
| $s$ | maximize | *spare* volume <u>free of the *k* rounds</u>, $s + k \cdot t = v$, | in bits/μframe |
| $t$ | $t = b + r$ | *total* amount of info handled per <u>each of *k* runs</u>, | in bits/round |
| $b$ | maximize | its portion that results simply from *bypass*, $b \ge 0$, | in bits/round |
| $r$ | minimize | its portion that results from *representation*, $r \ge 1$, | in bits/round |
| $C_r$ | $C_r = r \ln 2$ | (log) transport *capacity* of the represented portion, | in nepers |
| $M_r$ | $M_r = n_\mathrm{e} \ln N - b \ln 2$ | (log) data *modulus* represented per a round, | in nepers |
| U$w$ | minimize | unsigned ALU complexity class = required bit *width*, | in steps |

TABLE III
STUDIED MULTIPLEXING PROCESS

| Characteristics | First (Multi-)Pass | Second Pass |
|---|---|---|
| Run count of the pass | up to $n_s$ times per a μframe | once per a μframe |
| Trigger to run the pass | each of up to $n_s$ events | every new μframe |
| Echo-dealing behavior | (lossless) accumulation | cancellation |
| Special means employed | save box for $n_s$ transfer units | $w$-bit unsigned ALU |
| Pass execution structure | single chain over $n_\mathrm{p}$+$n_s$ units | $k$+$k_s$ bounded rounds |
| Multiplexing technique | (domino-like) substitution | representation |
| Example implementation | see Stage 2 Data, $n_s = 1$ | see Stage 4 Mixed |

The principal change lies in how the process deals with an event. In contrast to the process described in [2] that treats an event in the scope of the respective microframe time period, during which this event occurs, the process we propose treats the such in the scope of the transfer unit corresponding to the octet time period,[6] during which the event occurs. Expectedly, this impacts on the purpose of the spare bits.

[6] In the case of 10GBASE-T, bit time is 100 ps, octet time is 800 ps.

TABLE IV
TRANSFER UNIT VARIANTS

| $N$ | Decomposition | Bypassed Bits | Items' Type = Purpose {Count} |
|---|---|---|---|
| 256 | = $N_r \times N_b = 1 \times 2^8$ | n/a | DATA {256} only |
| 257 | = $257\times\ 1 = 2^8 + 2^0$ | $b/n_e = \log_2 2^0 = 0$ | DATA {256} + CTRL {1} |
| 258 | = $129\times\ 2 = 2^8 + 2^1$ | $b/n_e = \log_2 2^1 = 1$ | DATA {256} + CTRL {1} + event {1} |
| 260 | = $65\times\ 4 = 2^8 + 2^2$ | $b/n_e = \log_2 2^2 = 2$ | DATA {256} + CTRL {2} + event {2} |
| 264 | = $33\times\ 8 = 2^8 + 2^3$ | $b/n_e = \log_2 2^3 = 3$ | DATA {256} + CTRL {4} + event {4} |
| 272 | = $17\times 16 = 2^8 + 2^4$ | $b/n_e = \log_2 2^4 = 4$ | DATA {256} + CTRL {8} + event {8} |
| 279 | = $2^8+2^4+2^2+2^1+2^1$ | $b/n_e = \log_2 2^1 = 0$ | all above {272} + extra controls {7} |

TABLE V
PROCESS FASHION VARIANTS

| $n_e$ | 8 | 10 | 16 | 20 | 25 | 40 | 50 | 80 | 100 | 200 | 400 |
|---|---|---|---|---|---|---|---|---|---|---|---|
| $k$ | 50 | 40 | 25 | 20 | 16 | 10 | 8 | 5 | 4 | 2 | 1 |
| lim $s$ | 0 | ≤10 | ≤25 | ≤30 | ≤34 | ≤40 | ≤42 | ≤45 | ≤46 | ≤48 | ≤49 |

TABLE VI
SEARCH BEST RESULTS

| $n_e$ | $N$ | $C_r : M_r$ | $t$ = | $b$ + | $r$ | ALU | $k$ | $s$ |
|---|---|---|---|---|---|---|---|---|
| 8 | 279 | 1.00011 | 65 | — | 65 | U128 | 50 | — |
| 10 | 272 | 1.00307 | 81 | 40 | 41 | U64 | 40 | 10 |
| | max | | | | min | min | | enough |
| 16 | 264 | 1.00359 | 129 | 48 | 81 | U128 | 25 | 25 |
| 20 | 264 | 1.00111 | 161 | 60 | 101 | U128 | 20 | 30 |
| 25 | 260 | 1.00293 | 201 | 50 | 151 | U256 | 16 | 34 |
| 40 | 260 | 1.00044 | 321 | 80 | 241 | U256 | 10 | 40 |
| 50 | 258 | 1.00125 | 401 | 50 | 351 | U512 | 8 | 42 |
| 80 | 258 | 1.00018 | 641 | 80 | 561 | U1024 | 5 | 45 |
| 100 | 257 | 1.00055 | 801 | — | 801 | U1024 | 4 | 46 |
| 200 | 257 | 1.00055 | 1,602 | — | 1,602 | >U1024 | 2 | 46 |
| 400 | 257 | 1.00023 | 3,203 | — | 3,203 | >U1024 | 1 | 47 |
| 400 | 279 | 1.00011 | 3,250 | — | 3,250 | >U1024 | 1 | — |

TABLE VII
EVENT POSITION RANGE

| Event Positions in OT → | /0 | /1 | /2 | /3 | /4 | /5 | /6 | /7 |
|---|---|---|---|---|---|---|---|---|
| Coded as Bits $p_2,p_1,p_0$ → | 000 | 001 | 010 | 011 | 100 | 101 | 110 | 111 |

NOTE – Information bit time (BT) is 100 ps, then octet time (OT = 8 × BT) is 800 ps.

TABLE VIII
TRANSFER UNIT ENCODING

| Type | ↓Transfer Unit Content, Bits→ | $u_\varepsilon$ | $u_\delta$ | $u_\gamma$ | $u_\beta$ | $u_\alpha$ | $u_3$ | $u_2$ | $u_1$ | $u_0$ | Variety |
|---|---|---|---|---|---|---|---|---|---|---|---|
| DATA | one of $2^8$ possible data octets | 0 | $d_7$ | $d_6$ | $d_5$ | $d_4$ | $d_3$ | $d_2$ | $d_1$ | $d_0$ | 16 × 16 |
| CTRL | one of $2^3$ possible control codes | 1 | 0 | 0 | 0 | 0 | 0 | $c_2$ | $c_1$ | $c_0$ | 1/2 × 16 |
| event | one of $2^3$ relative positions [in OT] | 1 | 0 | 0 | 0 | 0 | 1 | $p_2$ | $p_1$ | $p_0$ | 1/2 × 16 |

TABLE IX
SAVE BOX KEEP

| State | ↓Save Box Content, Bits→ | $s_9$ | $s_8$ | $s_7$ | $s_6$ | $s_5$ | $s_4$ | $s_3$ | $s_2$ | $s_1$ | $s_0$ | Variety |
|---|---|---|---|---|---|---|---|---|---|---|---|---|
| empty | meaningless, except bit $s_9$ | 0 | – | – | – | – | – | – | – | – | – | — |
| DATA | 400th transfer unit with data | 1 | 0 | $d_7$ | $d_6$ | $d_5$ | $d_4$ | $d_3$ | $d_2$ | $d_1$ | $d_0$ | 16 × 16 |
| CTRL | 400th transfer unit with ctrl | 1 | 1 | – | – | – | – | – | $c_2$ | $c_1$ | $c_0$ | 1/2 × 16 |

According to the change, we need to substitute the existing transfer unit—whose position in the stream is associated with the octet time period during which the event occurs—by a new transfer unit that reflects the position of the event. Moreover, after done this, we also need to consecutively (avalanche-like) substitute all the rest transfer units, each by its preceding one, beginning from the unit that follows the substituted by one of the event, and then so on, again and again.

In the scope of the respective microframe, that leads for the last transfer unit just drops out of the microframe's contents, therefore we introduce what that we will further refer to as the save box, to store such a transfer unit, preventing it to be lost, and here, naturally to support for this, we seek for a dedicated space in the microframe's payload, that accounts for the need in those spare bits and also predefines their purpose: to be the dedicated space for the necessary save box.

So, we conduct that search the entire way described in [2], see Tables IV, V, and VI, making our choice for the variant with $N = 272$, $n_\mathrm{e} = 10$, $k = 40$, and $s = 10$, and then draw up the rest sensible details, see Tables VII, VIII, and IX, also introducing $N_\mathrm{data} + N_\mathrm{extra} = 256 + 16 = 272 = N = N_r{\cdot}N_b$ and $N_\mathrm{ctrl} + N_\mathrm{event} = 8 + 8 = 16 = N_\mathrm{extra}$ as well as $n_s = 1$, $t_s = 9$, $s_s = 1$, etc, as the equivalents of, respectively, $n_e$, $t$, $s$, etc, but for now in the view of the save box, that makes us equipped completely to go beyond.

## CODING SCHEME

Applied in the scope of the microframe, every microframe time period, the coding means runs thru its seven $(2 \times 2 + 3)$ stages, see Tables X toward XVI, coupled complexly but well traceable, together operating as the following.[7]

Event stage 1, see Table X, detects the relative position of an event, when and if it occurs during the currently processed microframe time period, and conditions (converts) that into a step-like mask consisting of $1 + 400 \cdot 8 = 3{,}201$ dependent bits that together depict the same-size choice, none or 1 of 3,200, and then travel further, into the next stages.[8]

Event stage 2, see Table XI, encodes the mask to generate a new transfer unit—whose own position and then the content, respectively, is associated precisely with the octet time period (within the currently processed microframe time period) and then reflects exactly the bit time period (within that octet time period of that microframe time period), during which the event occurs—in the defined form of $\lceil \log_2 N_r \rceil = 5$ dependent root and $\log_2 N_b = 4$ independent affix bits that together depict the new transfer unit and then travel further, too.

Data stage 1, see Table XII, conditions (converts), if it is of control, and then encodes all the data/control symbols met in all $50 \times 8 = 400$ octet time periods of the currently processed microframe time period, to generate so the respective transfer units, all in the same form as mentioned above.

[7]We describe the coding means capable to transfer just single event per a microframe time period. Hoverer, also possible are other variants capable to transfer multiple events per the same time, of course, for the cost of a coarser resolution due to lower $N_\mathrm{event}$, see Table XVII and the following notes.

[8]When supports for multiple events, this stage runs the respective number of times, dropping out all the before detected events, if any, on each next run during the current multi-pass, so, event stage 1 and data stage 2 are the only multi-passed stages in the studied multiplexing process, see Table III.

TABLE X
STAGE 1 EVENT

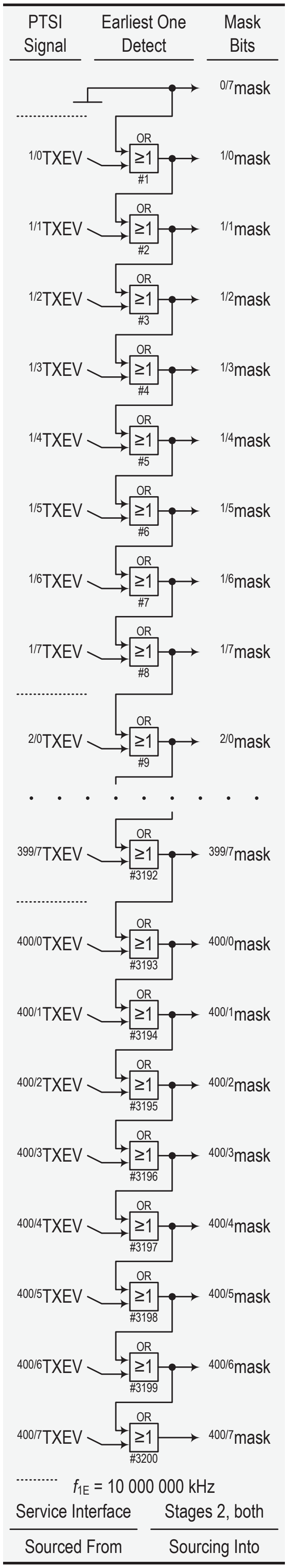

TABLE XI
STAGE 2 EVENT

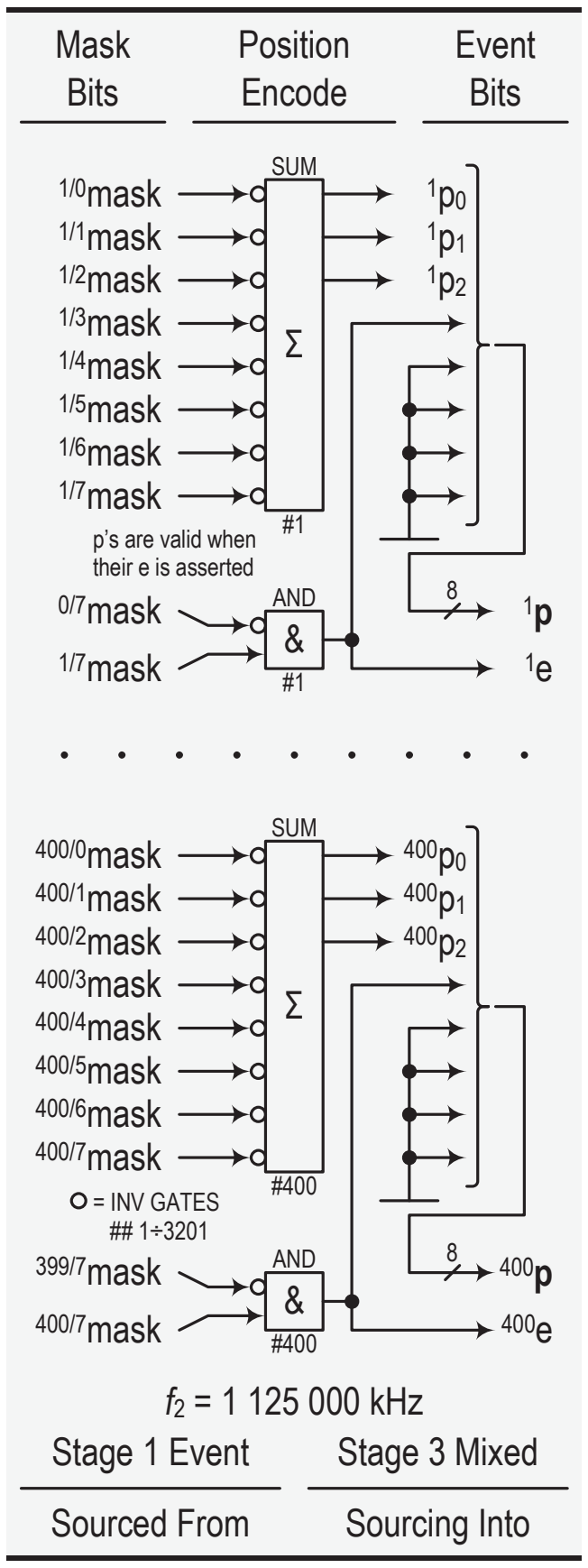

TABLE XII
STAGE 1 DATA

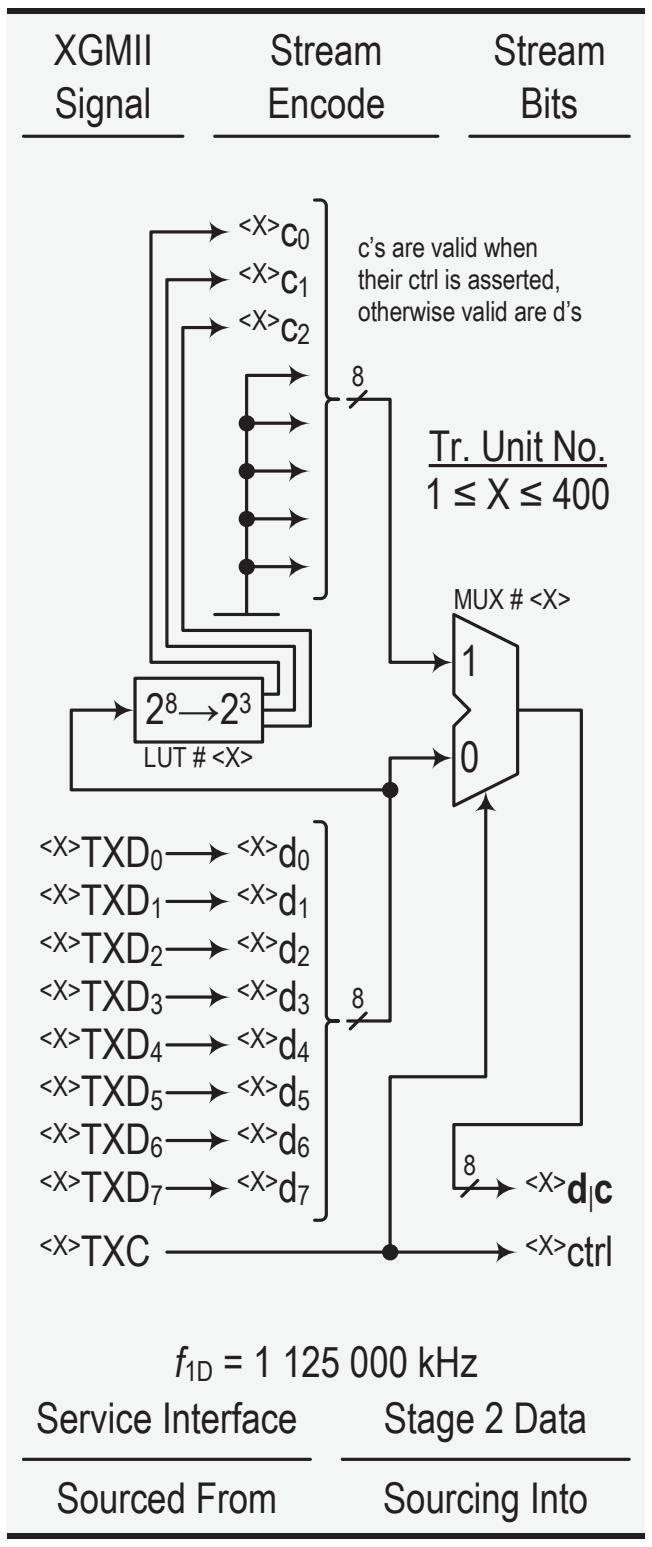

TABLE XIII
STAGE 2 DATA

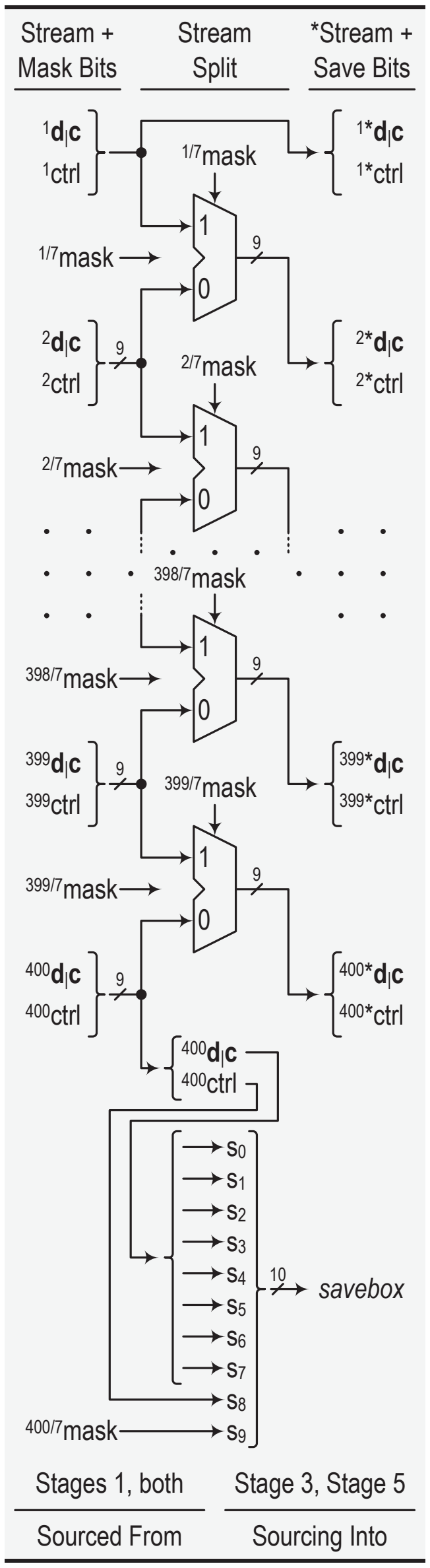

TABLE XIV
STAGE 3 MIXED

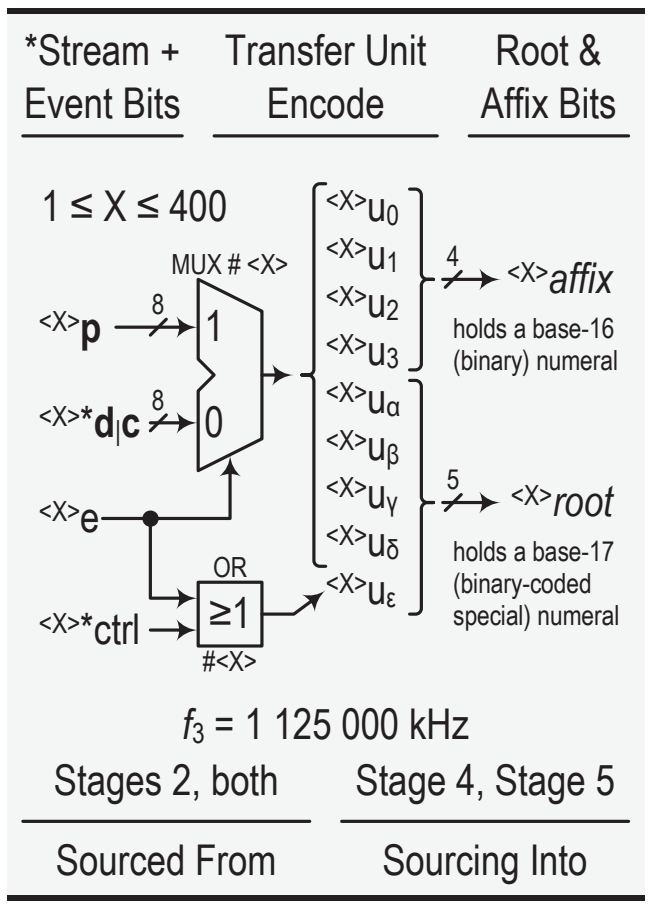

TABLE XV
STAGE 4 MIXED

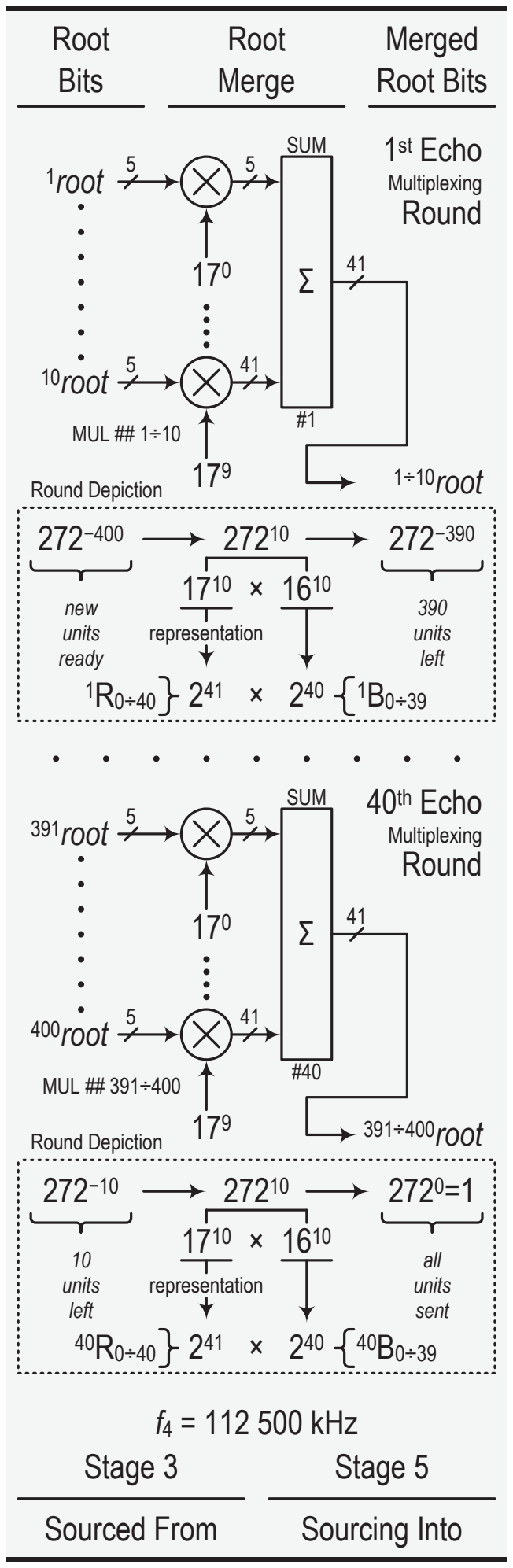

TABLE XVI
STAGE 5 MIXED

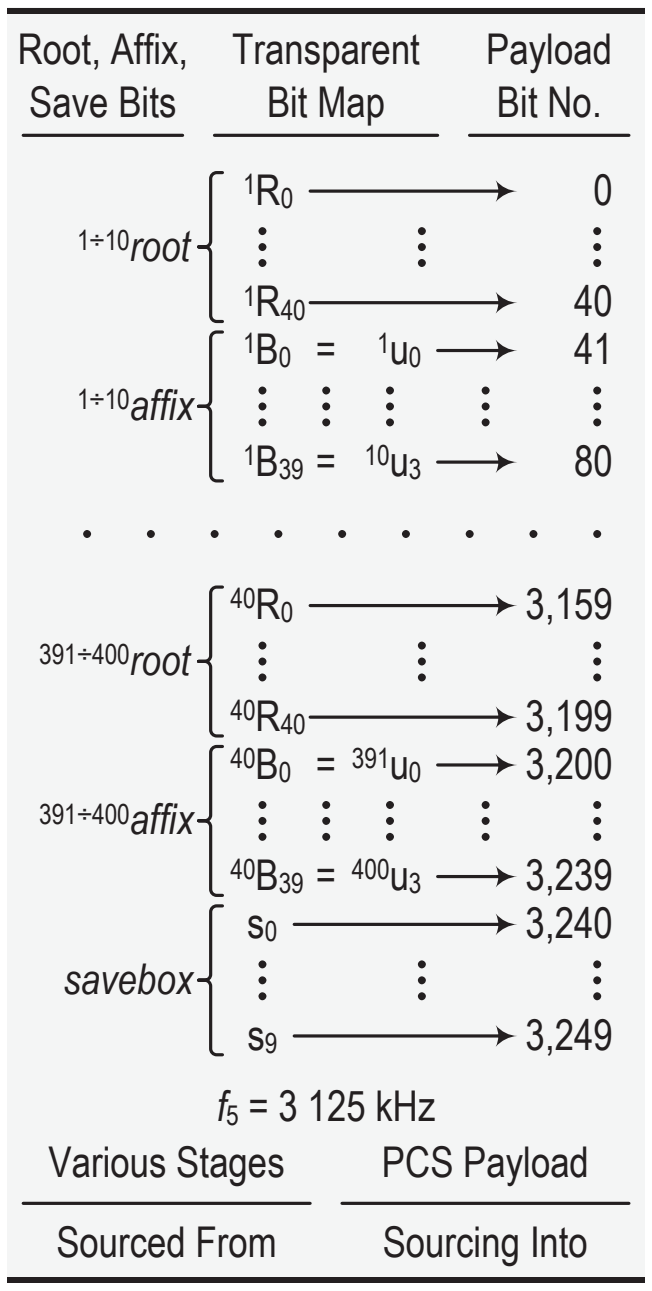

TABLE XVII
EXAMPLE APPLICABILITY INCLUDING BACKWARD

| μFrame | $n_\mathrm{p}$ | = | $n_\mathrm{e}$ | × | $k$ | $N$ | = | $N_r$ | × | $N_b$ | = | $N_\mathrm{data}$ | + | $N_\mathrm{extra}$ | $N_\mathrm{event}$ | $w$ | $s$ | = | $s_s$ | + | $t_s$ | $n_s$ | $N_s$ | $w_s$ | $k_s$ | Comment |
|---|---|---|---|---|---|---|---|---|---|---|---|---|---|---|---|---|---|---|---|---|---|---|---|---|---|---|
| 10GBASE-T | 400 | = | 10 | × | 40 | 272 | = | 17 | × | 16 | = | 256 | + | 16 | ≤15 | 64 | 10 | = | 1 | + | 1×9 | 1 | ≤512 | ≤16 | — | see Stages$^{1\div5}$ |
| 10GBASE-T | 400 | = | 16 | × | 25 | 264 | = | 33 | × | 8 | = | 256 | + | 8 | ≤7 | 128 | 25 | = | — | | 25 | 3 | ≤323 | ≤32 | 1 | up to 3 $^{\mathrm{ev}}/_{\mu\mathrm{frame}}$ |
| 10GBASE-T | 400 | = | 25 | × | 16 | 260 | = | 65 | × | 4 | = | 256 | + | 4 | ≤3 | 256 | 34 | = | 1 | + | 33 | 4 | ≤304 | ≤64 | 1 | up to 4 $^{\mathrm{ev}}/_{\mu\mathrm{frame}}$ |
| 10GBASE-T | 400 | = | 50 | × | 8 | 258 | = | 129 | × | 2 | = | 256 | + | 2 | ≤1 | 512 | 42 | = | 1 | + | 41 | 5 | ≤294 | ≤64 | 1 | up to 5 $^{\mathrm{ev}}/_{\mu\mathrm{frame}}$ |
| 1000BASE-T1 | 450 | = | 15 | × | 30 | 264 | = | 33 | × | 8 | = | 256 | + | 8 | ≤7 | 128 | 15 | = | 6 | + | 1×9 | 1 | ≤512 | ≤16 | — | up to 1 $^{\mathrm{ev}}/_{\mu\mathrm{frame}}$ |
| 1000BASE-T1 | 450 | = | 18 | × | 25 | 264 | = | 33 | × | 8 | = | 256 | + | 8 | ≤7 | 128 | 20 | = | 2 | + | 2×9 | 2 | ≤512 | ≤16 | — | up to 2 $^{\mathrm{ev}}/_{\mu\mathrm{frame}}$ |
| 1000BASE-T1 | 450 | = | 25 | × | 18 | 260 | = | 65 | × | 4 | = | 256 | + | 4 | ≤3 | 256 | 27 | = | — | | 3×9 | 3 | ≤512 | ≤16 | — | up to 3 $^{\mathrm{ev}}/_{\mu\mathrm{frame}}$ |
| 1000BASE-T1 | 450 | = | 45 | × | 10 | 258 | = | 129 | × | 2 | = | 256 | + | 2 | ≤1 | 512 | 35 | = | 2 | + | 33 | 4 | ≤304 | ≤64 | 1 | up to 4 $^{\mathrm{ev}}/_{\mu\mathrm{frame}}$ |

Data stage 2, see Table XIII, splits the stream consisting of the just encoded data/control transfer units, precisely—thanks to and based on the mask bits—in the given position of the new event transfer unit, causing substitution of the stream units in all its consecutive positions since the matching to the given—at which it inserts that new one—and till the last, all included, but the same time preventing the one in the last position to be lost, by storing it into the save box.[9]

Mixed stage 3, see Table XIV, finally encodes all the transfer units of the modified stream, see also Table VIII at this point, and then separates between all the root as well as all the affix bits of each just encoded transfer unit, to immediately bypass the latter—that all are independent—into the final stage, doing nothing on them, as well as to purposely prepare the former—that all are dependent, but still in the scope of the unit—to be further merged, before they reach the final stage.

Mixed stage 4, see Table XV, merges $10 \times 5$ root bits of a bundle of $10$ transfer units, repeating with each of $40$ bundles, into $10 \times 4 + 1$ merged, i.e., unambiguously reversibly coupled by arithmetic representation via prior weighting and following summation, ones—that makes them inter-dependent, i.e., such in the scope of all the units with the bundle—that corresponds to a generalized linguistic multiplexing process featuring $k = 40$ echo multiplexing rounds each treating over $n_\mathrm{e} = 10$ units of modulus $N = 272$ and resulting in $t = b + r = (10 \times 4) + (10 \times 4 + 1)$ bits defining a native echo sample.[10] [11]

Mixed stage 5, see Table XVI, maps the merged root bits, the bypassed affix bits, and the save box bits directly onto the bits—respectively designated to them—of the payload related to the currently processed microframe time period, finalizing by this the overall run of the coding means.

In the terms of static definition, application and characterization, and complexity estimation, the coding means seems to be very similar, if not the same, in all its essentials compared with those of the means described in [2].

[9] When supports for multiple events, this stage runs the respective number of times, inserting (the new transfer unit containing the encoded position of) each currently detected event, if any, into the stream, that causes avalanche (or domino-like) substitution of its units with the last going into the save box until it has no empty space for, each the mentioned time.

[10] When supports for multiple events, this stage may do an extra multiplexing round, on the transfer units stored in the save box. In Table XVII, such a case is easy found with $k_s = 1$ in the respective row, while a dash in that row labels a case when no extra round is involved.

[11] There are $2^t = 2^r \times 2^b$ distinguishable echo samples, $N^{n_\mathrm{e}} = N_r^{n_\mathrm{e}} \times 2^b$ out of those are considered native and intended to represent the result of an echo multiplexing round, and the rest $(2^r - N_r^{n_\mathrm{e}}) \times 2^b$ samples are free to use, e.g., to represent a forced echo containing a necessary extra signal.

## CONCLUSION

We proposed a way to embed an event passing ability inside a given, appropriate coding entity featuring a very long binary transport word as the microframe, and the same time preserve the measure of the data passing ability already existing in such the entity, qualitatively and quantitatively.

Studying the proposed way, we also described one possible coding means, in its very details, in regards to the 10GBASE-T PCS microframe and its payload. The described coding means is compliant with the original duty of the considered physical layer, moreover, it provides for completely separate and since behaviorally autonomous transfer units as well as for the fully sufficient numbers of control[12] and position[13] codes available per each of those units, instead of the original block encoding limits intrinsic to the regarded entity.

We believe the proposed way is applicable, at least in theory, in the field of the 10GBASE-T and other but similar physical layer protocols, including 1000BASE-T1 considered earlier in the preceding work [2], see Table XVII, and also has a visible potential for its further development.

[12] How many control codes, $N_\mathrm{ctrl}$, are necessary there? Let us consider all the standardized cases, except the six reserved, see [1], related to the 64B/65B block encoding used in 10GBASE-T : • • Idle, or /I/. Signaled by a dedicated control code mandatory to detect /S/ and /T/, see on them further. • Start, or /S/. Signaled by an idle to data (or error, if supported) transition, requiring no control code. 64B/65B does not allow for passing /S/ in an arbitrary selected octet period, only in each forth. The proposed design has no such limitation. • Terminate, or /T/. Signaled by a data (or error, if supported) to idle transition, requiring no code. 64B/65B passes /T/ in any octet period and the proposed design allows for the same. • Error propagation flag, or /E/. May be signaled by a dedicated control code, when and if supported. 64B/65B does not allow for passing /E/ in an arbitrary selected octet period, only in a dedicated block (occupies eight octet periods) of those instead, due to this, it may be signaled via a forced echo sample (occupies $n_\mathrm{e}$ octet periods) in the proposed design. • Sequence ordered set, or /Q/. May be signaled by a dedicated control code, if supported. 64B/65B does not allow for passing /Q/ in an arbitrary selected octet period, only once or twice per a block, therefore it may be signaled also by some forced echo sample. • Signal ordered set, or /Fsig/. May be signaled the same manner as /Q/. • Low power idle assert, or /LPI/. May be signaled the same manner as /Q/, too. • • Thus, supporting for one control code (idle) is enough while for two (idle and error) is very useful in the proposed design, i.e., $N_\mathrm{ctrl} \geq 1$, that also limits $N_\mathrm{event} \leq N_\mathrm{extra} - 1$, see Table XVII.

[13] The lower is the number of event position codes, $N_\mathrm{event} \geq 1$, the coarser is the event fixation scale, $f = N_\mathrm{event} \times 1.25$ GHz, and vice versa. Nominally, with $N_\mathrm{event} = 8$, see Table VII, it is $1/f = 100$ ps, i.e., $f = 10$ GHz.

# Data Coding Means and Event Coding Means Multiplexed Over the 10GBASE-KR PCS Payload

Alexander Ivanov

***Abstract*—We still continue to extend an appropriate protocol featuring a very long binary transport word, we propose to call the microframe, with an event transfer ability, now considering the 10GBASE-KR the point we focus on.**

***Index Terms*—Ethernet, linguistic multiplexing, multiplexing, precise synchronization, synchronization, 10GBASE-KR.**

## INTRODUCTION

TEN-GIGABIT Ethernet, type 10GBASE-KR [1] is a PHY layer intended to operate over an electrical backplane. It embodies a FEC[1] under its PCS,[2] reducing the service-related payload of the PCS down to $32 \times 65 = 2{,}080$ information bits, that all are purposed to transfer $32 \times 8 = 256$ separate units, comprising its original duty, see Table I.[3]

In the rest of this paper, we consider a way to introduce an extra ability—in a form of an event transfer means intended to solve a precise synchronization problem—into the mentioned physical layer, but the same time preserve the original duty of the data transfer means of the 10GBASE-KR PHY, primarily in the terms of the speed[4] [5] of its service.

## SPARE TRANSFER UNITS

Recalling the concept of the save box [2], that results from the change beyond the purpose of the spare bits [3], that itself results from the change in an excessively redundant and thus appropriate structure, it seems logic to make the next change, for further unification via suitable generalization, and expand over the so called initial scope, $n_\mathrm{p}$, defining the original duty, into a higher target scope, $*n_\mathrm{p} > n_\mathrm{p}$, see Table II.[6]

Conducting the cursory search the way it was done in both [2] and [3], we find its results promising, see Table III, after that, we first prepare for and then perform on the main search in a form of an arithmetical problem solvable via calculations

TABLE I
ORIGINAL MICROFRAME STRUCTURE

| Hierarchy and Elements | Contents, Comments | Size, bits |
|---|---|---|
| Microframe (or μframe) entirely | issue rate ≈ 4.88M μframes/s | Σ 2,112 |
| ├ Payload, visible at XGMII | max 256 data octets → 2,048 bits | 32 × 65 = 2,080 |
| ├ 64B/65B coded block #1 | first 8 data octets or controls | 8×8+1= 65 |
| . . . . . . . . . . . . . . . . . . . . | . . . . . . . . . . . . . . . . . . . . . . | . . . |
| └ 64B/65B coded block #32 | last 8 data octets or controls | 8×8+1= 65 |
| └ FEC parity, hidden by PHY | calculated on the payload bits | 32 |

TABLE II
GENERALIZED TERMINOLOGY

| Ref. | Objective | Definition, Description, Designation | Measured |
|---|---|---|---|
| U*w*, *w* | minimize | unsigned ALU complexity class = required bit *width*, | in steps |
| $N$ | $N = N_r \times N_b$ | data modulus of the transfer *unit*, $257 \le N \le 279$, | in items |
| $N_b$ | maximize | its multiplicative portion that *is* a power of two, | in items/root |
| $N_r$ | minimize | its multiplicative portion that *is not* a power of two, | in roots |
| $N_{\underline{d}\mathrm{ata}}$ | fixed to $2^8$ | its additive part that corresponds to *data* octets, | in items |
| $N_{\mathrm{e}\underline{\mathrm{x}}\mathrm{tra}}$ | $N_{\underline{\mathrm{d}}} + N_{\underline{\mathrm{x}}} = N$ | its additive part that corresponds to *extra* codes, | in items |
| $N_{\underline{\mathrm{c}}\mathrm{trl}}$ | $N_{\underline{\mathrm{c}}} \ge 1$ | its additive part that corresponds to *control* codes, | in items |
| $N_{\underline{\mathrm{e}}\mathrm{vent}}$ | $N_{\underline{\mathrm{c}}} + N_{\underline{\mathrm{e}}} \le N_{\underline{\mathrm{x}}}$ | its additive part that corresponds to *event* codes, | in items |
| $n_\mathrm{p}$ | fixed to 256 | *initial* scope of the payload, $n_\mathrm{p} = 32 \times 8 = 256$, | in units/μframe |
| $*n_\mathrm{p}$ | $*n_\mathrm{p} > n_\mathrm{p}$ | *target* scope of the payload, $*n_\mathrm{p} = k \cdot n_\mathrm{e} + *k \cdot *n_\mathrm{e}$, | in units/μframe |
| $k$ | $k \ge 1$ | *complete* run count to cover the most of it, | in rounds/μframe |
| $*k$ | $0 \le *k \le 1$ | *partial* run count to cover the rest of it, [if any] | in rounds/μframe |
| $n_\mathrm{e}$ | $n_\mathrm{e} \ge 1$ | scope of a *complete* echo multiplexing round, | in units/round |
| $*n_\mathrm{e}$ | $0 \le *n_\mathrm{e} < n_\mathrm{e}$ | scope of a *partial* echo multiplexing round, [if any] | in units/round |
| $v$ | fixed to 2,080 | information *volume* of the payload, $v = 32 \times 65$, | in bits/μframe |
| $s$ | $s \ge 0$ | its *spare* part, if present, $(*k \cdot *t +)\ k \cdot t + s = v$, | in bits/μframe |
| $^{(*)}t$ | $t = b + r$ | *total* amount of information handled per a run, | in bits/round |
| $^{(*)}b$ | maximize | its part that results simply from *bypass*, $b \ge 0$, | in bits/round |
| $^{(*)}r$ | minimize | its part that results from *representation*, $r \ge 1$, | in bits/round |
| $C_r$ | $C_r = r \ln 2$ | (log) transport *capacity* of the represented part, | in nepers |
| $M_r$ | $M_r = n_\mathrm{e} \ln N - b \ln 2$ | (log) data *modulus* represented per a run, | in nepers |

A manuscript of this work was submitted to IEEE Communications Letters December 26, 2022 and rejected for fair reasons during its peer review.

Please sorry for the author has no time to find this work a new home, peer reviewed or not, except of arXiv, and just hopes there it meets its reader, one or maybe various, whom the author beforehand thanks for their regard.

A. Ivanov is with JSC Continuum, Yaroslavl, the Russian Federation.

[1]Forward Error Correction, the means right under the 10GBASE-KR PCS, intended to guard on its payload, implements the Reed-Solomon (66, 65, $2^5$) systematic code after the 64b/66b to 65b transcoder, see [1], Clause 74.

[2]Physical Coding Sublayer, the highest sublayer of the entire 10GBASE-KR physical layer (PHY), implements the 10GBASE-R PCS, indirectly stimulates on the 10GBASE-KR PMD, see [1], Clauses 49 and 72, respectively.

[3]For the purpose of this paper, we assume for the PCS has a direct access to the FEC input, in an abstract form of some FEC service interface, completely bypassing the 64b/66b to 65b transcoder in the mentioned FEC.

[4]Considering almost any PCS, we distinguish the two speeds (or bit rates), accommodation and transportation, actual for the PCS service interface on the top and the PMA service interface on the bottom, respectively, of this PCS, that are different. Note, the PCS service interface implements the overall PHY service interface, if exposed, in a form of some appropriate MII, and its speed, estimated assuming pure data transmission in one selected direction, transmit or receive, is the accommodation speed we tell about. The transportation speed is addressed similarly. By default, mentioning a speed with its scope omitted assumes for the accommodation speed as it is defined above.

[5]In the case of 10GBASE-KR, considered in this paper, the accommodation speed or accommodation bit rate is exact 10 billion bits per every second, or shortly, 10 Gbps, while the transportation speed or transportation bit rate is $2{,}112/2{,}048 = 66/64 = 33/32$ times higher, providing for $33/32 \times 10 = 10.3125$ Gbps. For the comparison, the symbol rate or baud rate, observed at the 10GBASE-KR MDI, is 10.3125 billion NRZI symbols per every second, shortly 10.3125 Gbaud, i.e., $32/33 = 0.(96) < 1$ accommodation bits per a symbol and exact $33/33 = 1$ transportation bit per every baud, both rated, respectively. And, of course, all these bits are information bits.

[6]We construct the generalized terminology for the new purpose, but mainly based on the terminology provided in [2]'s Table II and [3]'s Table II.

TABLE III
CURSORY SEARCH RESULTS

| $n_e$ | $N$ | $C_r : M_r$ | $t$ = | $b$ + | $r$ | ALU | $k$ | $s$ |
|---|---|---|---|---|---|---|---|---|
| 8 | 279 | 1.00011 | 65 | — | 65 | U128 | 32 | — |
| 16 | 264 | 1.00359 | 129 | 48 | 81 | U128 | 16 | 16 |
| 32 | 260 | 1.00147 | 257 | 64 | 193 | U256 | 8 | 24 |
| 64 | 258 | 1.00063 | 513 | 64 | 449 | U512 | 4 | 28 |
| 128 | 257 | 1.00027 | 1025 | — | 1025 | >U1024 | 2 | 30 |
| 256 | 279 | 1.00011 | 2080 | — | 2080 | >U1024 | 1 | — |

TABLE IV
MAIN SEARCH TASK

| Given | Condition | Dependants | Objective | Find |
|---|---|---|---|---|
| $N$ | $t = 8n_e + 1$ | $N_r$, $N_b$, $b$, $r$, ALU, $v : t$ | $n_e \longrightarrow n_{e,max}$ | $n_{e,max}$ |

TABLE V
MAIN SEARCH FLOW

| $N$ = | $N_r$ × | $N_b$ | $8n_{e,max} + 1$ = | $b$ + | $r$ | ALU | $v : t$ @ | $n_{e,max}$ |
|---|---|---|---|---|---|---|---|---|
| 279 | 279 | 1 | 64 + 1 | — | 65 | U128 | =32 | 8 |
| 278 | 139 | 2 | 64 + 1 | 8 | 57 | U64 | =32 | 8 |
| 277 | 277 | 1 | 64 + 1 | — | 65 | U128 | =32 | 8 |
| 276 | 69 | 4 | 72 + 1 | 18 | 55 | U64 | <29 | 9 |
| 275 | 275 | 1 | 72 + 1 | — | 73 | U128 | <29 | 9 |
| 274 | 137 | 2 | 80 + 1 | 10 | 71 | U128 | <26 | 10 |
| 273 | 273 | 1 | 80 + 1 | — | 81 | U128 | <26 | 10 |
| 272 = | 17 × | 16 | $t$ = 88 + 1 = | 44 + | 45 | U64 | <24 @ | 11 |
| 271 | 271 | 1 | 96 + 1 | — | 97 | U128 | <22 | 12 |
| 270 | 135 | 2 | 104 + 1 | 13 | 92 | U128 | <20 | 13 |
| 269 | 269 | 1 | 104 + 1 | — | 105 | U128 | <20 | 13 |
| 268 | 67 | 4 | 120 + 1 | 30 | 91 | U128 | <18 | 15 |
| 267 | 267 | 1 | 128 + 1 | — | 129 | U256 | <17 | 16 |
| 266 | 133 | 2 | 144 + 1 | 18 | 127 | U128 | <15 | 18 |
| 265 | 265 | 1 | 160 + 1 | — | 161 | U256 | <13 | 20 |
| 264 | 33 | 8 | 176 + 1 | 66 | 111 | U128 | <12 | 22 |
| 263 | 263 | 1 | 200 + 1 | — | 201 | U256 | <11 | 25 |
| 262 | 131 | 2 | 232 + 1 | 29 | 204 | U256 | <9 | 29 |
| 261 | 261 | 1 | 280 + 1 | — | 281 | U512 | <8 | 35 |
| 260 | 65 | 4 | 352 + 1 | 88 | 265 | U512 | <6 | 44 |
| 259 | 259 | 1 | 472 + 1 | — | 473 | U512 | <5 | 59 |
| 258 | 129 | 2 | 712 + 1 | 89 | 624 | U1024 | <3 | 89 |
| 257 | 257 | 1 | 1416 + 1 | — | 1417 | >U1024 | <2 | 177 |

TABLE VI
SELECTED SOLUTION

| $N$ | $s$ | Run Type | $t$ = | $b$ + | $r$ | $k \cdot n_e$ ($k \cdot t$) | $k$ | $n_e$ |
|---|---|---|---|---|---|---|---|---|
| 272 | — | complete | 89 | 44 | 45 | 253 (2,047) | 23 | 11 |
| 17 | 16 | partial, $i \mid f$ | 33 | 16 \| 12 | 17 \| 21 | $4 \mid 3\frac{2}{2}$ (33) | 1 | 3+1 |
| $N_r$ × | $N_b$ | integral approach \| fractional approach | $*t$ = | $*b$ + | $*r$ | $*k \cdot *n_e$ (...) | $*k$ | $*n_e$ |

TABLE VII
TRANSFER UNIT ENCODING

| Type | ↓Transfer Unit Content, Bits→ | $u_\varepsilon$ | $u_\delta$ | $u_\gamma$ | $u_\beta$ | $u_\alpha$ | $u_3$ | $u_2$ | $u_1$ | $u_0$ | Variety |
|---|---|---|---|---|---|---|---|---|---|---|---|
| "zero" | intended to fill up a partial run | 0 | 0 | 0 | 0 | 0 | 0 | 0 | 0 | 0 | — |
| DATA | one of $2^8$ possible data octets | 0 | $d_7$ | $d_6$ | $d_5$ | $d_4$ | $d_3$ | $d_2$ | $d_1$ | $d_0$ | 16 × 16 |
| CTRL | one of $2^3$ possible control codes | 1 | 0 | 0 | 0 | 0 | 0 | $c_2$ | $c_1$ | $c_0$ | $^1/_2$ × 16 |
| event | one of $2^3$ relative positions [in OT] | 1 | 0 | 0 | 0 | 0 | 1 | $p_2$ | $p_1$ | $p_0$ | $^1/_2$ × 16 |
| NOTE – At 10 Gb/s, octet time [OT] is 800 ps. | | Root | | | | | Affix | | | | Σ 17 × 16 |

TABLE VIII
EVENT FIXATION SCALE

| $N_{event}$ | 1=$2^0$ | 2=$2^1$ | 3 | 4=$2^2$ | 5 | 6 | 7 | 8=$2^3$ | 9 | 10 | Unit |
|---|---|---|---|---|---|---|---|---|---|---|---|
| $f$ | 1.25 | 2.5 | 3.75 | 5 | 6.25 | 7.5 | 8.75 | 10 | 11.25 | 12.5 | GHz |
| $1/f$ | 800.0 | 400.0 | 266.(6) | 200.0 | 160.0 | 133.(3) | ~114.3 | 100.0 | 88.(8) | 80.0 | ps |

to answer the question what would be when $257 \le N \le 279$, first designating for its task and then logging on all its flow, i.e., output, see Tables IV and V, respectively.

Among the so logged output, we find the transfer unit with the data modulus of $N = 272 = 17 \times 16 = N_r \times N_b$ the most interesting, because it is balanced the most possible as well as the most optimal manner: it features the minimum difference between its root and its affix portions, i.e., $N_r = N_b + 1$, as well as the same number of the root (dependent) bits after the representation and of the affix (independent) bits anytime, per the unit, i.e., $\lfloor \log_2 N_r \rfloor = \log_2 N_b = 4$, respectively.

Having the above done, we make our choice for the variant with $N_r \times N_b = 17 \times 16 = 272$, $n_e \times k = 11 \times 23 = 253$, and $*n_e \times *k = 4 \times 1 = 4$, see Table VI, that provides with $*n_p = n_e \times k + *n_e \times *k = 257 = 256 + 1 = n_p + 1$, i.e., one spare transfer unit—single and side but extra and entire—over the original duty, per the microframe. Based on this, we then define how the coding means must encode the content of each of all the transfer units, including of the spare, met before the representation of their root bits, in the stream, see Table VII, and estimate forth what, in the view of synchronization, such encoding can support for, see Table VIII.

Considering the spare transfer unit either like a regular one (integer 1) or as a pair of halves (fraction $\frac{2}{2}$), we can construct the coding means the two accordant ways, see Table VI again, we further refer to as the integral approach and the fractional approach, respectively, so we immediately do.

## INTEGRAL APPROACH

Enabling this approach, we consider the spare transfer unit like any regular one that consists of $\log_2 N_b = 4$ independent (affix) bits, anytime, as well as of $\lceil \log_2 N_r \rceil = 5$ and then of $\lfloor \log_2 N_r \rfloor = 4$ dependent (root) bits, before and then after the representation, respectively. So, we point on that behavior and account for the spare unit as an integral quantity, 1, when we express the target scope, $*n_p$, see Table VI.

By this approach, inserting an event causes for a new single transfer unit appears into the stream—that new unit splits the stream in the position associated with the respective unit time period, during which the event occurs, and then fulfils the so exempted space by itself—and can be performed just once but per each microframe time[7] period, see Table IX.

Being selected, this approach improves on the multi-stage $(2 \times 2 + 3)$ coding scheme described in [2]. Except stages 4 and 5, all the earlier stages of the coding scheme we consider in this paper, are similar in their implementation to the respective stages of the coding scheme described in [2], therefore we do

[7] In the case of 10GBASE-KR, bit time (BT) is 100 ps, octet time (OT) is $8 \times \mathrm{BT} = 800$ ps, and microframe time ($\mu$FT) is $n_\mathrm{p} \times \mathrm{OT} = 204.8$ ns, with 1 spare unit per $\mu$FT, it gives up to $1/\mu\mathrm{FT} = 4{,}882{,}812.5$ events/s.

TABLE IX
EVENT INSERTION—INTEGRAL APPROACH

| Scenario | Ref. | Before Insertion | | | After Insertion | | | | Delta |
|---|---|---|---|---|---|---|---|---|---|
| event ↓ data | Root | ··· | $d_H$ | ··· | ··· | ⊠ | $d_H$ | ··· | +1 |
| | Affix | ··· | $d_L$ | ··· | ··· | e | $d_L$ | ··· | +1 |
| event ↓ control | Root | ··· | ⊠ | ··· | ··· | ⊠ | ⊠ | ··· | +1 |
| | Affix | ··· | c | ··· | ··· | e | c | ··· | +1 |
| Unit Period, $m$ → | | $m-1$ | $m$ | $m+1$ | $m-1$ | $m$ | $m^{bis}$ | $m+1$ | ··· |

TABLE XII
EVENT INSERTION—FRACTIONAL APPROACH

| Scenario | Before Insertion | | | After Insertion | | | | | Delta |
|---|---|---|---|---|---|---|---|---|---|
| event ↓ data | ··· | $d_H$ | ··· | ··· | ⊠ | $d_H$ | $d_L$ | ··· | +2 |
| | ··· | $d_L$ | ··· | ··· | e | | | ··· | — |
| event ↓ control | ··· | ⊠ | ··· | ··· | ⊠ | ⊠ | c | ··· | +2 |
| | ··· | c | ··· | ··· | e | | | ··· | — |
| $m$ → | $m-1$ | $m$ | $m+1$ | $m-1$ | $m$ | $m^{bis}$ | $m^{ter}$ | $m+1$ | ··· |

TABLE X
STAGE 4—INTEGRAL

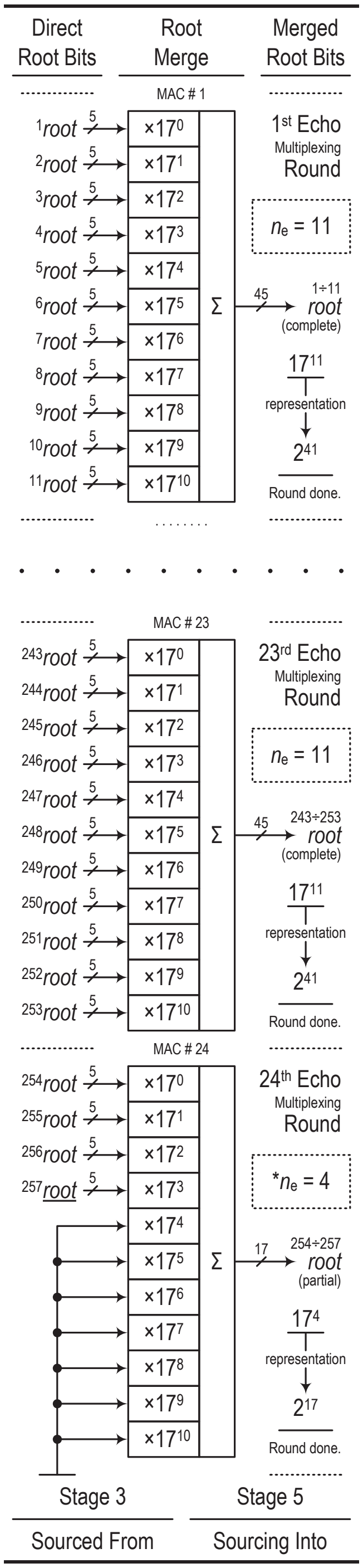

TABLE XI
STAGE 5—INTEGRAL

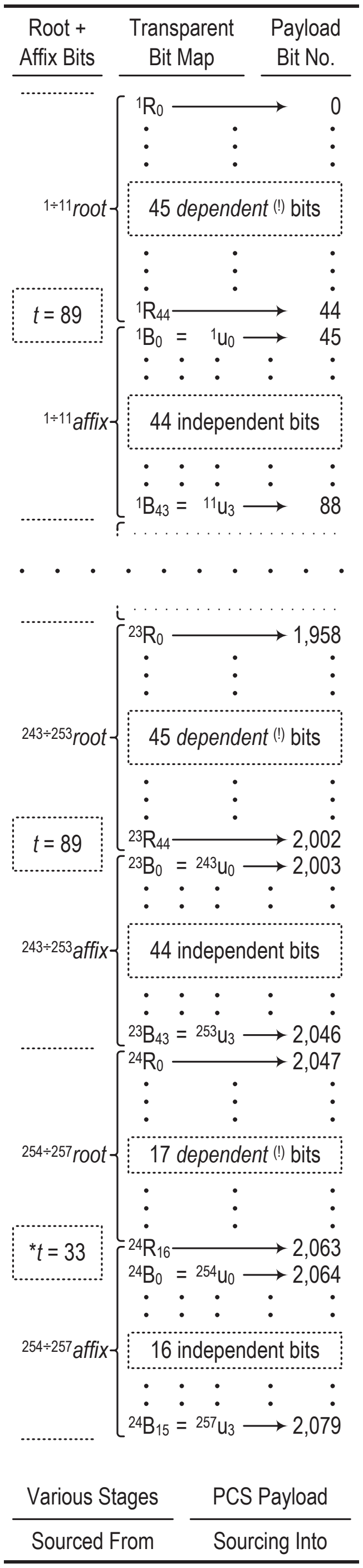

TABLE XIII
STAGE 4—FRACTIONAL

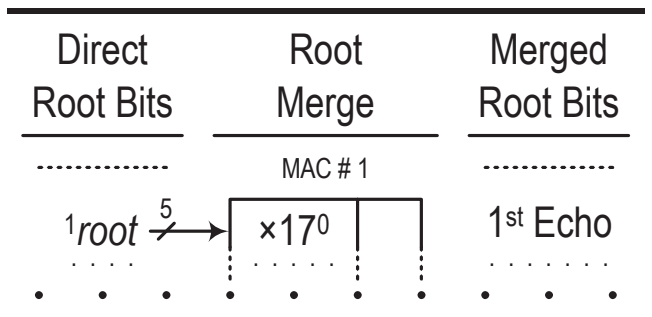

TABLE XIV
STAGE 5—FRACTIONAL

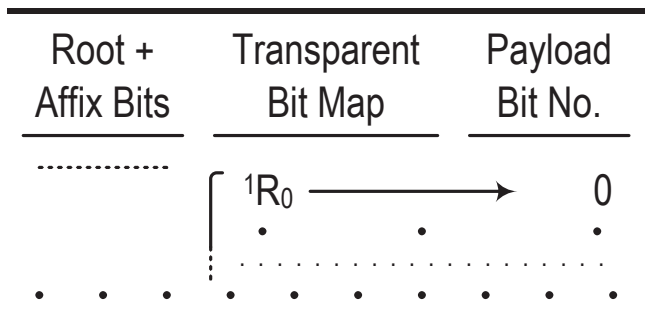

not focus on them. Regarding the rest stages, i.e., stages 4 and 5, the coding means performs as the following.

On stage 4, see Table X, it merges the root bits of $*n_\mathrm{p} = n_\mathrm{p} + 1 = 256 + 1 = 257$ transfer units, regular (256) plus spare (1), by $n_\mathrm{e} = 11$ (all regular) in each of $k = 23$ complete multiplexing rounds then $*n_\mathrm{e} = 4$ (3 regular and 1 spare) in the sole (of $*k = 1$) partial multiplexing round, representing $11 \times 23 = 253$ roots (ordered 1 to 253) then $3 + 1 = 4$ roots (ordered 254 to 257), respectively.

On stage 5, see Table XI, it maps the merged and bypassed bits, $r \times k + *r \times *k = 45 \times 23 + 17 \times 1 = 1{,}052$ (ref.) and $b \times k + *b \times *k = 44 \times 23 + 16 \times 1 = 1{,}028$ (ref.), respectively, or $1{,}052 + 1{,}028 = 2{,}047 + 33 + 0 = 89 \times 23 + 33 \times 1 + 0 = t \times k + *t \times *k + s = v =$ exact 2,080 sent bits, aggregately, onto the designated bits of the microframe payload.

Advantages of this approach lie in its clarity and naturality, especially in the view of the succession of the approaches described in [2] and [3]. It also ensures for the full multiplexing

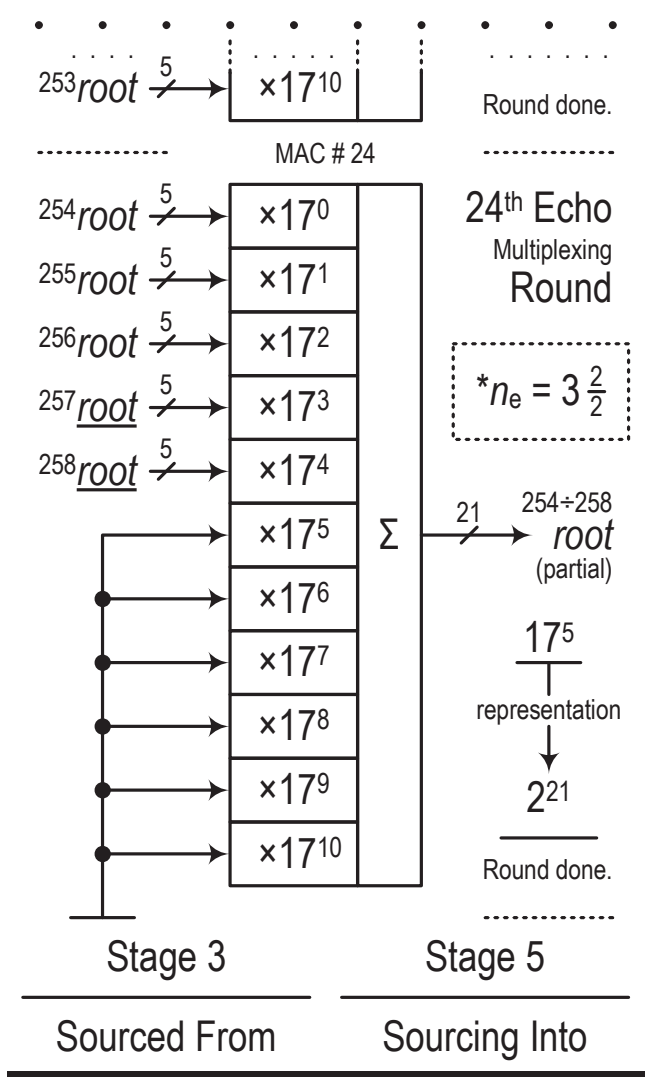

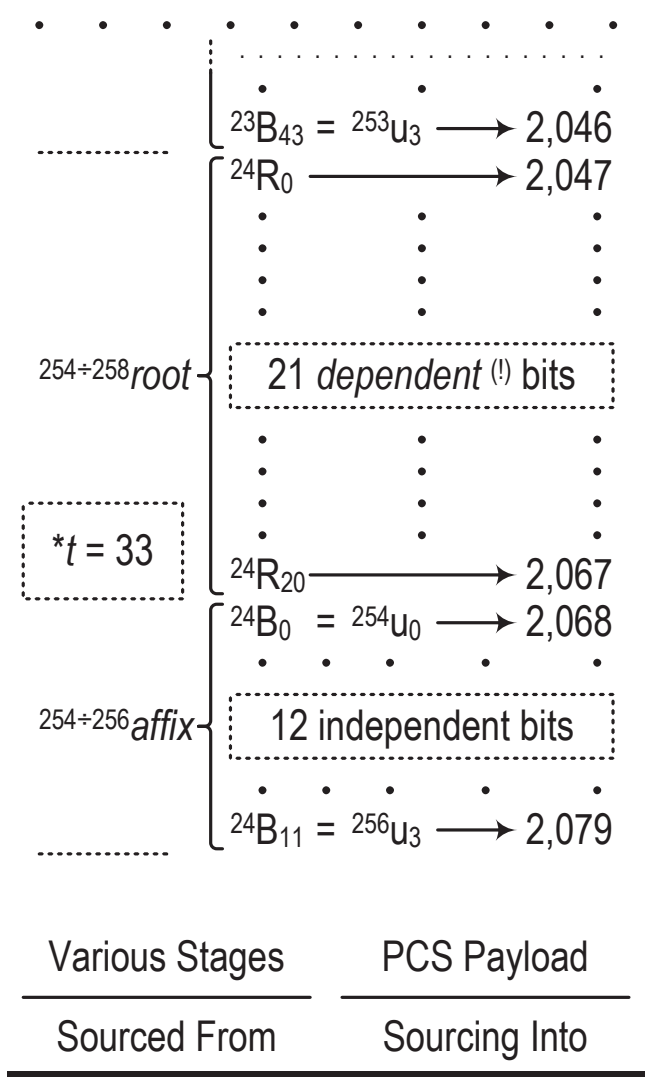

TABLE XV
EXAMPLE APPLICABILITY INCLUDING BACKWARD

| μFrame | $n_p$ | → | $*n_p$ | = | $n_e$ | × | $k$ | + | $*n_e$ | × | $*k$ | $N$ | = | $N_r$ | × | $N_b$ | = | $N_{data}$ | + | $N_{extra}$ | $N_{event}$ | $s$ | $t$ | = | $b$ | + | $r$ | $w$ | Comment |
|---|---|---|---|---|---|---|---|---|---|---|---|---|---|---|---|---|---|---|---|---|---|---|---|---|---|---|---|---|---|
| 10GBASE-KR | 256 | → | $256\frac{2}{2}$ | = | 11 | × | 23 | + | $3\frac{2}{2}$ | × | 1 | 272 | = | 17 | × | 16 | = | 256 | + | 16 | ≤15 | — | 89 | = | 44 | + | 45 | ≤64 | up to 1 ev/μframe |
| 10GBASE-KR | 256 | → | 257 | = | 11 | × | 23 | + | 4 | × | 1 | 272 | = | 17 | × | 16 | = | 256 | + | 16 | ≤15 | — | 89 | = | 44 | + | 45 | ≤64 | up to 1 ev/μframe |
| 10GBASE-KR | 256 | → | $256\frac{4}{2}$ | = | 22 | × | 11 | + | $14\frac{4}{2}$ | × | 1 | 264 | = | 33 | × | 8 | = | 256 | + | 8 | ≤7 | — | 177 | = | 66 | + | 111 | ≤128 | up to 2 ev/μframe |
| 10GBASE-KR | 256 | → | 258 | = | 22 | × | 11 | + | 16 | × | 1 | 264 | = | 33 | × | 8 | = | 256 | + | 8 | ≤7 | 4 | 177 | = | 66 | + | 111 | ≤128 | up to 2 ev/μframe |
| 10GBASE-KR | 256 | → | $256\frac{4}{2}$ | = | 33 | × | 7 | + | $25\frac{4}{2}$ | × | 1 | 260 | = | 65 | × | 4 | = | 256 | + | 4 | ≤3 | — | 265 | = | 66 | + | 199 | ≤256 | up to 2 ev/μframe |
| 10GBASE-KR | 256 | → | 259 | = | 37 | × | 7 | | — | | — | 260 | = | 65 | × | 4 | = | 256 | + | 4 | ≤3 | 1 | 297 | = | 74 | + | 223 | ≤256 | up to 3 ev/μframe |
| 10GBASE-T | 400 | → | $400\frac{2}{2}$ | = | 10 | × | 40 | + | $0\frac{2}{2}$ | | — | 272 | = | 17 | × | 16 | = | 256 | + | 16 | ≤15 | — | 81 | = | 40 | + | 41 | ≤64 | up to 1 ev/μframe |
| 10GBASE-T | 400 | → | 401 | = | 10 | × | 40 | + | 1 | | — | 272 | = | 17 | × | 16 | = | 256 | + | 16 | ≤15 | 1 | 81 | = | 40 | + | 41 | ≤64 | up to 1 ev/μframe |
| 10GBASE-T | 400 | → | $400\frac{3}{2}$ | = | 12 | × | 33 | + | $4\frac{3}{2}$ | × | 1 | 264 | = | 33 | × | 8 | = | 256 | + | 8 | ≤7 | 1 | 97 | = | 36 | + | 61 | ≤64 | up to $\frac{3}{2}$ ev/μframe |
| 10GBASE-T | 400 | → | 403 | = | 16 | × | 25 | + | 3 | × | 1 | 264 | = | 33 | × | 8 | = | 256 | + | 8 | ≤7 | — | 129 | = | 48 | + | 81 | ≤128 | up to 3 ev/μframe |
| 10GBASE-T | 400 | → | $400\frac{6}{2}$ | = | 29 | × | 13 | + | $23\frac{6}{2}$ | × | 1 | 260 | = | 65 | × | 4 | = | 256 | + | 4 | ≤3 | — | 233 | = | 58 | + | 175 | ≤256 | up to 3 ev/μframe |
| 10GBASE-T | 400 | → | 405 | = | 41 | × | 9 | + | 36 | × | 1 | 260 | = | 65 | × | 4 | = | 256 | + | 4 | ≤3 | — | 329 | = | 82 | + | 247 | ≤256 | up to 5 ev/μframe |
| 1000BASE-T1 | 450 | → | $450\frac{1}{2}$ | = | 11 | × | 40 | + | $10\frac{1}{2}$ | × | 1 | 272 | = | 17 | × | 16 | = | 256 | + | 16 | ≤15 | — | 89 | = | 44 | + | 45 | ≤64 | up to $\frac{1}{2}$ ev/μframe |
| 1000BASE-T1 | 450 | → | 451 | = | 13 | × | 34 | + | 9 | × | 1 | 268 | = | 67 | × | 4 | = | 256 | + | 8+4 | ≤8+3 | 2 | 105 | = | 26 | + | 79 | ≤128 | up to 1 ev/μframe |
| 1000BASE-T1 | 450 | → | 452 | = | 18 | × | 25 | + | 2 | | — | 266 | = | 133 | × | 2 | = | 256 | + | 8+2 | ≤8+1 | 2 | 145 | = | 18 | + | 127 | ≤128 | up to 2 ev/μframe |
| 1000BASE-T1 | 450 | → | $450\frac{4}{2}$ | = | 16 | × | 28 | + | $2\frac{4}{2}$ | × | 1 | 264 | = | 33 | × | 8 | = | 256 | + | 8 | ≤7 | 1 | 129 | = | 48 | + | 81 | ≤128 | up to 2 ev/μframe |
| 1000BASE-T1 | 450 | → | 453 | = | 22 | × | 20 | + | 13 | × | 1 | 264 | = | 33 | × | 8 | = | 256 | + | 8 | ≤7 | — | 177 | = | 66 | + | 111 | ≤128 | up to 3 ev/μframe |
| 1000BASE-T1 | 450 | → | $450\frac{5}{2}$ | = | 31 | × | 14 | + | $16\frac{5}{2}$ | × | 1 | 260 | = | 65 | × | 4 | = | 256 | + | 4 | ≤3 | — | 249 | = | 62 | + | 187 | ≤256 | up to $\frac{5}{2}$ ev/μframe |
| 1000BASE-T1 | 450 | → | 454 | = | 35 | × | 12 | + | 34 | × | 1 | 260 | = | 65 | × | 4 | = | 256 | + | 4 | ≤3 | — | 281 | = | 70 | + | 211 | ≤256 | up to 4 ev/μframe |

process takes place inside the microframe, i.e., no echo passes across the microframe boundaries.

Disadvantages of this approach derive from the complexity of its stage splitting the stream to insert an event, because it necessitates for (avalanche-like) substitution of all the bits of the transfer unit, i.e., $\lceil \log_2 N_r \rceil + \log_2 N_b = 5 + 4 = 9$ bits, root plus affix, per each unit in the stream.

## Fractional Approach

Enabling this approach, we consider the spare transfer unit as a pair of "twin" halves, each consisting of five and then of four dependent bits, before and then after the representation, respectively, i.e., both treated as the root bits. Thence, we point on that behavior and account for the spare unit as a fractional quantity, $\frac{2}{2}$, in the target scope, $*n_\text{p}$, see Table VI.

By this approach, inserting an event causes for a two new roots appear in the stream[8]—one for the substituted root plus one for the substituted affix, since the latter fits into, too—and also can be performed at the same[7] rate, see Table XII.

Being compared, this and the integral approaches share the earlier stages and differ clear in stages 4 and 5, because now the coding means performs as the following.

On stage 4, see Table XIII, it merges the root bits of $n_\text{p} + 2 \cdot 1 = 256 + 2 = 258$ roots of $*n_\text{p} = 256\frac{2}{2}$ transfer units, regular (256) plus spare ($\frac{2}{2} = 1$), by $n_\text{e} = 11$ (all regular) in each of $k = 23$ complete multiplexing rounds then $*n_\text{e} = 3\frac{2}{2}$ (3 regular and $\frac{2}{2} = 1$ spare) in the partial multiplexing round, now representing $11 \times 23 = 253$ roots (ordered 1 to 253) then $3 + 2 = 5$ roots (ordered 254 to 258), respectively.

On stage 5, see Table XIV, it maps the merged and bypassed bits, $r \times k + *r \times *k = 45 \times 23 + 21 \times 1 = 1{,}056$ $(+4)$ and $b \times k + *b \times *k = 44 \times 23 + 12 \times 1 = 1{,}024$ $(-4)$, respectively, onto the designated bits of the microframe payload.

Disadvantages of this approach are opposite to the advantages of the integral approach, as they were mentioned above, except the behavior when the echo can propagate across the microframe boundaries, that we assume actually giving agility and thus may be useful, see Table XV.

Advantages of this approach are opposite to the disadvantages of the integral approach, as they were mentioned above, too, because now it necessitates for substitution of just the root (and no affix) bits of the transfer unit, i.e., $\lceil \log_2 N_r \rceil = 5$ bits, root only,[9] per each unit in the stream.

## Conclusion

In this paper, we considered a way to embed a new extra ability—in a form of an event transfer means intended to solve a precise synchronization problem—into the PCS of a given physical layer featuring a very long binary transport word, we call the microframe, and the same time preserve the original duty of the data transfer ability of the given layer.

We also proposed the two concepts, we call the integral and fractional approaches, both based on the linguistic multiplexing model and intended to implement, at least in theory, the coding means suitable for 10GBASE-KR, we focus on in this paper, as well as 10GBASE-T and 1000BASE-T1, considered earlier in [2] and [3], respectively, see Table XV.

[8] Essentially, we divide it into the two streams: affix-related and *root*-related, never affected and *affected* each time an event occurs, respectively. While in the integral approach, it is unit-related and thus affected entirely.

[9] It is fair for any variant with $N_r > N_b$, while for a variant with $N_r < N_b$, we need to consider the two streams vise versa: root-related and *affix*-related, never affected and *affected* each time an event occurs, respectively.

# Data Coding Means and Event Coding Means Multiplexed Over the MultiGBASE-T1 PCS Payload

Alexander Ivanov

***Abstract*—We proceed to consider existing protocols featuring a very long binary transport word, we refer to as the microframe, and the fresh point we currently stand in is the MGBASE-T1 or MultiGBASE-T1 family (scalable) applying to the 2.5GBASE-T1 (has a scale of 1/4), 5GBASE-T1 (scale of 1/2), and 10GBASE-T1 (scale of 1/1, maximum and reference) physical layers.**

***Index Terms*—Ethernet, linguistic multiplexing, multiplexing, precise synchronization, synchronization, 2.5G/5G/10GBASE-T1, MGBASE-T1, MultiGBASE-T1.**

## Introduction

MULTI-GIGABIT Ethernet, type 2.5GBASE-T1 ("quarter speed"), type 5GBASE-T1 ("half speed"), and type 10GBASE-T1 ("full speed"), collectively referenced to as the MultiGBASE-T1 (or MGBASE-T1) family[1] [1], introduce the respective physical layers all intended to operate over a single balanced pair of conductors in a harsh environment, primarily but expectedly not limited to, like of automotive.

Regardless of its respective scale defining its speed as a part of the reference (and maximum) speed, 10 Gbps, any member of this family implements the same microframe, the structure that is intended to convey $50 \times 8 = 400$ quasi-separate transfer units, collected in blocks of $8$ per each, see Table I, and that predefines the original duty of the considered physical layer, as the initial scope of its data passing ability.

In this paper, we describe a way to re-discover, re-perceive, and re-employ, if seen, the excessiveness of the informational redundancy pre-established since the initial design inside the microframe, after that, apply the linguistic multiplexing model, propose the coding means capable to perform such the multiplexing, and finally refine the considered physical layer, trying to embed an event passing ability into it.

TABLE I
ORIGINAL MICROFRAME STRUCTURE

| Hierarchy and Elements | Contents, Comments | Size, bits |
|---|---|---|
| Microframe (or μframe) entirely | rate = Scale×3.125M μframes/s | Σ 3,600 |
| ├ Payload, visible at XGMII | max 400 data octets → 3,200 bits | 50 × 65 = 3,250 |
| ├ 64B/65B coded block #1 | (first) 8 data octets or controls | 8×8+1= 65 |
| . . . . . . . . . . . . . . . . . . . . | . . . . . . . . . . . . . . . . . . . . . . . | . . . |
| └ 64B/65B coded block #50 | (last) 8 data octets or controls | 8×8+1= 65 |
| ├ OAM bits, hidden by PHY | purposed for link management | 10 |
| └ FEC parity, hidden by PHY | calculated over all the bits above | 340 |

NOTE – Scale is 1. / .5 / .25 for 10G / 5G / 2.5GBASE-T1 PHY layer, respectively.

TABLE II
ADDITIONAL TERMINOLOGY

| Ref. | Objective | Definition, Description, Designation | Measured |
|---|---|---|---|
| $\xi$ | $t = 8n_e + \xi$ | *extra* with $t$ compared to pure data octets, $\xi \geq 1$, | in bits/round |
| $g$ | $g \geq 1$ | *gathering* ("linking") factor, defines how many transfer units are considered ("linked") together | in units/unit |

TABLE III
HEAD TRANSFER UNIT VARIANTS

| $N$ | = | $N_r$ | × | $N_b$ | $\xi$ | $8n_{e,max} + \xi$ | = | $b$ | + | $r$ | ALU | @$n_{e,max}$ |
|---|---|---|---|---|---|---|---|---|---|---|---|---|
| 257 | | 257 | > | 1 | 1 | 1,416 + 1 | | — | | 1,417 | >U1024 | 177 |
| 258 | | 129 | > | 2 | 1 | 712 + 1 | | 89 | | 624 | U1024 | 89 |
| 260 | | 65 | > | 4 | 1 | 352 + 1 | | 88 | | 265 | U1024 | 44 |
| 264 | | 33 | > | 8 | 1 | 176 + 1 | | 66 | | 111 | U256 | 22 |
| 272 | | 17 | > | 16 | 1 | 88 + 1 | | 44 | | 45 | U64 | 11 |
| 288 | | 9 | < | 32 | 1 | 40 + 1 | | 25 | | 16 | U16 | 5 |
| | | | | | 2 | 88 + 2 | | 55 | | 35 | U64 | 11 |
| 320 | | 5 | < | 64 | 1 | 24 + 1 | | 18 | | 7 | U8 | 3 |
| | | | | | 2 | 48 + 2 | | 36 | | 14 | U16 | 6 |
| 384 | | 3 | < | 128 | 1 | 8 + 1 | | 9 | | — | — | 1 |
| | | | | | 2 | 24 + 2 | | 21 | | 5 | U8 | 3 |
| | | | | | 3 | 40 + 3 | | 35 | | 8 | U8 | 5 |

## Transfer Unit Gathering

Succeeding the work [2] in the way and in the terminology, we develop its idea further the following manner.

We gather a number, $g \geq 2$, of consecutive transfer units in the stream into what that we will further refer to as a transfer span. In the general case, when $g > 2$, the earliest transfer unit in a transfer span is its head unit,[2] that resides in the position $m$ in the stream, while the rest $g - 1 \geq 1$ transfer units are its body units, they reside in the $g - 1$ positions next to the one of the head, $m + 1$ to $m + g - 1$, the latest of those may also be called its tail unit, that resides in the position $m + g - 1$, respectively. In the least case, when $g = 2$, we distinguish just the head unit and the tail unit, in a transfer span.

A manuscript of this work was submitted to IEEE Communications Letters December 26, 2022 and rejected for fair reasons during its peer review.

Please sorry for the author has no time to find this work a new home, peer reviewed or not, except of arXiv, and just hopes there it meets its reader, one or maybe various, whom the author beforehand thanks for their regard.

A. Ivanov is with JSC Continuum, Yaroslavl, the Russian Federation.

[1]The following are the major characteristics of the members within it:

| | | | |
|---|---|---|---|
| • MGBASE-T1 protocol = | 10GBASE-T1 • | 5GBASE-T1 • | 2.5GBASE-T1 • |
| • Accommodation speed = | [ref] 10 Gbps • | 5 Gbps • | 2.5 Gbps • |
| • Speed scale factor, S = | $1/1 = 1$ • | $1/2 = 0.5$ • | $1/4 = 0.25$ • |
| • Bit time, BT = | $100/\mathrm{S} = 100$ ps • | 200 ps • | 400 ps • |
| • Octet time, OT = | $8 \times \mathrm{BT} = 0.8$ ns • | 1.6 ns • | 3.2 ns • |
| • $\mu$Frame time = | $n_\mathrm{p} \times \mathrm{OT} = 0.32\ \mu s$ • | $0.64\ \mu s$ • | $1.28\ \mu s$ • |

[2]Because in the scope of a transfer span of $g$ units, $m$ advances by $g$, i.e., $\Delta m = g$, the position $m$ points both on the $m$-th span as well as on its head unit in the stream, unambiguously and simultaneously.

TABLE IV
STANDARD GATHERING APPROACH

| ↓ Span Content, Span Bits → | $^{(m)}u_\alpha$ | $^{(m)}u_\beta$ | $^{(m)}u_\gamma$ | $^{(m)}u_{\delta\|2}$ | $^{(m)}u_{\varepsilon\|3}$ | $^{(m)}u_4$ | $^{(m)}u_5$ | $^{(m)}u_6$ | $^{(m)}u_7$ | $^{(m+1)}u_0$ ••• $^{(m+1)}u_7$ | $^{(m+2)}u_0$ • • • $^{(m+6)}u_7$ | $^{(m+7)}u_0$ ••• $^{(m+7)}u_7$ |
|---|---|---|---|---|---|---|---|---|---|---|---|---|
| Eight DATA octets purely | 0 | $^{(m)}b_1$ | $^{(m)}b_2$ | $^{(m)}b_3$ | $^{(m)}b_4$ | $^{(m)}b_5$ | $^{(m)}b_6$ | $^{(m)}b_7$ | $^{(m)}b_8$ | $^{(m)}b_9$ ••• $^{(m)}b_{16}$ | $^{(m)}b_{17}$ • • • $^{(m)}b_{56}$ | $^{(m)}b_{57}$ ••• $^{(m)}b_{64}$ |
| At least one CTRL, $N = 5 \times 64$ | 1 | 0 | 0 | *m*-th 1-of-15 transcoded 6-bit block type field | | | | | | | | |
| At least one CTRL, $N = 9 \times 32$ | 1 | 0 | 0 | 0 | ········ 5-bit ········ | | | | | | | |
| At least one CTRL, $N = 17 \times 16$ | 1 | 0 | 0 | 0 | 0 | ········ 4-bit ········ | | | | | | |
| NOTE – In SGA, $\Delta m = g = 8$. | *m*-th Root Bits (*to be represented*) | | | | | *m*-th Affix Bits (*to be simply bypassed into the respective payload bits*) | | | | | | |

TABLE V
STANDARD GATHERING APPROACH (SGA) EXAMPLE APPLICABILITY INCLUDING BACKWARD

| μFrame | $n_p$ | $v$ | $N_r$ | × | $N_b$ | $g$ | $*n_p$ | = | $n_e$ | × | $k$ | + | $*n_e$ | × | $*k$ | $n_e : g$ | $\xi$ | $t$ | = | $b$ | + | $r$ | $w$ | $s$ | Comment |
|---|---|---|---|---|---|---|---|---|---|---|---|---|---|---|---|---|---|---|---|---|---|---|---|---|---|
| MGBASE-T1, | 400 | 3,250 | 17 | × | 16 | 8 | 400 | = | 88 | × | 4 | + | 48 | × | 1 | 11 | 1 | 705 | = | 660 | + | 45 | ≤64 | 45 | uses 64B/65B |
| 10GBASE-T | | | 17 | × | 16 | 8 | 400 | = | 80 | × | 5 | | — | | — | 10 | 1 | 641 | = | 600 | + | 41 | ≤64 | 45 | coded blocks |
| (as they are similiar) | | | 9 | × | 32 | 8 | 400 | = | 80 | × | 5 | | — | | — | 10 | 2 | 642 | = | 610 | + | 32 | ≤32 | 40 | |
| | | | 9 | × | 32 | 8 | 400 | = | 40 | × | 10 | | — | | — | 5 | 1 | 321 | = | 305 | + | 16 | ≤16 | 40 | |
| | | | 5 | × | 64 | 8 | 400 | = | 24 | × | 16 | + | 16 | × | 1 | 3 | 1 | 193 | = | 186 | + | 7 | ≤8 | 33 | |
| 10GBASE-KR | 256 | 2,080 | 17 | × | 16 | 8 | 256 | = | 80 | × | 3 | + | 16 | × | 1 | 10 | 1 | 641 | = | 600 | + | 41 | ≤64 | 28 | uses 64B/65B |
| | | | 17 | × | 16 | 8 | 256 | = | 64 | × | 4 | | — | | — | 8 | 1 | 513 | = | 480 | + | 33 | ≤64 | 28 | coded blocks |
| | | | 9 | × | 32 | 8 | 256 | = | 40 | × | 6 | + | 16 | × | 1 | 5 | 1 | 321 | = | 305 | + | 16 | ≤16 | 25 | |
| | | | 9 | × | 32 | 8 | 256 | = | 32 | × | 8 | | — | | — | 4 | 1 | 257 | = | 244 | + | 13 | ≤16 | 24 | |
| | | | 5 | × | 64 | 8 | 256 | = | 24 | × | 10 | + | 16 | × | 1 | 3 | 1 | 193 | = | 186 | + | 7 | ≤8 | 21 | |
| 1000BASE-T1 | 450 | 3,645 | 17 | × | 16 | 8 | 452\|448 | = | 56 | × | 8 | + | 4\|0 | | — | 7 | 1 | 449 | = | 420 | + | 29 | ≤32 | 20\|53 | using 64B/65B |
| (if replacing 80B/81B) | | | 9 | × | 32 | 8 | 452\|448 | = | 32 | × | 14 | + | 4\|0 | | — | 4 | 1 | 257 | = | 244 | + | 13 | ≤16 | 14\|47 | coded blocks, |
| | | | 5 | × | 64 | 8 | 452\|448 | = | 16 | × | 28 | + | 4\|0 | | — | 2 | 1 | 129 | = | 124 | + | 5 | ≤8 | 0\|33 | full and halved |

TABLE VI
EXAMPLE LINGUISTIC MULTIPLEXING PROCESS

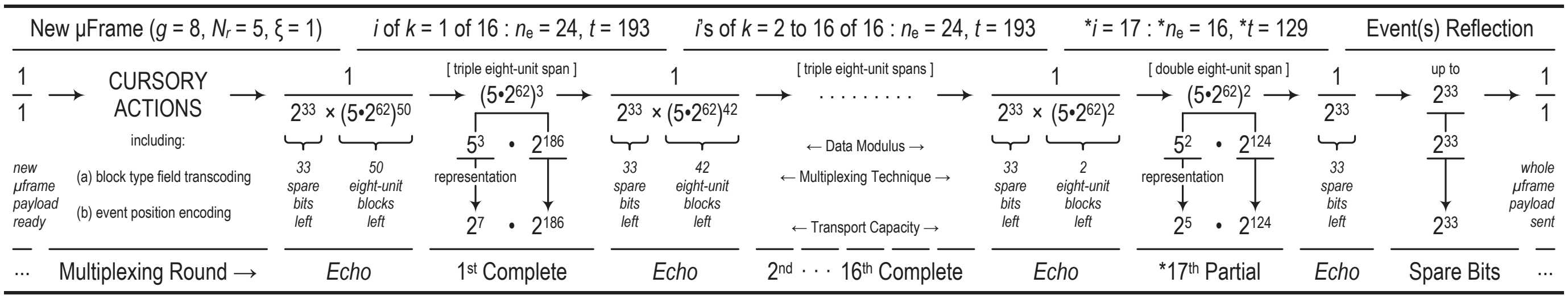

We assume the main goal of such the gathering is to reduce the total number of the transfer units, whose root (dependent) bits do participate in the representation during the underlying linguistic multiplexing process, down onto $\lceil n_\mathrm{p} : g \rceil < n_\mathrm{p}$ units per each microframe time period, all residing in the positions of the head units only, i.e., just a unit per a span, and further, if it's possible, also reduce the number of the mentioned root bits as much as we expect to enable the multiplexing variants with $N_r < N_b$,[3] or leastwise with $N_r = N_b + 1$ as the lower considerable limit, see Tables II and III.[4]

Searching for such variants, we incipiently accept for they may freely implement the spare transfer units [2], the save box cells [3], and/or the spare bits [4].

## STANDARD GATHERING APPROACH

We introduce the so called standard gathering approach, or SGA in short, to address the ability multiplexing problem in deep regard of the interior block encoding, 64B/65B, as well as the exterior service interface, XGMII,[5] both standardly present in the considered physical layer.

With SGA, we look on each 64B/65B coded block consisting of $65 = 1 \times 9 + 7 \times 8$ bits, $b_0$ and $b_1$ to $b_8$ then $b_9$ to $b_{64}$, very straight: as a transfer span consisting of $g = 8 = 1 + 7$ head then body units, by 9 then by 8 bits in the former then in (each of) the latter, respectively. When $b_0 = 0$, all the bits of the block are simply copied into the respective bits of the span. When $b_0 = 1$, the coding means just transcodes the next eight bits[6] of the block, $b_1$ to $b_8$, into the eight bits of the head transfer unit, $u_\beta$ to $u_7$, as well as simply copies all of the rest bits, $b_0$ then $b_9$ to $b_{64}$, too, see Table IV.

Thus, SGA drastically increase the number of the bypassed (affix) bits, $b$, over (v.) the number of the merged (root) ones, $r << b$, simplifying the ALU[7] required to perform the further representation, that makes its (SGA's) application reasonable, suitable, and adaptable, see Table V.

[3]No multiplexing variant with $N_r < N_b$ was enabled in [2], [3], or [4].
[4]Tables II and III success on [2]'s Tables II and V, respectively.
[5]10 Gigabit Media Independent Interface, a duplex parallel highway.
[6]MGBASE-T1 uses 15 out of 16 permitted 8-bit control codes, see [1].
[7]We assume an unsigned arithmetic logic unit, $w$ bits in width, see [2].

TABLE VII
ADVANCED GATHERING APPROACH

| ↓ Span Content, Span Bits → | $^{(m)}u_\alpha$ | $^{(m)}u_\beta$ | $^{(m)}u_{\gamma\vert 1}$ | $^{(m)}u_{\delta\vert 2}$ | $^{(m)}u_{\varepsilon\vert 3}$ | $^{(m)}u_4$ | $^{(m)}u_5$ | $^{(m)}u_6$ | $^{(m)}u_7$ | $^{(m+1)}u_0$ | • • • | $^{(m+g-1)}u_7$ |
|---|---|---|---|---|---|---|---|---|---|---|---|---|
| All ($g$) DATA octets purely | 0 | $^{(m)}d_0$ | $^{(m)}d_1$ | $^{(m)}d_2$ | $^{(m)}d_3$ | $^{(m)}d_4$ | $^{(m)}d_5$ | $^{(m)}d_6$ | $^{(m)}d_7$ | $^{(m+1)}d_0$ | • • • | $^{(m+g-1)}d_7$ |
| At least one CTRL, $N = 3 \times 128$ ‖ per span | 1 | 0 | 1 \| 0 | $^{(m)}p_0 \vert {}^{(\dots)}c_\alpha$ | $^{(m)}p_1 \vert {}^{(\dots)}c_\beta$ | $^{(m)}p_2 \vert {}^{(\dots)}c_\gamma$ | $^{(m)}p_3 \vert {}^{(\dots)}c_\delta$ | $^{(m)}p_4 \vert {}^{(\dots)}c_\varepsilon$ | $^{(m)}p_5 \vert {}^{(\dots)}c_\zeta$ | 8×($g$−1) of $^{(\dots)}$d's and/or $^{(\dots)}$c's | | |
| ‖ per unit | 1 | 0 | $^{(\dots)}x_\alpha$ | $^{(\dots)}x_\beta$ | $^{(\dots)}x_\gamma$ | $^{(\dots)}x_\delta$ | $^{(\dots)}x_\varepsilon$ | $^{(\dots)}x_\zeta$ | $^{(\dots)}x_\eta$ | 8×($g$−1) of $^{(\dots)}$d's and/or $^{(\dots)}$x's | | |
| • • • | • | • | • | • | • | • | • | • | • | • • • | | |
| At least one CTRL, $N = 17 \times 16$ ‖ per span | 1 | 0 | 0 | 0 | 0 | 1 \| 0 | $^{(m)}p_0 \vert {}^{(\dots)}c_\alpha$ | $^{(m)}p_1 \vert {}^{(\dots)}c_\beta$ | $^{(m)}p_2 \vert {}^{(\dots)}c_\gamma$ | 8×($g$−1) of $^{(\dots)}$d's and/or $^{(\dots)}$c's | | |
| ‖ per unit | 1 | 0 | 0 | 0 | 0 | $^{(\dots)}x_\alpha$ | $^{(\dots)}x_\beta$ | $^{(\dots)}x_\gamma$ | $^{(\dots)}x_\delta$ | 8×($g$−1) of $^{(\dots)}$d's and/or $^{(\dots)}$x's | | |
| Event Fixation Scope ↑ | $m$-th Root Bits (*to be arithmetically represented*) | | | | | $m$-th Affix Bits (*to be simply bypassed into the respective payload bits*) | | | | | | |

TABLE VIII
EXAMPLE PER-SPAN AGA-BASED ENCODING

| Span $m$ Scenario | Bits of Unit at Period $m+0$: α | β | γ | δ | ε | 4 | 5 | 6 | 7 | Bits of Unit at Period $m+1$: 0 | 1 | 2 | 3 | 4 | 5 | 6 | 7 |
|---|---|---|---|---|---|---|---|---|---|---|---|---|---|---|---|---|---|
| d → d → | 0 | $^{(m+0)}d_{0\div7}$ | | | | | | | | $^{(m+1)}d_{0\div7}$ | | | | | | | |
| c → d → | 1 | 0 | 0 | 0 | 0 | 0 | 0 | $^{(m+0)}c_{\gamma,\delta}$ | | $^{(m+1)}d_{0\div7}$ | | | | | | | |
| d → c → | 1 | 0 | 0 | 0 | 0 | 0 | 1 | $^{(m+1)}c_{\gamma,\delta}$ | | $^{(m+0)}d_{0\div7}$ | | | | | | | |
| c → c → | 1 | 0 | 0 | 0 | 0 | 0 | — | 0 | 0 | — | | $^{(m+0)}c_{\gamma,\delta}$ | | — | | $^{(m+1)}c_{\gamma,\delta}$ | |
| event → | 1 | 0 | 0 | 0 | 0 | 1 | $^{(m)}p_{0,1,2}$ | | | *the same as before insertion* | | | | | | | |

| +0 | +1 | $g = 2$ | 000 | 001 | 010 | 011 | 100 | 101 | 110 | 111 |
|---|---|---|---|---|---|---|---|---|---|---|
| Unit Disp. | | $N_b = 16$ | Event Position Code (*in the time of the current transfer span*) | | | | | | | |

Please remember that ⁞ this code, if there is an event, is present *explicitly* in the affix bits ...

| $p_0$ | $p_1$ | $p_2$ | $p_{0:2}$ | Definition, Description, Designation |
|---|---|---|---|---|
| 0 | 0 | 0 | 000 | it is an event in the 1st (earliest) one eighth of this ST |
| 0 | 0 | 1 | 001 | ........................ 2nd ........................................ |
| • | • | • | • | • • • • • • • • • • • • • |
| 1 | 1 | 0 | 110 | ........................ 7th ........................................ |
| 1 | 1 | 1 | 111 | ........................ 8th (latest) ............................... |

$|p| = 8 = \frac{1}{2} N_b$ $|p| = 8 = g \bullet N_{\text{event}}$ : (because of) $N_{\text{event}} \le \frac{1}{2} N_b : g$

... that are conveyed by the *post-gathered* span head transfer unit related to the ST during which the event happens.

| Place | Ref. | Before Insertion | | | After Insertion | | | | Delta |
|---|---|---|---|---|---|---|---|---|---|
| event ↓ head$_m$ | head | $H_{m-g}$ | $H_m$ | $H_{m+g}$ | $H_{m-g}$ | ev | $H_m$ | $H_{m+g}$ | +1 |
| | tail | $T_{m-g}$ | $T_m$ | $T_{m+g}$ | $T_{m-g}$ | $T_m$ | $T_{m+g}$ | ··· | — |
| event ↓ tail$_m$ | head | $H_{m-g}$ | $H_m$ | $H_{m+g}$ | $H_{m-g}$ | ev | $H_m$ | $H_{m+g}$ | +1 |
| | tail | $T_{m-g}$ | $T_m$ | $T_{m+g}$ | $T_{m-g}$ | $T_m$ | $T_{m+g}$ | ··· | — |
| Span Period → | | $m-g$ | $m$ | $m+g$ | $m-g$ | $m$ | $m^{\text{bis}}$ | $m+g$ | . . . |

NOTE – In a per-span case, event insertion is done AFTER data/control encoding.
NOTE – Bit time (BT) = Scale•100 ps, span time (ST) = $g$ × 8•BT = Scale•1600 ps.

TABLE IX
EXAMPLE PER-UNIT AGA-BASED ENCODING

| Span $m$ Scenario | Bits of Unit at Period $m+0$: α | β | γ | δ | ε | 4 | 5 | 6 | 7 | Bits of Unit at Period $m+1$: 0 | 1 | 2 | 3 | 4 | 5 | 6 | 7 |
|---|---|---|---|---|---|---|---|---|---|---|---|---|---|---|---|---|---|
| d → d → | 0 | $^{(m+0)}d_{0\div7}$ | | | | | | | | $^{(m+1)}d_{0\div7}$ | | | | | | | |
| x → d → | 1 | 0 | 0 | 0 | 0 | 0 | $^{(m+0)}x_{\beta,\gamma,\delta}$ | | | $^{(m+1)}d_{0\div7}$ | | | | | | | |
| d → x → | 1 | 0 | 0 | 0 | 0 | 1 | $^{(m+1)}x_{\beta,\gamma,\delta}$ | | | $^{(m+0)}d_{0\div7}$ | | | | | | | |
| x → x → | 1 | 0 | 0 | 0 | 0 | — | 0 | 0 | 0 | — | $^{(m+0)}x_{\beta,\gamma,\delta}$ | | | — | $^{(m+1)}x_{\beta,\gamma,\delta}$ | | |

| +0 | +1 | $g = 2$ | 00 | 01 | 10 | 11 | 00 | 01 | 10 | 11 |
|---|---|---|---|---|---|---|---|---|---|---|
| Unit Disp. | | $N_b = 16$ | Event Position Code (*in the time of the respective transfer unit*) | | | | | | | |

Please remember that ⁞ this code, if there is an event, is present *implicitly* in the extra bits ...

| $x_\beta$ | $x_\gamma$ | $x_\delta$ | $c_{\gamma:\delta}$ | Def. | $p_{1:2}$ | Definition, Description, Designation |
|---|---|---|---|---|---|---|
| 0 | 0 | 1 | 01 | error | | |
| 0 | 1 | 0 | 10 | idle | | |
| 0 | 1 | 1 | 11 | lpi | | |
| 1 | 0 | 0 | | | 00 | it is an event in the 1st quarter of this UT |
| 1 | 0 | 1 | | | 01 | ........................ 2nd ...................... |
| 1 | 1 | 0 | | | 10 | ........................ 3rd ...................... |
| 1 | 1 | 1 | | | 11 | ........................ 4th ...................... |

$|x| = 7 = N_{\text{extra}}$ $|c| = 3 = N_{\text{ctrl}}$ $|p| = 4 = N_{\text{event}}$ : $N_{\text{event}} + N_{\text{ctrl}} \le N_{\text{extra}}$ , $N_{\text{ctrl}} \ge 1$

... that all are conveyed by the respective *pre-gathered* transfer unit related to the UT during which the event happens.

| Place | Ref. | Before Insertion | | | After Insertion | | | | Delta |
|---|---|---|---|---|---|---|---|---|---|
| event ↓ head$_m$ | future head | $Y_{m-g}$ | $Y_m$ | $Y_{m+g}$ | $Y_{m-g}$ | ev | $Z_m$ | $Z_{m+g}$ | + 1/$g$ |
| | future tail | $Z_{m-g}$ | $Z_m$ | $Z_{m+g}$ | $Z_{m-g}$ | $Y_m$ | $Y_{m+g}$ | ··· | |
| event ↓ tail$_m$ | future head | $Y_{m-g}$ | $Y_m$ | $Y_{m+g}$ | $Y_{m-g}$ | $Y_m$ | $Z_m$ | $Z_{m+g}$ | + 1/$g$ |
| | future tail | $Z_{m-g}$ | $Z_m$ | $Z_{m+g}$ | $Z_{m-g}$ | ev | $Y_{m+g}$ | ··· | |
| Span Period → | | $m-g$ | $m$ | $m+g$ | $m-g$ | $m$ | $m^{\text{bis}}$ | $m+g$ | . . . |

NOTE – In a per-unit case, event insertion is done BEFORE data/extra encoding.
NOTE – Unit time (UT) = octet time (OT) = eight bit times (8x BTs) = Scale•800 ps.

TABLE X
AGA LIMITS @ $N = 17 \times 16$

| $g$ | $N_{\text{ctrl}}$ | $N_{\text{event}}$ | $N_{\text{extra}}$ | $N_{\text{event}}$ |
|---|---|---|---|---|
| 1 | 8 | 8 | 16 | 15 |
| 2 | 3 | 4 | 7 | 6 |
| 3 | 2 | 2⅔ | 5 | 4 |
| 4 | 1 | 2 | 3 | 2 |
| 5 | 1 | 1⅗ | 3 | 2 |
| 6 | 1 | 1⅓ | 2 | 1 |
| 7 | 1 | 1⅐ | 2 | 1 |
| 8 | — | 1 | 1 | — |
| → | Per-Span | | Per-Unit | |

TABLE XI
AGA LIMITS @ $N = 9 \times 32$

| $g$ | $N_{\text{ctrl}}$ | $N_{\text{event}}$ | $N_{\text{extra}}$ | $N_{\text{event}}$ |
|---|---|---|---|---|
| 1 | 16 | 16 | 32 | 31 |
| 2 | 7 | 8 | 15 | 14 |
| 3 | 5 | 5⅓ | 10 | 9 |
| 4 | 3 | 4 | 7 | 6 |
| 5 | 3 | 3⅕ | 6 | 5 |
| 6 | 2 | 2⅔ | 5 | 4 |
| 7 | 2 | 2 2/7 | 4 | 3 |
| 8 | 1 | 2 | 3 | 2 |
| → | Per-Span | | Per-Unit | |

TABLE XII
AGA LIMITS @ $N = 5 \times 64$

| $g$ | $N_{\text{ctrl}}$ | $N_{\text{event}}$ | $N_{\text{extra}}$ | $N_{\text{event}}$ |
|---|---|---|---|---|
| 1 | 32 | 32 | 64 | 63 |
| 2 | 15 | 16 | 30 | 29 |
| 3 | 10 | 10⅔ | 19 | 18 |
| 4 | 7 | 8 | 14 | 13 |
| 5 | 6 | 6⅖ | 11 | 10 |
| 6 | 5 | 5⅓ | 9 | 8 |
| 7 | 4 | 4 4/7 | 8 | 7 |
| 8 | 3 | 4 | 7 | 6 |
| → | Per-Span | | Per-Unit | |

TABLE XIII
AGA LIMITS @ $N = 3 \times 128$

| $g$ | $N_{\text{ctrl}}$ | $N_{\text{event}}$ | $N_{\text{extra}}$ | $N_{\text{event}}$ |
|---|---|---|---|---|
| 1 | 64 | 64 | 128 | 127 |
| 2 | 30 | 32 | 57 | 56 |
| 3 | 19 | 21⅓ | 37 | 36 |
| 4 | 14 | 16 | 27 | 26 |
| 5 | 11 | 12⅘ | 21 | 21 |
| 6 | 9 | 10⅔ | 17 | 16 |
| 7 | 8 | 9⅐ | 15 | 14 |
| 8 | 7 | 8 | 13 | 12 |
| → | Per-Span | | Per-Unit | |

TABLE XIV
ACHIEVABLE RESOLUTION

| 10GBASE-T1 @$N_{event}$ = | | 2 | 4 | 8 | 16 | 5GBASE-T1 @$N_{event}$ = | | 2 | 4 | 8 | 16 | 2.5GBASE-T1 @$N_{event}$ = | | 2 | 4 | 8 | 16 |
|---|---|---|---|---|---|---|---|---|---|---|---|---|---|---|---|---|---|
| Frequency, $f$ | GHz | 2.5 | 5 | 10 | 20 | Frequency, $f$ | GHz | 1.25 | 2.5 | 5 | 10 | Frequency, $f$ | GHz | .625 | 1.25 | 2.5 | 5 |
| $1/f \rightarrow$ | ps | 400 | 200 | 100 | 50 | $1/f \rightarrow$ | ps | 800 | 400 | 200 | 100 | $1/f \rightarrow$ | ps | 1600 | 800 | 400 | 200 |

TABLE XV
ADVANCED GATHERING APPROACH (AGA) EXAMPLE APPLICABILITY INCLUDING BACKWARD

| μFrame | $n_p$ | $v$ | $N_r \times N_b$ | $g$ | $*n_p$ | = $n_e$ | × $k$ | + $*n_e$ | × $*k$ | $n_e : g$ | ξ | $t$ | = $b$ | + $r$ | $w$ | $s$ | Comment |
|---|---|---|---|---|---|---|---|---|---|---|---|---|---|---|---|---|---|
| MGBASE-T1, | 400 | 3,250 | 17 × 16 | 2 | 402 | 14 | 28 | 10 | 1 | 7 | 1 | 113 | 84 | 29 | ≤32 | 5 | up to 2 $^{ev}/_{\mu frame}$ |
| 10GBASE-T (as they are similiar) | | | 9 × 32 | 2 | 401 | 10 | 40 | 1 | — | 5 | 1 | 81 | 65 | 16 | ≤16 | 1 | up to 1 $^{ev}/_{\mu frame}$ |
| | | | 9 × 32 | 3 | 402 | 15 | 26 | 12 | 1 | 5 | 1 | 121 | 105 | 16 | ≤16 | 7 | up to 2 $^{ev}/_{\mu frame}$ |
| | | | 9 × 32 | 4 | 403 | 20 | 20 | 3 | — | 5 | 1 | 161 | 145 | 16 | ≤16 | 3 | up to 3 $^{ev}/_{\mu frame}$ |
| | | | 5 × 64 | 4 | 402 | 12 | 33 | 6 | 1 | 3 | 1 | 97 | 90 | 7 | ≤8 | — | up to 2 $^{ev}/_{\mu frame}$ |
| | | | 5 × 64 | 8 | 404 | 24 | 16 | 20 | 1 | 3 | 1 | 193 | 186 | 7 | ≤8 | 1 | up to 4 $^{ev}/_{\mu frame}$ |
| | | | 3 × 128 | 7 | 401 | 21 | 19 | 2 | — | 3 | 2 | 170 | 165 | 5 | ≤8 | 2 | up to 1 $^{ev}/_{\mu frame}$ |
| | | | 3 × 128 | 8 | 402 | 24 | 16 | 18 | 1 | 3 | 2 | 194 | 189 | 5 | ≤8 | 1 | up to 2 $^{ev}/_{\mu frame}$ |
| 10GBASE-KR | 256 | 2,080 | 17 × 16 | 2 | 257 | 14 | 18 | 5 | — | 7 | 1 | 113 | 84 | 29 | ≤32 | 1 | up to 1 $^{ev}/_{\mu frame}$ |
| | | | 9 × 32 | 4 | 258 | 20 | 12 | 18 | 1 | 5 | 1 | 161 | 145 | 16 | ≤16 | 3 | up to 2 $^{ev}/_{\mu frame}$ |
| | | | 5 × 64 | 4 | 257 | 12 | 21 | 5 | 1 | 3 | 1 | 193 | 186 | 7 | ≤8 | 2 | up to 1 $^{ev}/_{\mu frame}$ |
| | | | 3 × 128 | 8 | 257 | 24 | 10 | 17 | 1 | 3 | 2 | 194 | 189 | 5 | ≤8 | 3 | up to 1 $^{ev}/_{\mu frame}$ |
| 1000BASE-T1 | 450 | 3,645 | 17 × 16 | 2 | 452 | 16 | 28 | 4 | 1 | 8 | 1 | 129 | 96 | 33 | ≤64 | — | up to 2 $^{ev}/_{\mu frame}$ |
| | | | 9 × 32 | 3 | 451 | 15 | 30 | 1 | — | 5 | 1 | 121 | 105 | 16 | ≤16 | 6 | up to 1 $^{ev}/_{\mu frame}$ |
| | | | 5 × 64 | 4 | 451 | 24 | 18 | 19 | 1 | 6 | 2 | 194 | 180 | 14 | ≤16 | — | up to 1 $^{ev}/_{\mu frame}$ |
| | | | 3 × 128 | 8 | 451 | 24 | 18 | 19 | 1 | 3 | 2 | 194 | 189 | 5 | ≤8 | — | up to 1 $^{ev}/_{\mu frame}$ |

Implementing SGA, the coding means implements a variant of the so modeled multi-round linguistic multiplexing process, see Table VI (example with $N_r = 5$ and $\xi = 1$).

## ADVANCED GATHERING APPROACH

We introduce the so called advanced gathering approach, or AGA in short, to address the one problem but now regardless of the most of the specifics detected in the considered physical layer, to obtain an advanced solution, still appropriate but now more agile, when compared to SGA.

With a per-span case of AGA, see Tables VII and VIII, the coding means always locates the bits keeping the ("in-span") relative position of an event, p's,[8] in the head transfer unit of the span associated with that span time period, during which the event occurs, substituting the initial bits of this head unit with those position bits and so pushing the former out of this head unit into the next head unit in the stream, and so on and so on, that causes a new or appends an extra over the already accumulated echo existing until it is fully canceled thanks to the provisionally established spare unit(s).

Such the case is similar to the fractional approach [2], but separates the stream into the two of transfer units, head and other, instead of their bits in the predecessor, root and affix, affected and never affected by an event, respectively.

With a per-unit case of AGA, see Tables VII afore IX, the coding means locates the bits—keeping the ("in-unit") relative position of an event—now in the transfer unit associated with that transfer unit time period, during which the event occurs, inserting a new transfer unit with all of its bits, x's,[9] encoded according to its content, into the stream, that by design relies on the echo to be properly managed, too.

On succession again, such the case is similar to the integral approach [2], because they both manipulate on the stream in the scope of its entire transfer unit.

Although AGA demonstrates the obvious limits, see Tables X, XI,[10] XII,[10] XIII,[10] they hamper for neither its destination nor application, see Tables XIV and XV, respectively.

Implementing AGA, the coding means implements another but still an adoption of the so modeled process.

## CONCLUSION

In this paper, we described a way to multiplex a data passing ability, preexisting in a given layer featuring a very long binary transport word, with an event passing ability, new to the layer, that primarily fits for the MultiGBASE-T1 family of Ethernet physical layers as well as for similar protocols.[11]

[8] Independent bits are indexed using Arabic numerals, e.g., 0, 1, 2, etc.

[9] Dependent bits are indexed using small Greek letters, e.g., $\alpha$, $\beta$, $\gamma$, etc.

[10] For the given $N = N_r \times N_b$, also technically feasible but just not shown explicitly are the respective multiplexing variants with $g > 8$.

[11] It also fits well back for 10GBASE-KR, 10GBASE-T, and 1000BASE-T1, considered in [2], [3], and [4], respectively, see Tables V and XV.

# Data Coding Means and Event Coding Means Multiplexed Over the 25G/40GBASE-T FEC Input

Alexander Ivanov

***Abstract*—We reach the peak point in our running consideration of Ethernet protocols featuring a very long binary transport word, inside of that we try to multiplex two flows of information: about a data stream and about an event train. On this point are the 25GBASE-T and 40GBASE-T physical layers.**

***Index Terms*—Ethernet, linguistic multiplexing, multiplexing, precise synchronization, synchronization, 25G/40GBASE-T.**

## Introduction

MULTI-GIGABIT Ethernet, type 25GBASE-T and type 40GBASE-T, together known with the MultiGBASE-T, MGBASE-T, or 25G/40GBASE-T family of electrical physical layers [1], both evolutionary develop on the common 10 Gbps predecessor, 10GBASE-T, still operating over a similar media (four twisted pairs), but improving in their speed many times, to up to 25 Gbps and 40 Gbps, respectively.

However, this improvement costs them the strong reduction in the informational redundancy of their microframe payload, down to the need to use the 512B/513B block encoding, thus we think that—to make the multiplexing possible, as it is our goal—instead of the payload, one should work directly on the FEC input, see Table I, considering it the user accessible part of the FEC frame,[1] we will further refer short to as the frame, as well as longer than the payload exactly by some three bits, one auxiliary and two zero-padding [1].

To reach the goal, we plan to develop on the gathering [2], primarily, but also do not exclude the use of the spare transfer units [3], the save box cells [4], and/or even the spare bits [5], in the expectation they could help us, too.

Before we throw ourselves into, we anticipatorily add some new measures, see Table II, to the terminology of [2] and so to the generalized terminology of [3], as well as one new entity, see Table III, to the shortened block type codes standardized for the 512B/513B block encoding [1].

## Paired Multiplexing Scheme

We begin with the standard gathering approach, or SGA [2], gathering each $g = 8$ consecutive transfer units in the stream, into a span of the same length, and next examine in how many of the spare bits [5] this way could result,[2] when it is grounded on the accordant transcoding of the head unit within each span in the frame, see Tables IV and V, respectively.

Based on this, we consider the initial stream comprising all the transfer units, like eight (because $g = 8$) pseudo-separate (or parallel) streams, the first of which consists exclusively of the head units of all the spans in the initial stream, while the next consists exclusively of the earliest body units in the same spans, and so on and so on, see Table VI.

We also resolve that inserting an event causes the respective substitution in the stream of the head transfer units, i.e., related to the (imaginary) row the substitution happens in, generating

TABLE I
ORIGINAL FRAMING

| Hierarchy and Elements | Contents, Comments | | Size, bits |
|---|---|---|---|
| FEC [= Forward Error Correction] frame | 7.8125 or 12.5 Mio frames/s | 7 × 512 = Σ | 3,584 |
| ├ FEC input, visible by user | max 400 data octets → 3,200 bits | + 11 = | 3,211 |
| │ ├ RS-FEC input bits | hold 186 $GF(2^8)$ input symbols | 186×8= | 1,488 |
| │ └ RS-LDPC input bits | hold 237⅙ $GF(2^6)$ input symbols | 237×6+1= | 1,723 |
| └ FEC parity, hidden by PHY | calculated over FEC input bits | | 373 |
| ├ RS-FEC parity bits | hold 6 $GF(2^8)$ parity symbols | 6×8= | 48 |
| └ RS-LDPC parity bits | hold 54⅙ $GF(2^6)$ parity symbols | 54×6+1= | 325 |

TABLE II
ADDITIONAL TERMINOLOGY

| Ref. | Objective | Description, Definition, Designation | Measured |
|---|---|---|---|
| $s_F$ | $s_F \geq 0$ | spare part in the view of the FEC input user, | in bits/frame |
| $s_p$ | $0 \leq s_p \leq s_F$ | spare part in the view of the PCS payload user, | in bits/frame |
| $T_m$ ($\Theta_m$) | 1-of-15(16) | 8(4)-bit code in the *m*-th block (span) type field, | in masks |
| $E$ | $E \geq 2$ | modulus of the echo caused from event insertion, | in items |
| $K_E$ | $K_E = \log_2 E / s$ | FEC run count to fully cancel the echo of $1/E$, | in frames |

TABLE III
BLOCK TYPE FIELD CODE MATCH

| $T_m$ ↔ | $\Theta_m$ | $T_m$ ↔ | $\Theta_m$ | $T_m$ ↔ | $\Theta_m$ | $T_m$ ↔ | $\Theta_m$ |
|---|---|---|---|---|---|---|---|
| (8-bit) | (4-bit) | 0x55 | 1011 | 0x87 | 1100 | 0xCC | 0011 |
| 0x1E | 1000 | 0x66 | 1101 | 0x99 | 1010 | 0xD2 | 0110 |
| 0x2D | 0100 | 0x78 | 0111 | 0xAA | 1001 | 0xE1 | 0000 |
| 0x33 | 1110 | 0x4B | 0001 | 0xB4 | 0101 | 0xFF | 1111 |
| — | 0010 → | signals that an event occurs during the *m*-th span time period | | | | | |

A manuscript of this work was submitted to IEEE Communications Letters December 26, 2022 and rejected for fair reasons during its peer review.

Please sorry for the author has no time to find this work a new home, peer reviewed or not, except of arXiv, and just hopes there it meets its reader, one or maybe various, whom the author beforehand thanks for their regard.

A. Ivanov is with JSC Continuum, Yaroslavl, the Russian Federation.

[1]FEC frame is the standardized term to refer to such objects, see [1].

[2]25G/40GBASE-T encodes $n_{\mathrm{p}} = 400$ ("quasi-separate") transfer units into $400/8 = 50$ collected 64/65B coded blocks, next transcodes $50 - 2 = 48$ of them into $48/8 = 6$ re-collected 512B/513B coded blocks, and then passes all these $2 \times 65 + 6 \times 513 = 3{,}208$ bits, among yet 1 auxiliary and 2 zero-padding bits, i.e., $3{,}208 + 1 + 2 = 3{,}211$ bits total, into the FEC input, once per every (FEC) frame time period of 128 ns at 25 Gbps or 80 ns at 40 Gbps, respectively. (FEC frame time and microframe time are the same.)

TABLE IV
UNIVERSAL SOLUTIONS

| Protocol | $v$ | $N_r \times N_b$ | $g$ | $*n_p = n_e \times k + *n_e \times *k$ | $n_e : g$ | $\xi$ | $t = b + r$ | $w$ | $s_F$ | $s_p$ | Gathering |
|---|---|---|---|---|---|---|---|---|---|---|---|
| 25GBASE-T, | 3,211 | 17 × 16 | 8 | 400 = 56 × 7 + 8 × 1 | 7 | 1 | 449 = 420 + 29 | ≤32 | 3 | — | SGA or AGA |
| 40GBASE-T |  | 17 × 16 | 8 | 400 = 88 × 4 + 48 × 1 | 11 | 1 | 705 = 660 + 45 | ≤64 | 6 | 3 | SGA or AGA |
| ($n_p$ = 400) |  | 9 × 32 | 8 | 400 = 40 × 10 — — | 5 | 1 | 321 = 305 + 16 | ≤16 | 1 | — | SGA or AGA |
|  |  | 9 × 32 | 8 | 400 = 88 × 4 + 48 × 1 | 11 | 2 | 706 = 671 + 35 | ≤64 | 2 | — | SGA or AGA |

TABLE V
SPAN TRANSCODING RULE

| ↓ Span Content, Span Bits → | $^{(m)}u_\alpha$ | $^{(m)}u_\beta$ | $^{(m)}u_\gamma$ | $^{(m)}u_\delta$ | $^{(m)}u_{\varepsilon\|3}$ | $^{(m)}u_4$ | $^{(m)}u_5$ | $^{(m)}u_6$ | $^{(m)}u_7$ | $^{(m+1)}u_0$ ••• $^{(m+1)}u_7$ | $^{(m+2)}u_0$ • • • $^{(m+6)}u_7$ | $^{(m+7)}u_0$ ••• $^{(m+7)}u_7$ |
|---|---|---|---|---|---|---|---|---|---|---|---|---|
| Pure DATA octets, no event | 0 | $^{(m)}b_1$ | $^{(m)}b_2$ | $^{(m)}b_3$ | $^{(m)}b_4$ | $^{(m)}b_5$ | $^{(m)}b_6$ | $^{(m)}b_7$ | $^{(m)}b_8$ |  |  |  |
| Event or CTRL(s), $N$ = 9 × 32 | 1 | 0 | 0 | 0 | $^{(m)}\theta_0$ | $^{(m)}\theta_1$ | $^{(m)}\theta_2$ | $^{(m)}\theta_3$ | $^{(m)}$C/D | $^{(m)}b_9$ ••• $^{(m)}b_{16}$ | $^{(m)}b_{17}$ • • • $^{(m)}b_{56}$ | $^{(m)}b_{57}$ ••• $^{(m)}b_{64}$ |
| Event or CTRL(s), $N$ = 17 × 16 | 1 | 0 | 0 | 0 | 0 | $^{(m)}\theta_0$ | $^{(m)}\theta_1$ | $^{(m)}\theta_2$ | $^{(m)}\theta_3$ |  |  |  |

TABLE VI
EVENT INSERTION RULES

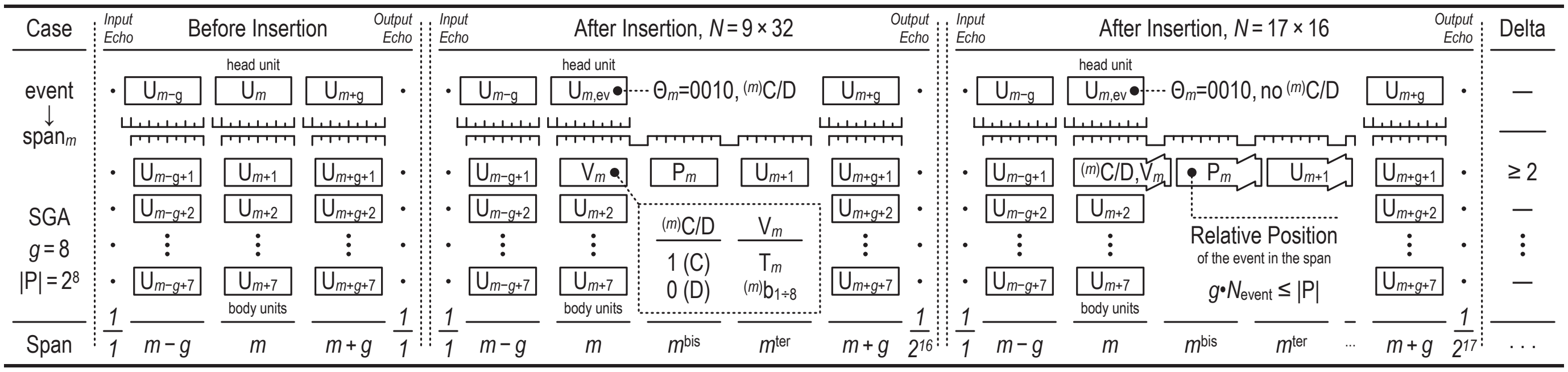

TABLE VII
MULTIPLEXING CONFIGURATION

| Case | $g$•$N_{event}$ | $N_r \times N_b$ | $k$ + *$k$ | $E$ | $K_E$ |
|---|---|---|---|---|---|
| $s_F$ = 1 (FEC Input) | ≤256 | 9 × 32 | 10 + 0 | $2^{16}$ | 16 |
| $s_p$ = 3 (Payload) | ≤256 | 17 × 16 | 4 + 1 | $2^{17}$ | 3⅖ |
| $s_F$ = 3 (FEC Input) | ≤256 | 17 × 16 | 7 + 1 | $2^{17}$ | 5⅔ |

TABLE VIII
PAIRED MULTIPLEXING SCHEME

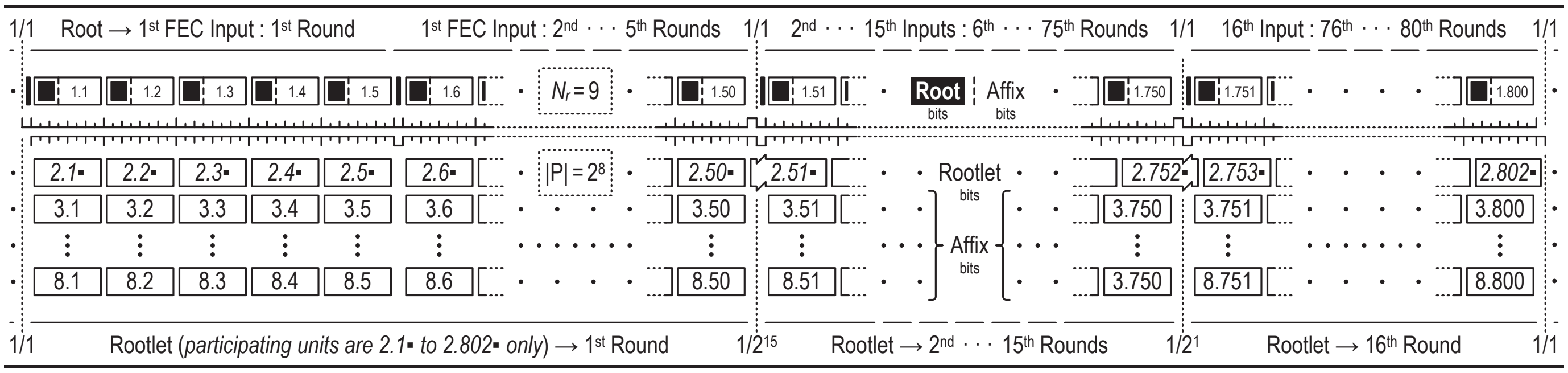

TABLE IX
ACHIEVABLE RESOLUTION

| 40GBASE-T @ | $g$•$N_{event}$ = | 8•½ | 8•1 | 8•2 | 8•4 | 8•8 | 8•16 | 8•32 |
|---|---|---|---|---|---|---|---|---|
| Frequency, $f$ | GHz | 2.5 | 5 | 10 | 20 | 40 | 80 | 160 |
| 1 / $f$ → | ps | 400 | 200 | 100 | 50 | 25 | 12.5 | 6.25 |

| 25GBASE-T @ | $g$•$N_{event}$ = | 8•½ | 8•1 | 8•2 | 8•4 | 8•8 | 8•16 | 8•32 |
|---|---|---|---|---|---|---|---|---|
| Frequency, $f$ | GHz | 1.5625 | 3.125 | 6.25 | 12.5 | 25 | 50 | 100 |
| 1 / $f$ → | ps | 640 | 320 | 160 | 80 | 40 | 20 | 10 |

TABLE X
RESTRICTED SOLUTIONS

| Protocol | $v$ | $N_r \times N_b$ | $g$ | $*n_p = n_e \times k + *n_e \times *k$ | $n_e : g$ | $\xi$ | $t = b + r$ | $w$ | $s_F$ | $s_p$ | Gathering |
|---|---|---|---|---|---|---|---|---|---|---|---|
| 25GBASE-T, | 3,211 | 9 × 32 | 10 | 400 = 50 × 8 — — | 5 | 1 | 401 = 385 + 16 | ≤16 | 3 | — | AGA peculiar |
| 40GBASE-T | | 5 × 64 | 20 | 400 = 60 × 6 + 40 × 1 | 3 | 1 | 481 = 474 + 7 | ≤8 | 4 | 1 | AGA peculiar |
| ($n_p$ = 400) | | 3 × 128 | 40 | 400 = 200 × 2 — — | 5 | 3 | 1,603 = 1,595 + 8 | ≤8 | 5 | 2 | AGA peculiar |

TABLE XI
PECULIAR SPAN ENCODING

| Format $N_r$ | $g$ | Unit Type | Bits of the Head Transfer Unit in the Current Span, $m$: $^{(m)}u_\alpha$ | $^{(m)}u_\beta$ | $^{(m)}u_{\gamma\|1}$ | $^{(m)}u_{\delta\|2}$ | $^{(m)}u_3$ | $^{(m)}u_4$ | $^{(m)}u_5$ | $^{(m)}u_6$ | $^{(m)}u_7$ | … | Bits of a Body Transfer Unit in the Current Span, $0 < j < g$: $^{(m+j)}u_0$ | $^{(m+j)}u_1$ | $^{(m+j)}u_2$ | $^{(m+j)}u_3$ | $^{(m+j)}u_4$ | $^{(m+j)}u_5$ | $^{(m+j)}u_6$ | $^{(m+j)}u_7$ | … |
|---|---|---|---|---|---|---|---|---|---|---|---|---|---|---|---|---|---|---|---|---|---|
| any | any | DATA | 0 | $^{(m)}d_0$ | $^{(m)}d_1$ | $^{(m)}d_2$ | $^{(m)}d_3$ | $^{(m)}d_4$ | $^{(m)}d_5$ | $^{(m)}d_6$ | $^{(m)}d_7$ | … | $^{(\#)}d_0$ | $^{(\#)}d_1$ | $^{(\#)}d_2$ | $^{(\#)}d_3$ | $^{(\#)}d_4$ | $^{(\#)}d_5$ | $^{(\#)}d_6$ | $^{(\#)}d_7$ | … |
| 9 | 10 | CTRL | 1 | 0 | 0 | 0 | $^{(\#)}x_\alpha$ | $^{(\#)}x_\beta$ | $^{(\#)}x_\gamma$ | $^{(\#)}x_\delta$ | $^{(\#)}x_5$ | … | omitted | omitted | $^{(\#)}x_0$ | $^{(\#)}x_\alpha$ | $^{(\#)}x_\beta$ | $^{(\#)}x_\gamma$ | $^{(\#)}x_\delta$ | $^{(\#)}x_5$ | … |
| | | event | 1 | 0 | 0 | 0 | 1 | 1 | 1 | 1 | $^{(m)}$C/D | … | *signaling an event in a span is possible within its head unit only* | | | | | | | | |
| 5 | 20 | CTRL | 1 | 0 | 0 | $^{(\#)}x_\alpha$ | $^{(\#)}x_\beta$ | $^{(\#)}x_\gamma$ | $^{(\#)}x_\delta$ | $^{(\#)}x_5$ | $^{(\#)}x_6$ | … | omitted | $^{(\#)}x_0$ | $^{(\#)}x_\alpha$ | $^{(\#)}x_\beta$ | $^{(\#)}x_\gamma$ | $^{(\#)}x_\delta$ | $^{(\#)}x_5$ | $^{(\#)}x_6$ | … |
| | | event | 1 | 0 | 0 | 1 | 1 | 1 | 1 | $^{(m)}$C/D | $^{(m)}p_0$ | … | . . . . . . . . . . | | | | | | | | |
| 3 | 40 | CTRL | 1 | 0 | $^{(\#)}x_\alpha$ | $^{(\#)}x_\beta$ | $^{(\#)}x_\gamma$ | $^{(\#)}x_\delta$ | $^{(\#)}x_5$ | $^{(\#)}x_6$ | $^{(\#)}x_7$ | … | $^{(\#)}x_0$ | $^{(\#)}x_\alpha$ | $^{(\#)}x_\beta$ | $^{(\#)}x_\gamma$ | $^{(\#)}x_\delta$ | $^{(\#)}x_5$ | $^{(\#)}x_6$ | $^{(\#)}x_7$ | … |
| | | event | 1 | 0 | 1 | 1 | 1 | 1 | $^{(m)}$C/D | $^{(m)}p_0$ | $^{(m)}p_1$ | … | . . . . . . . . . . | | | | | | | | |

TABLE XII
PREPARATIVE UNIT SEPARATION

| *m*-th Case | Definition, $m \le \# < m + g$ | *m*-th Separation Result Description = *m*-th Ready-to-Encode Transfer Span Structure Designation | Meta Information Kept |
|---|---|---|---|
| Pure DATAs | every #-th unit is DATA | [no CTRL unit followed by] $g_{m,\mathrm{D}} = g$ DATA units, *sorted originally* | — |
| Mixture of | both types are present | $g_{m,\mathrm{C}} \ge 1$ CTRL unit(s), *sorted anyway*, followed by $g_{m,\mathrm{D}} = g - g_{m,\mathrm{C}}$ DATA unit(s), *sorted by # ascending* | unit's # for each CTRL |
| Pure CTRLs | every #-th unit is CTRL | $g_{m,\mathrm{C}} = g$ CTRL units, *sorted anyway* [followed by no DATA unit] | unit's # for each CTRL |

TABLE XIII
NESTED MULTIPLEXING SCHEME

| Nested Root Bits | CTRL Meta Information and its | Value | Modulus | Value | of the CTRL Self Information and its Description, Definition, Designation | | Process Depiction |
|---|---|---|---|---|---|---|---|
| $^{(\#)}x_{\alpha\div\delta}$ ↕ 4 | reflects the #, kept as (# − *m*), of the CTRL unit whose meta is encoded in this transfer unit, either head, $U_m$, or body, $U_{m+j}$, of the current span, *m*, and so defines the original placement of that CTRL unit in the span before its separation | 0 … $g - 1$ | $g \times 3$ | 0 (error) | reflects and defines the purpose of that CTRL unit (error) as well as signals about this transfer unit is the last one conveying a CTRL in the current span | $N_{\mathrm{ctrl}} = 2$ | $g = 10$ : $g{\times}3 = 5{\bullet}3{\bullet}2^1$; 20: $5{\bullet}3{\bullet}2^2$; 40: $5{\bullet}3{\bullet}2^3$ |
| MUX = | | | | 1 (idle) | reflects and defines the purpose of that CTRL unit (idle) as well as signals about this transfer unit is the last one conveying a CTRL in the current span | $\frac{1}{E_{\mathrm{nested}}} = \frac{1}{g} \times \frac{1}{3}$ | → [ single round ] $5{\bullet}3{\bullet}2^{\ge 1}$ → $\frac{1}{1}$ |
| ↓ ≥1 $^{(\#)}x_{5\cdots}$ | $10 = 5{\bullet}2^1$  $20 = 5{\bullet}2^2$  $40 = 5{\bullet}2^3$ | | $< N_b$ | 2 (—) | signals that the purpose of that CTRL unit (error or idle) is encoded in the unit following this transfer unit | see NOTE | $5{\bullet}3$ (representation) → $2^4$ • $2^{\ge 1}$ → $2^{\ge 1}$ |
| | | | | | NOTE – In the scope of a given CTRL unit, except the encoded in the head transfer unit of a span, the value of the zeroth bit depends on the preceding CTRL unit, reflecting and defining its purpose. | $2^{\le 2}$ $2^1$ | |
| Nested Affix Bit(s) | Choice 1 of $g$ ($2^2 < g / 2^{\ge 1} < 2^3$) | | Transport Capacity | Choice 1 of 3 ($2^1 < 3 < 2^2$) ⟶ $N_{\mathrm{extra}} = 3 = (N_{\mathrm{ctrl}} + 1)$ | | omitted (if any) $^{(\#)}x_0$ | $^{(\#)}x_{\alpha\div\delta}$  $^{(\#)}x_{5\cdots}$ |

TABLE XIV
COMPLEXITY COMPARISON

| Case | $g$ | $n_e : g$ | $w$ | Vector ALU Modes | Case | $g$ | $n_e : g$ | $w$ | Vector ALU Modes | Case | $g$ | $n_e : g$ | $w$ | Vector ALU Modes |
|---|---|---|---|---|---|---|---|---|---|---|---|---|---|---|
| $N_b = 32$ (5 bits) | 10 | 5 | 16 | U16 X 1, U4 X 4 | $N_b = 64$ (6 bits) | 20 | 3 | 8 | U8 X 1, U4 X 2 | $N_b = 128$ (7 bits) | 40 | 5 | 8 | U8 X 1, U4 X 2 |

TABLE XV
RESOLUTION ACHIEVABILITY CONDITIONS

| Case | Bits in the Head | Bits [to be transferred] in the Echo | | $g{\bullet}N_{\mathrm{event}}$ = | $g{\bullet}1$ | $g{\bullet}2$ | $g{\bullet}4$ | $g{\bullet}5$ | $g{\bullet}8$ | $g{\bullet}10$ | $g{\bullet}16$ | $g{\bullet}20$ | $g{\bullet}32$ |
|---|---|---|---|---|---|---|---|---|---|---|---|---|---|
| $g = 10$ | $^{(m)}$C/D | up to 8 (DATA) | $^{(m)}p_0$, $^{(m)}p_1$, $^{(m)}p_2$, ... | | | | | | | | | | |
| $g = 20$ | $^{(m)}$C/D, $^{(m)}p_0$ | substituted bits + | $^{(m)}p_1$, $^{(m)}p_2$, ... | $\log_2 E$ | ≥12 (= 8 + 4) | ≥13 (= 8 + 5) | ≥14 (= 8 + 6) | ≥14 (= 8 + 6) | ≥15 (= 8 + 7) | ≥15 (= 8 + 7) | ≥16 (= 8 + 8) | ≥16 (= 8 + 8) | ≥17 (= 8 + 9) |
| $g = 40$ | $^{(m)}$C/D, $^{(m)}p_0$, $^{(m)}p_1$ | of the head unit | $^{(m)}p_2$, ... | | | | | | | | | | |

TABLE XVI
ECHO BUDGET MEMO

| Budget Subline | Samples Within | |
|---|---|---|
| Total Echo Pool Available—Paired | pow ( 2, $t$ ) = pow ( 2, $r$ ) × pow ( 2, $b$ ) | |
| ├ Native Echo Pool | pow ( $N_r$, $n_e$ : $g$ ) × pow ( 2, $b$ ) | |
| └ Rest if any, including forced, if needed, and free | [ pow ( 2, $r$ ) − pow ( $N_r$, $n_e$ : $g$ ) ] × pow ( 2, $b$ ) | |
| Total Echo Pool Available—Nested | Head Unit: $N_b$ | Body Units: $2^8 = 256$ |
| ├ Native Echo Pool + Event [head only] | $m$ $16/16 \cdot N_b$ | $m + j$ $2 \cdot 15/16 \cdot N_b$ |
| └ Rest if any, including forced, if needed, and free | — | $256 - 2 \cdot 15/16 \cdot N_b$ |

NOTE – Forced echo pool may cover extra signaling, e.g., OAM, LPI, ordered sets.

the known echo, whose (encoded) content corresponds to the (original) content of that so substituted unit as well as to the (relative) position of the event, $\log_2 E > 0$ bits in its total, but the respective representation in the stream of the earliest body units, i.e., related to the row next to the row of the head unit, performing the echo cancellation, whose (multi-round) process affects across many of the consecutive frames, $K_E$ runs in its total, by up to $s_{\mathrm{F}}$ (or $s_{\mathrm{p}}$) bits per a run, see Table VI again, then Tables VII and VIII (example with $N_r = 9$ and $s_{\mathrm{F}} = 1$), and after then Table IX, conclusively.

Thus, we say that we employ a pair of some inter-dependent linguistic multiplexing processes, one intended to multiplex an event into the initial stream while other intended to cancel the so caused echo, respectively. While the former operates on quantities of $\lceil \log_2 N_r \rceil > 1$ bits in their length each,[3] the latter operates on quantities of just a single bit in their length each,[4] referenced to as roots vs. rootlets to be clearly distinguishable, respectively, see Table VIII again.

Because every rootlet consists of a bit, we can consider the respective stream as a stream of single-bit-wide transfer units, and each of up to $s_{\mathrm{F}}$ (or $s_{\mathrm{p}}$) extra among them—in the view of the fractional approach [3]—as an instance of the spare unit, managed like a cell of the save box [4].

## NESTED MULTIPLEXING SCHEME

We then continue with the advanced gathering approach, or AGA [2], trying to collate for $g > 8$ consecutive transfer units into a span, and find out some very attractive,[5] usable results, see Table X compared to Table IV.

In the support of such the variants, we define the so called peculiar encoding capable to address transfer spans consisting of $g > 8$ units, that now replaces the former encoding limited to the 64B/65B-inherited spans consisting of $g = 8$ units only, see Table XI compared to Table V, respectively, but by design necessitates for the proper separation of the transfer units to be done in each span, see Table XII alone, as well as for the further representation of the related information to be done in each of the control units, too, enabling a small since interior, nested linguistic multiplexing process, that—however—never replaces but always accompanies its exterior thus big, paired multiplexing not-a-twin sisters, see Table XIII in conjunction with Table VIII (example of SGA with $N_r = 9$ and $s_{\mathrm{F}} = 1$, with $s_{\mathrm{F}} = 3$, AGA, and the peculiar encoding, it will end up about three times faster,[6] sure), respectively.

Hence, there are the three multiplexing processes operating the following cooperative manner.

One of the paired processes multiplexes an event, inserting the information about that into the stream of the head transfer units, but cannot cancel the so caused echo. This process never penetrates across the boundaries of the respective multiplexing round whether or not the event affects it.

Other of the paired processes multiplexes the echo content bits into the stream of the rootlet bits enabled over the bits of the earliest body units in the spans then over the spare bits in the FEC input, as the basis to release the duty and as the extra to cancel the echo, jointly. This process propagates across the boundaries of all the consecutive inputs conveying the spare bits affected by the echo, since the input that conveys the span affected by the event that caused the echo.

In its turn listed last but performed first, the nested process multiplexes some unit-related information during the peculiar encoding. This process also never penetrates across the boundaries of the span it currently performs over.

Maybe it looks some comprehensive applied along [1], but it is really nothing beyond except a mixture of the known from [2], [3], [4], and [5], complimentarily.

Based on this, we consider the so composed coding means appropriate, see Table XIV, suitable, see Table XV in conjunction with Table IX, and extensible, see Table XVI.

## CONCLUSION

We proposed a way that fits for a coding entity featuring a comparatively not so excessive redundancy, like the respective means in 25G/40GBASE-T and similar protocols. Preserving the existing, data passing ability of such a means, the proposed way enables an extra, event passing ability inside it.

Among the necessary coding recipes, we considered the so called peculiar encoding for a span of up to 40 units.

[3] It invokes an event-related substitution but no event-related representation, because it has no means to cancel the echo caused by that substitution.

[4] We say it implements either a series of avalanche-like bit-per-bit explicit substitutions, or a base-$(2^{400} \times 2^s)$ into base-$2^{400+s}$ implicit representation, because any of thereof cancels the same echo. Anyway, it causes for no echo by itself, but affects on itself with the echo coming from aside.

[5] In the terms of coding delays and of complexity, the variant with $N_r = 5$ ($t = 481$) is comparable to the 512B/513B transcoding originally present in 25G/40GBASE-T. Moreover, the variants with $N_r = 5$ ($r = 7$) and $N_r = 3$ ($r = 8$) are mappable to the RS-FEC GF($2^8$) operations also used in it.

[6] I.e., in less than $\lceil 16/3 \rceil = 6$ complete frame periods. And about four and five times faster with $s_{\mathrm{F}} = 4$ ($N_r = 5$) and with $s_{\mathrm{F}} = 5$ ($N_r = 3$), i.e., in exact $16/4 = 4$ and in less than $\lceil 16/5 \rceil = 4$ such periods, respectively.

# Data Coding Means and Event Coding Means Multiplexed Over the 2.5G/5GBASE-T FEC Input

Alexander Ivanov

***Abstract*—We stand in some very interesting point today, still considering existing coding means featuring a very long binary transport word we call a microframe. This interest grows on the fact that the respective protocols, 2.5GBASE-T and 5GBASE-T, embody a very excessive redundancy, compared to other similar specifications we considered earlier.**

## INTRODUCTION

MULTI-GIGABIT Ethernet, type 2.5GBASE-T and type 5GBASE-T, together known with the MultiGBASE-T, MGBASE-T, or 2.5G/5GBASE-T family of electrical physical layers [1], both rateably downscale from the common 10 Gbps predecessor, 10GBASE-T, still operating over a similar media (four twisted pairs), but dropping down a number of times in the speed, into 2.5 Gbps and 5 Gbps, respectively.

In the view of the practical coding redundancy available to the user, the microframe payload of this family does not look somehow unusual, see Table I, but the naked FEC input does and moreover it seems very attractive, see Table II, compared along the similar multi- and one-gigabit protocols considered earlier in [2], [3], [4], [5], and [6], see Table III.[1]

In the view of the coding means staying behind the physical coding sublayer of this family, it has a duty to transfer 1,600 accommodation bits it consumes from the interface on its top, i.e., from the service interface of the whole physical layer, per every 2,048 transportation bits it generates onto the interface on its bottom, i.e., onto the service interface of the underlying physical medium attachment sublayer, with only 325 of them dedicated to transfer the FEC parity output.

That so excessive redundancy enables the layer designer to introduce a set of renewed useful features—never achievable before the respective design provides with such degree of the excessiveness—into the MGBASE-T family, among those are seamless preemption and quality of service, e.g., in the manner and by the way we describe further in this paper.

A manuscript of this work was submitted to IEEE Communications Letters December 26, 2022 and rejected for fair reasons during its peer review.

Please sorry for the author has no time to find this work a new home, peer reviewed or not, except of arXiv, and just hopes there it meets its reader, one or maybe various, whom the author beforehand thanks for their regard.

A. Ivanov is with JSC Continuum, Yaroslavl, the Russian Federation.

[1]For the common terminology we mention in this paper, one should trace in [2]'s Table II after [3]'s Table II after [4]'s Table II. The latter defines the generalized terminology grown on [5]'s Table II and [6]'s Table II.

TABLE I
ORIGINAL MICROFRAME STRUCTURE

| Hierarchy and Elements | | Contents, Comments | Size, bits | |
|---|---|---|---|---|
| Microframe (or µframe) entirely | | rate = S × 3.125 Mio µframes/s | | Σ 2,048 |
| ├ Payload, visible at XGMII | | max 200 data octets → 1,600 bits | | 25 × 65 = 1,625 |
| ├ 64B/65B coded block #1 | | ├ first 8 data octets or controls | | ├ 8×8+1= 65 |
| . . . . . . . . . . . . . . . . . . . . | | . . . . . . . . . . . . . . . . . . . . . . . . | | . . . . . . . |
| └ 64B/65B coded block #25 | | └ last 8 data octets or controls | | └ 8×8+1= 65 |
| ├ Auxiliary bit | hidden by PHY | has no standardized purpose | Σ423 | 1 |
| ├ Padded run | | vendor-filled random data | | 97 |
| └ FEC parity | | calculated on all the bits above | | 325 |

TABLE II
NAKED FEC FRAME

| Hierarchy and Elements | Contents, Comments | Size, bits |
|---|---|---|
| FEC [= Forward Error Correction] frame | 1.5625 or 3.125 Mio frames/s | 4 × 512 = Σ 2,048 |
| ├ FEC input, visible by user | max 200 data octets → 1,600 bits | + 123 = 1,723 |
| └ RS-LDPC input bits | hold 237⅙ GF($2^6$) input symbols | 237×6+1= 1,723 |
| └ FEC parity, hidden by PHY | calculated over FEC input bits | 325 |
| └ RS-LDPC parity bits | hold 54⅙ GF($2^6$) parity symbols | 54×6+1= 325 |

TABLE III
REDUNDANCY COMPARISON

| Protocol | Scope | $v$ | $8n_p$ | $(v-8n_p)/v$ | Comment |
|---|---|---|---|---|---|
| 2.5GBASE-T, 5GBASE-T | payload | 1,625 | 1,600 | ≈ 1.54 % | 0 % |
| | FEC input | *1,723* | 1,600 | ≈ *7.14 %* | (max) +364 % |
| 25GBASE-T, 40GBASE-T | payload | 3,208 | 3,200 | ≈ 0.25 % | (min) −84 % |
| | FEC input | 3,211 | 3,200 | ≈ 0.34 % | −78 % |
| MGBASE-T1 (M=2.5/5/10) | payload | 3,250 | 3,200 | ≈ 1.54 % | 0 % |
| | FEC input | 3,260 | 3,200 | ≈ 1.84 % | +19 % |
| 10GBASE-KR | any | 2,080 | 2,048 | ≈ 1.54 % | Ref. = 100 % |
| 10GBASE-T | payload | 3,250 | 3,200 | ≈ 1.54 % | 0 % |
| | FEC input | 3,259 | 3,200 | ≈ 1.81 % | +18 % |
| 1000BASE-T1 | payload | 3,645 | 3,600 | ≈ 1.23 % | −20 % |
| | FEC input | 3,654 | 3,600 | ≈ 1.48 % | −4 % |

## SYNTHESIS OBJECTIVES

We begin with a preparatory examination of the considered physical layers in their details, see Tables IV and V.[2] Based on that, we ground and word the following objectives.

Because there is the obvious, highest bit rate denominator, see Table IV again, it is suitable to support the nominal event fixation scale of 10 GHz (i.e., 100 ps in time domain).

Because there are some different layers but any of them is just a scaled version of the reference one, see Table V again,

[2]2.5GBASE-T (scale of 1/2) and 5GBASE-T (reference) are the only real members of the MGBASE-T family; "10GBASE-T" (scale of 2/1) is one we just imagine to clearly compare with their predecessor, 10GBASE-T.

TABLE IV
BASIC TERMS

| PCS Interface | Prefix | Bit Time | Visual Comarison | | Bit Rate |
|---|---|---|---|---|---|
| Accommodation | A- | 200 ps / S | A-Bit Time | A-Speed | S × 5 Gbps |
| Transportation | T- | 156¼ ps / S | T-Bit Time | T-Speed | S × 6.4 Gbps |
| Increase(+)/Decrease(−) | | −22.375 % | Delta | | +28 % |
| Relative Performance<br>Scale Factor, S | "HALF-SPEED"<br>1/2 | "FULL-SPEED"<br>1/1 (ref.) | "DOUBLE-SPEED"<br>2/1 | | shown for COMAPARISON only |
| Parameter | 2.5GBASE-T | 5GBASE-T | "10GBASE-T" | | 10GBASE-T |

TABLE V
PARAMETRIC COMPARISON

| Parameter | 2.5GBASE-T | 5GBASE-T | "10GBASE-T" | 10GBASE-T |
|---|---|---|---|---|
| Protocol Family | MGBASE-T | MGBASE-T | MGBASE-T | 10GBASE-T |
| Protocol Original | Clause 126 | Clause 126 | *imaginary* | Clause 55 |
| A-Interface (MII) | XGMII | XGMII | XGMII | XGMII |
| T-Interface | abstract | abstract | abstract | abstract |
| Scale Factor, S | 1/2 = 0.5 | 1/1 = 1.0 | 2/1 = 2.0 | — |
| Scaled A-Speed | 2.5 Gbps | 5.0 Gbps | 10.0 Gbps | 10.0 Gbps |
| Scaled T-Speed | 3.2 Gbps | 6.4 Gbps | 12.8 Gbps | 11.2 Gbps |
| Duty/μFrame, $n_p$ | 200 units | 200 units | 200 units | 400 units |
| FEC Input Size | 1,723 bits | 1,723 bits | 1,723 bits | 3,259 bits |
| FEC Frame Size | 2,048 bits | 2,048 bits | 2,048 bits | 3,584 bits |
| A-Bit Time, ABT | 400 ps | 200 ps | 100 ps | 100 ps |
| T-Bit Time, TBT | 312½ ps | 156¼ ps | 78⅛ ps | 89²⁄₇ ps |
| T/A Bit Time Ratio | $^{1600}/_{2048} = ^{25}/_{32}$ | $^{1600}/_{2048} = ^{25}/_{32}$ | $^{1600}/_{2048} = ^{25}/_{32}$ | $^{3200}/_{3584} = ^{25}/_{28}$ |
| T/A Bit Rate Ratio | (exact) 1.28 | (exact) 1.28 | (exact) 1.28 | (exact) 1.12 |
| OT = UT = 8•ABT | 3.2 ns | 1.6 ns | 0.8 ns | 0.8 ns |
| (μ,fec)FT = $n_p$•OT | 0.64 μs | 0.32 μs | 0.16 μs | 0.32 μs |
| TX Least Delay | 140 ns | 70 ns | 35 ns | 34²⁄₇ ns |
| TX LD Verified @32b | 140.8 ns | 70.4 ns | 35.2 ns | 35.2 ns |
| Encode Buffer @32b | ≥11×32 bits | ≥11×32 bits | ≥11×32 bits | ≥11×32 bits |
| TX LD Verified @64b | 153.6 ns | 76.8 ns | 38.4 ns | 38.4 ns |
| Encode Buffer @64b | ≥6×64 bits | ≥6×64 bits | ≥6×64 bits | ≥6×64 bits |
| RX Least Delay | 0.64 μs | 0.32 μs | 0.16 μs | 0.32 μs |
| Decode Buffer | ≥2,048 bits | ≥2,048 bits | ≥2,048 bits | ≥3,584 bits |
| MDI Symbol Code | 4-bit Gray | 4-bit Gray | 4-bit Gray | DQS-128 |
| MDI Symbol Type | PAM-16 | PAM-16 | PAM-16 | PAM-16 |
| MDI Symbol Rate | 200 MBd | 400 MBd | 800 MBd | 800 MBd |

NOTE – OT, UT, FT are octet time, transfer unit time, and frame time, respectively.

TABLE VI
TRANSMIT LEAST DELAY VERIFIED

| PCS Interface | Current Cycle Time Period @ A-Interface | Next Cycle . . . |
|---|---|---|
| Transportation (output) | $^{TX}$LD → time lost due to the cycle time of the a-interface<br>$^{TX}$LDV → | $^{TX}$LDV ≥ $^{TX}$LD |

NOTE – Cycle time is S×32•ABT (S×64•ABT) for 32-bit (64-bit) transfers at XGMII.

TABLE VII
TRANSMIT LEAST DELAY AND ENCODE MINIMAL BUFFER

| PCS Interface | Current Frame Time Period | | Next Frame Time Period |
|---|---|---|---|
| (input) Accommodation | 200 × 8 = 1,600 accommodation bits<br>1,600 × ABT = $^{32}/_{32}$ × FT | (32−25=7) | Transmit Least Delay<br>$^{TX}$LD = $^{7}/_{32}$ × FT |
| ↓ TX ENCODE | | | |
| Transportation (output) | $^{TX}$LD ↔ 1,600 × TBT = $^{25}/_{32}$ × FT<br>1,600 transportation bits | rest<br>448 t-bits | Encode Minimal Buffer<br>$^{ENC}$B ≥ $^{TX}$LD / ABT |

NOTE – We assume zero encode delay and a systematic forward error correction.

We need (or prefer at least) a (suite of) respectively scalable coding means we can resolve into a single design capable to (abstractly) implement all the multiple scales.

Because a coding means based upon a systematic FEC may transmit an input portion of the FEC frame before encode its parity portion finally and/or completely, see Tables VI and VII, a particular abstract design of such a coding means can affect the transmit delay (transmit side latency) and therefore must minimize it as much as possible. This objective is especially important, meeting the ground of the next.

Because a coding means based on a systematic FEC needs to receive an entire frame before decode it, see Table VIII, an abstract design cannot manage the receive delay (receive side latency) so as it can do against the transmit delay, that allows to exclude the former from the consideration.

## DETACHED DUTY DELIVERY

Given such the name and function, detached duty delivery, or DDD in short, is a transfer unit collection that, by design, is free of (literally, detached from) any information except of related to the original duty of the considered physical layers: each its element is an instance of the regular transfer unit and corresponds to either a data octet value or a control character code signal observed at the accommodation interface during the respective octet time period. Technically, we assume there is some single-round first-pass echo multiplexing process that brings the events beyond the units, preventing any substitution but/by accumulating the so generated echo.

Intended to support for quality of service (QoS), there are four distinct (user-distinguishable) traffic classes each related to the respective class of service (CoS), also one among four, all are apart from idle, see Table IX. Those traffic classes may correspond with either stream-processing or packet-bounding fragments of successive transfer units, depending on the goal of the implementer. As an extra, idle also may correspond with a separate stream of units, packetized or not.

It is expected that, preempting the traffic, the coding means mandatory follows all the preemption rules established in the traffic fragmentation analysis, see Tables X and XI. Following those rules ensures that, in its content, the accumulated (first-pass) echo will never exceed over five and six or six and five preempt and release fragmentation events, respectively, per the FEC input, i.e., this limits the possible income of that echo to up to five preempt (p) and one mixed (p/r) and five release (r) such the events per every FEC frame time period.

Purposely, the second-pass echo multiplexing process does on DDD to fulfill the DDD section inside the FEC input.

TABLE VIII
RECEIVE LEAST DELAY AND DECODE MINIMAL BUFFER

| PCS Interface | Current Frame Time Period | Next Frame Time Period |
|---|---|---|
| (output) Accommodation | Receive Least Delay<br>$^{RX}$LD = FT ↔ | 1,600 accommodation bits<br>1,600 × ABT = FT |
| ↑ RX DECODE | | |
| Transportation (input) | 2,048 × TBT = FT<br>1,600 + 448 = 2,048 transportation bits | Encode Minimal Buffer<br>$^{DEC}$B = $^{RX}$LD / TBT |

NOTE – We assume zero decode delay and a systematic forward error correction.

TABLE IX
CLASSES OF SERVICE

| ↓Who can preempt Whom→ | | Idle | Pp | Px | Xp | Xx |
|---|---|---|---|---|---|---|
| (fragment ≥ 64 units) Fragment of Traffic Class 36 ≤ A + B + C < 64 | Pp | anytime | never | never | never | never |
| | Px | anytime | sent ≥ A & left ≥ A | never | never | never |
| | Xp | anytime | anytime | sent ≥ B & left ≥ B | never | never |
| | Xx | anytime | anytime | anytime | sent ≥ C & left ≥ C | never |

NOTE – Sent / left numbers are in transfer units and of the preempted traffic class.

TABLE X
TRAFFIC PREEMPTION ANALYSIS

| Least Σ of Units → | 1 | 2 | 3 | 4 ÷ 67 | 68 ÷131 | 132 ÷ 195 | 196 | Remark |
|---|---|---|---|---|---|---|---|---|
| Tr. Unit Stream | ▌□ | ▌□ | ▌□ | ▌□ ▌⋯▌□ | ▌□ ▌⋯▌□ | ▌□ ▌⋯▌□ | ▌□ ▌ | 0 ≤ ▌ ≤ 4 |
| Fragmentation Events: Pp | | Σ | ⊏ | | | | | |
| Fragmentation Events: Xp | Σ | | | ⊏ ⋯ Σ | ⊏ ⋯ Σ | ⊏ ⋯ Σ | ⊏ | Σ▌⋮▌ = 4 |

NOTE – This is one among the equivalent worst cases excluding intra-preemption.

TABLE XI
TRAFFIC INTRA-PREEMPTION ANALYSIS

| Displacement → | +0 | +1 | +2 | +3 | ≥+(A\|B\|C) | ≤−(A\|B\|C) | −3 | −2 | −1 | −0 |
|---|---|---|---|---|---|---|---|---|---|---|
| Tr. Unit Stream | □ | □ | □ | □ | · · · □ | □ · · · | □ | □ | □ | □ |
| F-Events { Pp \| Px \| Xp / Px \| Xp \| Xx } | ⊏ | • | • | • | • • • • | • • • • | • | • | • | Σ |
| | ( Pp \| Px \| Xp units *sent* ≥ A \| B \| C ) ⊏ • • • Σ ( Pp \| Px \| Xp units *left* ≥ A \| B \| C ) | | | | | | | | | |

NOTE – Intra-preemption relates to the shaded periods in the preemption analysis.

TABLE XII
SUITABLE F-EVENT ENCODING

| F-Event | Content Structure / Content Modulus | Root Bits | Affix Bits |
|---|---|---|---|
| <P> PREEMPT | { empty, or choice 1 of 35 } × { P□ / X□ } × { □p / □x } × { ⊏ } | α,β,γ,δ ($N_r$ = 9) once/event | 3,4,5,6,– (4 bits) once/event |
| | $2^5 < 9{\bullet}2^2 < 2^6$ — $2^1$ — $2^1$ — $2^0$ | | |
| <R> RELEASE | { empty, or choice 1 of 35 } × { always implies it is the current fragment } × { Σ } | α,β,γ,δ ($N_r$ = 9) once/event | 3,4,–,–,– (2 bits) once/event |
| | $2^5 < 9{\bullet}2^2 < 2^6$ — $2^0$ — $2^0$ | | |
| <P/R> PREEMPT/ RELEASE | { empty, or choice 1 of 35 } × { ⊏ / Σ } × { P□ / X□ } × { □p / □x } | α,β,γ,δ ($N_r$ = 9) once/event | 3,4,5,6,7 (5 bits) once/event |
| | $2^5 < 9{\bullet}2^2 < 2^6$ — $2^1$ — $2^1$ — $2^1$ | | |

NOTE – We need to embed 5P + P/R + 5R fragmentation events into the FEC input.

TABLE XIII
TRANSFER UNIT ENCODING

| Tr. Unit | Content Structure / Content Modulus | Root Bits | Affix Bits |
|---|---|---|---|
| <D/C> DATA/CTRL | { control code choice 1 of 32 (assuming that 32 = 1 × 32 ), or data octet value choice 1 of 256 (assuming that 256 = 8 × 32) } | α,β,γ,δ ($N_r$ = 9) once/event | 3,4,5,6,7 (5 bits) once/event |
| | $2^8 < 9{\bullet}2^5 < 2^9$ | | |

NOTE – We need to embed $n_p$ = 200 transfer units (of the duty) into the FEC input.

TABLE XIV
SUITABLE S-EVENT ENCODING

| S-Event | Content Structure / Content Modulus | Root Bits | Affix Bits |
|---|---|---|---|
| <TS> TIMESTAMP | { empty, or choice 1 of up to 6,560 } × { relates to the frame time period of the input embedding the stamp } | α,β,γ,δ ($N_r$ = 9) four times/event | –,–,–,–,– (unused) |
| | $2^{12} < 9^4 < 2^{13}$ — $2^0$ | | |

NOTE – We need to embed just a single synchronization event into the FEC input.

TABLE XV
ACHIEVABLE RESOLUTION

| Protocol | 2.5GBASE-T | 5GBASE-T | "10GBASE-T" | $(N_r)^4 - 1$ |
|---|---|---|---|---|
| *f* @ M 1 / *f* → | 10 GHz @ 6,400 100 ps @ 2/1×Mref | 10 GHz @ 3,200 100 ps @ 1/1×Mref | 10 GHz @ 1,600 100 ps @ 1/2×Mref | 6,560 (Mmax) |

## EXPOSED ECHO EXPRESS

Given such the name, exposed echo express as it performs in general, or exposed event express as it is for a case like of ours, when the first-pass echo content consists only of event-related information, or EEE in short for both such functions, is a concomitant event collection inextricably accompanying the respective DDD collection that is collected during the transfer unit time periods, with which the same time observed events, when and if present, are directly associated.

Intended to support for seamless fragmentation, the so called fragmentation events, or f-events in short, comprising preempt (p), release (r), and mixed (p/r), correspond to such the stream switching signals as the backstage of the introduced classes of service, see Table XII in conjunction with Table IX and Tables X and XI. Their encoding is root-matched to the encoding of the regular transfer unit, see Table XIII.

Intended to support for precise synchronization, the also so called synchronization event, or s-event in short, corresponds to a pulse-per-interval signal acting like a relative timestamp, see Table XIV. Its encoding is root-matched to the encoding of the regular transfer unit and, therefore, of the f-events, too, see again Tables XIII and XII, respectively.

Thus, all the items passable into the FEC input, or passables in short, exhibit the same shape of content organization, in the view of their roots, all whose bits further must be represented, not affixes, all whose bits are simply bypassed.

Purposely, the second-pass echo multiplexing process does on EEE to fulfill the EEE section inside the FEC input.

## CONCLUSION

So, we highlighted a scenario enabling a coding means that simultaneously supports for precise synchronization, seamless preemption, and quality of service via differential classes of service, all resolved inside the Physical Layer of the ISO OSI seven-layer Basic Reference Model, in the manner and by the way we also described above and below, in a form of the single design we abstractly synthesized, proposed, and then specified in its major details, important to understand it.

To reach the objectives, we made the proposed design fully operable at the multiple scales we are interested in, beginning with the task of precise synchronization, see Table XV.

Regardless of the selected among those multiple scales, the proposed design determines the FEC input by the single plan, occupying that input completely and therefore leaving no free transportation bits in its whole, see Table XVI.

At its heart, the proposed design embodies an implementation of the linguistic multiplexing operations, in a form of the

TABLE XVI
FEC INPUT PLAN

| Passable | Mux Process | | Merged Root Bits | Bypassed Affix Bits |
|---|---|---|---|---|
| D/C # 1 | ↦ $(9•32)^5$ | $^{1÷5}\alpha$, | $^{1}\beta$, $^{1}\gamma$, $^{1}\delta$ | $^{1}3$, $^{1}4$, $^{1}5$, $^{1}6$, $^{1}7$ |
| D/C # 2 | $9^5$ • $2^{25}$ | Echo | $^{2}\beta$, $^{2}\gamma$, $^{2}\delta$ | $^{2}3$, $^{2}4$, $^{2}5$, $^{2}6$, $^{2}7$ |
| D/C # 3 | representation | Muxing | $^{3}\beta$, $^{3}\gamma$, $^{3}\delta$ | $^{3}3$, $^{3}4$, $^{3}5$, $^{3}6$, $^{3}7$ |
| D/C # 4 | ↓ ↓ | Round | $^{4}\beta$, $^{4}\gamma$, $^{4}\delta$ | $^{4}3$, $^{4}4$, $^{4}5$, $^{4}6$, $^{4}7$ |
| D/C # 5 | $2^{16}$ • $2^{25}$ | # 1 | $^{5}\beta$, $^{5}\gamma$, $^{5}\delta$ | $^{5}3$, $^{5}4$, $^{5}5$, $^{5}6$, $^{5}7$ |
| D/C # 6 | $(9•32)^5$ | $^{6÷10}\alpha$, | $^{6}\beta$, $^{6}\gamma$, $^{6}\delta$ | $^{6}3$, $^{6}4$, $^{6}5$, $^{6}6$, $^{6}7$ |
| D/C # 7 | . . . . . . . . . | . . . . | $^{7}\beta$, $^{7}\gamma$, $^{7}\delta$ | $^{7}3$, $^{7}4$, $^{7}7$, $^{7}6$, $^{7}7$ |
| . . . | . . . | . . . | . . . | . . . |
| D/C # 194 | ↓ ↓ | . . . . | $^{194}\beta$, $^{194}\gamma$, $^{194}\delta$ | $^{194}3$, $^{194}4$, $^{194}5$, $^{194}6$, $^{194}7$ |
| D/C # 195 | $2^{16}$ • $2^{25}$ | # 39 | $^{195}\beta$, $^{195}\gamma$, $^{195}\delta$ | $^{195}3$, $^{195}4$, $^{195}5$, $^{195}6$, $^{195}7$ |
| D/C # 196 | $(9•32)^5$ | $^{196÷200}\alpha$, | $^{196}\beta$, $^{196}\gamma$, $^{196}\delta$ | $^{196}3$, $^{196}4$, $^{196}5$, $^{196}6$, $^{196}7$ |
| D/C # 197 | $9^5$ • $2^{25}$ | Echo | $^{197}\beta$, $^{197}\gamma$, $^{197}\delta$ | $^{197}3$, $^{197}4$, $^{197}5$, $^{197}6$, $^{197}7$ |
| D/C # 198 | representation | Muxing | $^{198}\beta$, $^{198}\gamma$, $^{198}\delta$ | $^{198}3$, $^{198}4$, $^{198}5$, $^{198}6$, $^{198}7$ |
| D/C # 199 | ↓ ↓ | Round | $^{199}\beta$, $^{199}\gamma$, $^{199}\delta$ | $^{199}3$, $^{199}4$, $^{199}5$, $^{199}6$, $^{199}7$ |
| D/C # 200 | $2^{16}$ • $2^{25}$ | # 40 | $^{200}\beta$, $^{200}\gamma$, $^{200}\delta$ | $^{200}3$, $^{200}4$, $^{200}5$, $^{200}6$, $^{200}7$ |
| P # 1 (F) | $(9•16)^5$ | $^{201÷205}\alpha$, | $^{201}\beta$, $^{201}\gamma$, $^{201}\delta$ | $^{201}3$, $^{201}4$, $^{201}5$, $^{201}6$ — |
| P # 2 (F) | $9^5$ • $2^{20}$ | Echo | $^{202}\beta$, $^{202}\gamma$, $^{202}\delta$ | $^{202}3$, $^{202}4$, $^{202}5$, $^{202}6$ — |
| P # 3 (F) | representation | Muxing | $^{203}\beta$, $^{203}\gamma$, $^{203}\delta$ | $^{203}3$, $^{203}4$, $^{203}5$, $^{203}6$ — |
| P # 4 (F) | ↓ ↓ | Round | $^{204}\beta$, $^{204}\gamma$, $^{204}\delta$ | $^{204}3$, $^{204}4$, $^{204}5$, $^{204}6$ — |
| P # 5 (F) | $2^{16}$ • $2^{20}$ | # 41 | $^{205}\beta$, $^{205}\gamma$, $^{205}\delta$ | $^{205}3$, $^{205}4$, $^{205}5$, $^{205}6$ — |
| R # 1 (F) | $(9•4)^5$ | $^{206÷210}\alpha$, | $^{206}\beta$, $^{206}\gamma$, $^{206}\delta$ | $^{206}3$, $^{206}4$ — |
| R # 2 (F) | $9^5$ • $2^{10}$ | Echo | $^{207}\beta$, $^{207}\gamma$, $^{207}\delta$ | $^{207}3$, $^{207}4$ — |
| R # 3 (F) | representation | Muxing | $^{208}\beta$, $^{208}\gamma$, $^{208}\delta$ | $^{208}3$, $^{208}4$ — |
| R # 4 (F) | ↓ ↓ | Round | $^{209}\beta$, $^{209}\gamma$, $^{209}\delta$ | $^{209}3$, $^{209}4$ — |
| R # 5 (F) | $2^{16}$ • $2^{10}$ | # 42 | $^{210}\beta$, $^{210}\gamma$, $^{210}\delta$ | $^{210}3$, $^{210}4$ — |
| P/R (F) | $(9•1)^5$ →\| | $^{211÷215}\alpha$, | $^{211}\beta$, $^{211}\gamma$, $^{211}\delta$ | $^{211}3$, $^{211}4$, $^{211}5$, $^{211}6$, $^{211}7$ |
| TS (S) ¼ | $9^5$ • $2^0$ | Echo | $^{212}\beta$, $^{212}\gamma$, $^{212}\delta$ | — |
| TS (S) 2/4 | representation | Muxing | $^{213}\beta$, $^{213}\gamma$, $^{213}\delta$ | — |
| TS (S) ¾ | ↓ ↓ | Round | $^{214}\beta$, $^{214}\gamma$, $^{214}\delta$ | — |
| TS (S) 4/4 | $2^{16}$ • $2^0$ | # 43 | $^{215}\beta$, $^{215}\gamma$, $^{215}\delta$ | — |
| $\sum$... = $211\frac{4}{4}$ | $\sum$ Rounds = 43 | | $\sum$ Root Bits = 688 | $\sum$ Affix Bits = 1,035 |

TABLE XVII
LINGUISTIC MULTIPLEXING PROPERTIES

| Passable | ${}^{pass}n_p$ = ${}^{pass}n_e$ × ${}^{pass}k$ | ${}^{pass}(N_r \times N_b)$ | ${}^{pass}t$ = ${}^{pass}b$ + ${}^{pass}r$ | $w$ |
|---|---|---|---|---|
| D/C (duty) | 200 = 5 × 40 | 9 × 32 | 41 = 25 + 16 | ≤16 |
| P-area (F) | 5 = 5 × 1 | 9 × 16 | 36 = 20 + 16 | ≤16 |
| R-area (F) | 5 = 5 × 1 | 9 × 4 | 26 = 10 + 16 | ≤16 |
| P/R (F) + TS | $1\frac{4}{4}$ = $1\frac{4}{4}$ × 1 | 9×32, 9×1 | 21 = 5 + 16 | ≤16 |
| | $\sum n_p$ = $211\frac{4}{4}$   $\sum k$ = 43 | | $t_{max}$ = ${}^{D/C}t$ = 41   $r_{max}$ = 16 | ≤16 |

TABLE XVIII
ECHO BUDGET MEMO

| Passable | Budget Subline | Samples Within |
|---|---|---|
| D/C (duty) | Total Echo Pool Available . . . . . | $2^{16} \times 2^{25} = 65{,}536 \times 2^{25}$ |
| | ├ Native Echo Pool, pre-planned | $9^5 \times 2^{25} = 59{,}049 \times 2^{25}$ |
| | └ Rest, i.e., free to any other use | $(2^{16} - 9^5) \times 2^{25} = 6{,}487 \times 2^{25}$ |
| P-area (F) | Total Echo Pool Available . . . . . | $2^{16} \times 2^{25} = 65{,}536 \times 2^{20}$ |
| R-area (F) | Total Echo Pool Available . . . . . | $2^{16} \times 2^{10} = 65{,}536 \times 2^{10}$ |
| P/R (F) + TS | Total Echo Pool Available . . . . . | $2^{16} \times 2^{5} = 65{,}536 \times 2^{5}$ |

NOTE – In-rest echo pools may cover extra signaling, e.g., OAM, LPI, ordered set.

TABLE XIX
DATA ENCODING FLOW

| Passable | $N$ = $N_{data}$ + $N_{extra}$ | $8•n_e$ | | ${}^{ENC}B@32b$ ≥ | | ${}^{ENC}B@64b$ ≥ |
|---|---|---|---|---|---|---|
| D/C (duty) | 288 = 256 + 32 | 40 | << | (11×32=) 352 | < | (6×64=) 384 |
| Comment → | 9•32 = 8•32 + 1•32 | (8•5) | | with XGMII half cycles | | with XGMII full cycles |

TABLE XX
FEC FRAME BITS

| FEC Input: 2,048 − 325 = 1,723 bits in total, accessible by user | | | FEC Parity |
|---|---|---|---|
| D/C (duty) | P-area (F) + R-area (F) + P/R (F) + TS (S) | | hidden by PHY |
| Detached Duty Delivery | | EEE | Parity |
| 1,640 transportation bits | | 83 t-bits | 325 t-bits |

respective poly-round second-pass echo multiplexing process that covers root-uniformly over all the diverse passables into the FEC input, see Tables XVII, XVIII, and XIX.

Regardless of the selected scale, again, the proposed design results in a suite of the respective transportation bits fulfilling both the DDD section and the EEE section of the FEC frame, as its input, ready and complete, see Table XX in conjunction with Table XVI taken together afore Table II.

All these major details, i.e., on those we are focused in this paper, are enough for us to consider the proposed design, as well as the synthesis principles it was grounded on, applicable in circumstances of the original MGBASE-T physical layers, 2.5GBASE-T and 5GBASE-T [1], as well as of other similar protocols featuring a very long binary transport word then so exhibiting a comparatively excessive redundancy, respectively, with some restrictions and/or assumptions.[3] [4]

In the rest of its details, the proposed design is a mixture of the recipes characterized in [2], [3], [4], [5], and [6], applied cumulatively and complimentarily.

[3] In the constraining expression of $36 \leq \mathrm{A} + \mathrm{B} + \mathrm{C} < 64$, see Table IX, $64 = n_{\mathrm{fragment,min}}$ is the least fragment duration and $36 = n_{\mathrm{offset,max}}$ is the longest relative offset, including the "empty" variant, directly codable by a fragmentation event, all measured in transfer unit periods.

[4] There are up to $5 + 1 + 5 = 11$ (p + p/r + r = total) fragmentation events allowed (sensible) during each frame time period of $n_{\mathrm{p}} = 200$ transfer unit times, see Table XVI, that gives the following number of their combinations possible theoretically, however, not so many of those are practically valid:

$$\langle\text{diversity}\rangle = C_{200}^{0} + \cdots + C_{200}^{11} = \ldots\ldots\ldots\ldots = 411{,}473{,}991{,}891{,}891{,}896 < 0.42 \times 10^{18},$$

where $C_x^y$ is the number of $y$-combinations in a set of $x$ elements, $C_x^y = \binom{x}{y}$, while the transport capacity provided by the respective areas in the FEC input, excluding related to the CoS bits of the p-events, is about two and a half times greater than even that number of the possible event combinations:

$$\langle\text{capacity}\rangle = 36^{11} \times 2^3 = \ldots\ldots\ldots\ldots = 1{,}052{,}973{,}630{,}738{,}137{,}088 > 1.05 \times 10^{18},$$

where $2^3$ is the modulus related to the non-offset bits of the p/r-event. Thus, we conclude that all the event combinations can be unambiguously encoded in the proposed design, somehow by a particular implementation, one or another, keeping the focus, we consider any out of the scope of this paper.

# Data Coding Means and Event Coding Means Multiplexed Over the 100GBASE-R FEC Inputs

Alexander Ivanov

***Abstract*—We consider existing Ethernet protocols featuring a very long binary transport word, in the scope and hope that we can re-use their redundancy and, by re-distributing it, embed an extra, event transfer ability into them. In this paper, we consider the 100GBASE-R family of Ethernet physical layers.**

## Introduction

ONE-HUNDRED Gigabit Ethernet physical layers, types 100GBASE-KP4 and 100GBASE-KR4 for board-reach electrical backplanes and type 100GBASE-CR4 for in-cabinet twinax cable interlinks [1], all use the common 100GBASE-R Forward Error Correction (FEC) sublayer [1], that is based on the Reed–Solomon (RS) systematic FEC options of $(528, 514)$ and $(544, 514)$ GF($2^{10}$) symbols, dependently.

Every FEC frame time (FT) period of $51.2 = 512 \times 0.1$ ns, nominally, that 100GBASE-R FEC sublayer issues a frame of one of the two possible types, we further refer to as the usual FEC frame and the FEC frame with alignment markers (AMs), normal or rapid, depending on should the payload of AMs be absent (usually) or present (rarely) in the currently issued FEC frame, see Tables I and II, respectively.[1]

Given such the problem, we perform a search for an appropriate coding scheme, based on the already proven recipes, in each of their turns, developed in [2], [3], [4], [5], [6], and [7], into the left from the right, successively.

Due to the two FEC frame types, mostly similar but distinct enough to necessitate for a dedicated treatment upon each, we plan also two linguistic multiplexing processes, each featuring the so called fixed part of $7 \times 513 + 10 = 3{,}601$ transportation bits, expectedly common for the both, and a so called varying part of $5{,}140 - 3{,}601 = 1{,}539$ transportation bits, peculiar for the usual FEC frame as well as for the FEC frame with AMs, respectively, see Tables III and IV.

Assuming the rest seemed trivial at the rich background of the known design recipes mentioned earlier, the most share of our current work lies in the area of representation procedures, primarily, of their implementation techniques and performance characteristics, actual in the new circumstances.[2]

TABLE I
USUAL FEC FRAME

| Hierarchy and Elements | Contents, Comments | Size, t-bits |
|---|---|---|
| FEC [Forward Error Correction] frame | nominal 19,531,250 frames/s | Σ 5,280 or 5,440 |
| ├ **Input, accessible by user** | **max 640** data octets→5,120 a-bits | 20 × 257 = 5,140 |
| ├ 256B/257B coded block #1 | ├ first 32 data octets or controls | ├ 32×8+1= 257 |
| . . . . . . . . . . . . . . . . . . . . | . . . . . . . . . . . . . . . . . . . . . . . . | . . . . . . . . . |
| └ 256B/257B coded block #20 | └ last 32 data octets or controls | └ 32×8+1= 257 |
| └ Parity, RS(528 or 544, 514) | calculated over all the bits above | 140 or 300 |

TABLE II
FEC FRAME WITH AMS

| Hierarchy and Elements | Contents, Comments | Size, t-bits |
|---|---|---|
| FEC [Forward Error Correction] frame | from every 2nd (4,096th) frame for RAMs (NAMs) | Σ 5,280 or 5,440 |
| ├ Input, dedicated to AMs | includes 1280 payload and 5 padding t-bits | 5 × 257 = 1,285 |
| ├ **Input, accessible by user** | **max 480** data octets→3,840 a-bits | 15 × 257 = 3,855 |
| ├ 256B/257B coded block #16 | ├ first 32 data octets or controls | ├ 32×8+1= 257 |
| . . . . . . . . . . . . . . . . . . . . | . . . . . . . . . . . . . . . . . . . . . . . . | . . . . . . . . . |
| └ 256B/257B coded block #20 | └ last 32 data octets or controls | └ 32×8+1= 257 |
| └ Parity, RS(528 or 544, 514) | calculated over all the bits above | 140 or 300 |

TABLE III
PROPOSED MULTIPLEXING APPROACH

| Frame | Varying Part of the FEC Input | | Fixed Part of the FEC Input | |
|---|---|---|---|---|
| Usual | 1st, 2nd, and 3rd Complete Echo Multiplexing Rounds | | 4th to 10th Complete Echo Multiplexing Rounds | Spare Bits |
| w/AMs | RAM/NAM Payload + Rest Padding | 3rd Partial Round | | |
| Size → | 5×257−3 = 1,280+2 t-bits | 1×257 t-bits | 7×513 transportation bits | 10 t-bits |

TABLE IV
LINGUISTIC MULTIPLEXING PROPERTIES

| Frame | $v$ | $^*n_\mathrm{p}$ | = | $n_\mathrm{e}$ | × | $k$ | + | $^*n_\mathrm{e}$ | × | $^*k$ | $t$ | $^*t$ | $s$ |
|---|---|---|---|---|---|---|---|---|---|---|---|---|---|
| Usual | 5,140 | 640 | = | 64 | × | 10 | | — | | — | 513 | — | 10 |
| w/AMs | 3,855 +3 | 480 | = | 64 | × | 7 | + | 32 | × | 1 | 513 | 257 | 7 +3 |
| Measured in | t-bits/frame | units/frame | | units/round | | rounds/frame | | units/round | | rounds/frame | t-bits/round | t-bits/round | t-bits/frame |

A manuscript of this work was submitted to IEEE Communications Letters December 26, 2022 and rejected for fair reasons during its peer review.

Please sorry for the author has no time to find this work a new home, peer reviewed or not, except of arXiv, and just hopes there it meets its reader, one or maybe various, whom the author beforehand thanks for their regard.

A. Ivanov is with JSC Continuum, Yaroslavl, the Russian Federation.

[1]During its "normal" mode of operation, the 100GBASE-R FEC sublayer issues a frame with normal alignment markers (FEC frame with NAMs) once from every 4,096th frame, that gives a bit stream of about 1.526 Mbps for each per-lane payload of NAMs, measured at the service interface of the respective physical medium attachment (PMA) sublayer. During a "transitional" mode, e.g., recovering from a low-power idle, instead of the former, the same entity issues a frame with rapid alignment markers (FEC frame with RAMs) once from every second frame. The latter gives a bit stream of exact 3.125 Gbps for each per-lane payload of RAMs, though, it ends up as fast as the so supported mode of the 100GBASE-R FEC operation ends up, too.

[2]At such high speeds of operation, like the considered speed of 100 Gbps, an arithmetic-based implementation of the representation procedure begins to seem very costly in the terms of power dissipation and silicon complexity, so now we need to discover for an appropriate logic-based one.

TABLE V
REPRESENTATION FUNCTION DEMONSTRATION

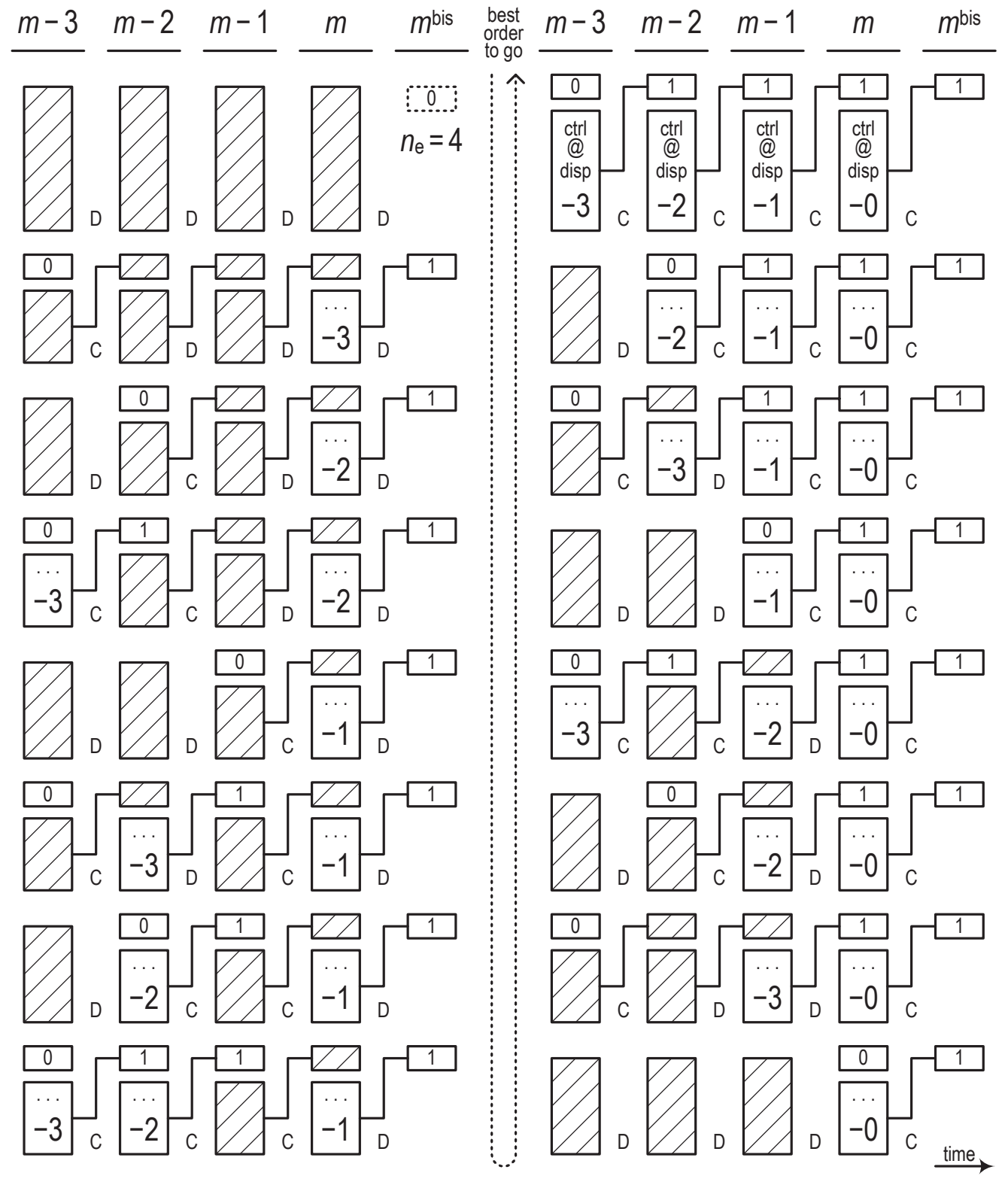

TABLE VI
EQUIVALENT RELAYING FRAMEWORK

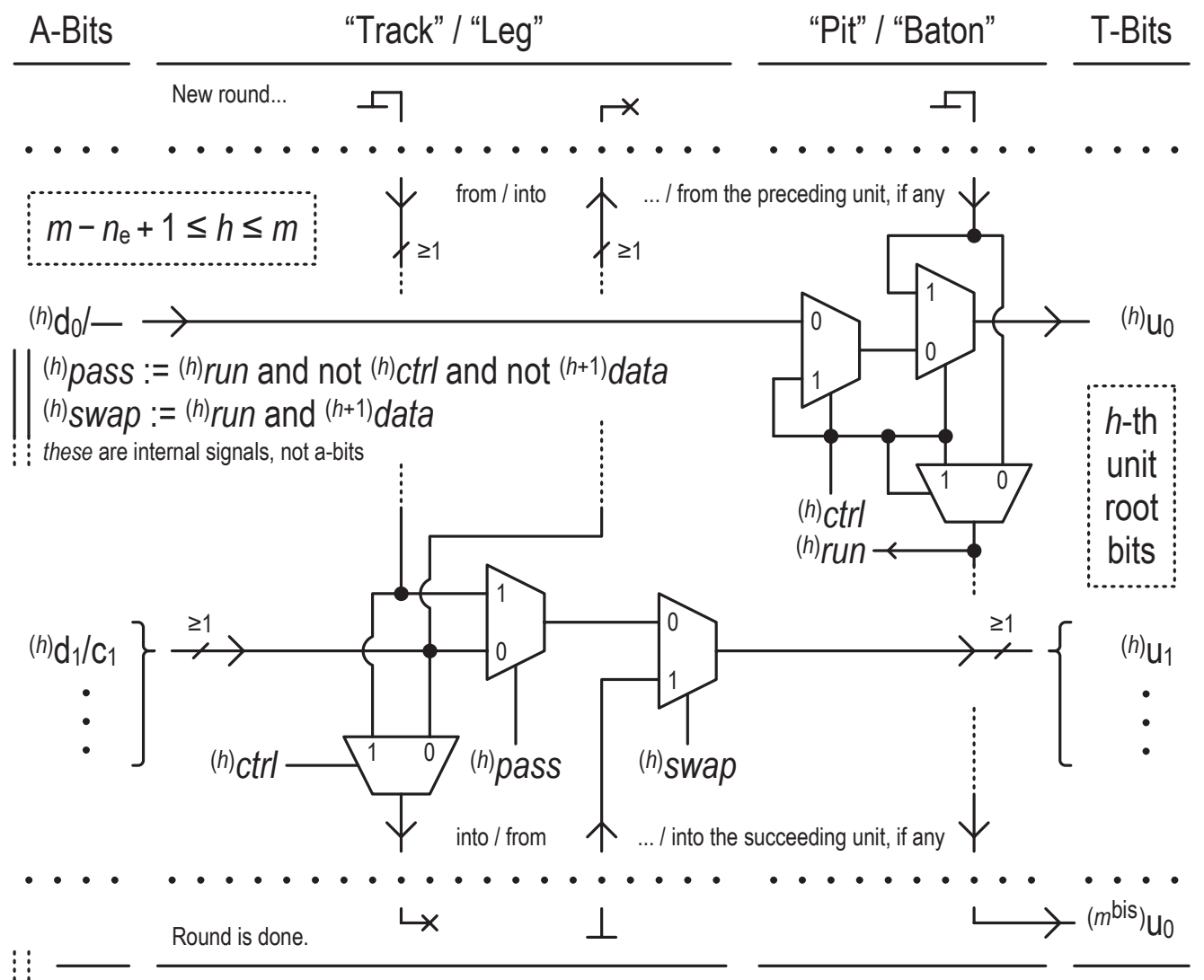

$^{(h)}ctrl$ := 1, if $h$-th is a control unit within the current round, and 0 otherwise.
$^{(h+1)}data$ := 1, ($h$+1)-th is a data unit within the current round, and 0 otherwise.

TABLE VII
CONTROL MANDATORY CONTENT

| Content Structure and Content Modulus | Constraints | Upper Limit |
|---|---|---|
| {displacement ($m-h$): choice 1 of $n_e$} × {$h$-th control code: choice 1 of $N_{ctrl}$} | $1 \le n_e \bullet N_{ctrl} \le N_{extra}$ | $N_{extra} = 2^{8-1}$ |

TABLE VIII
TRANSFER UNIT ENCODING—BINARY-PURE APPROACH

| Bit Role @ $n_e$ | | Bit Role @ $N_{ctrl}$ | | D/C A-Bits | | Bit Role @ $n_e$ | | Bit Role @ $N_{ctrl}$ | |
|---|---|---|---|---|---|---|---|---|---|
| "pit" (static) | any | "pit" (static) | any | $d_0$ | — | "baton" | any | "baton" | any |
| "track" | any | "track" | any | $d_1$ | $c_1$ | "leg" | any | "leg" | any |
| "track" | $=2^{\ge 2}$ | "fan" (static) | $=64$ | $d_2$ | $c_2$ | "leg" | $=2^{\ge 2}$ | "fan" (static) | $=64$ |
| "track" | $=2^{\ge 3}$ | "fan" (static) | $=2^{>4}$ | $d_3$ | $c_3$ | "leg" | $=2^{\ge 3}$ | "fan" (static) | $=2^{>4}$ |
| "track" | $=2^{\ge 4}$ | "fan" (static) | $=2^{>3}$ | $d_4$ | $c_4$ | "leg" | $=2^{\ge 4}$ | "fan" (static) | $=2^{>3}$ |
| "track" | $=2^{\ge 5}$ | "fan" (static) | $=2^{>2}$ | $d_5$ | $c_5$ | "leg" | $=2^{\ge 5}$ | "fan" (static) | $=2^{>2}$ |
| "track" | $=2^{\ge 6}$ | "fan" (static) | $=2^{>1}$ | $d_6$ | $c_6$ | "leg" | $=2^{\ge 6}$ | "fan" (static) | $=2^{>1}$ |
| "track" | $=128$ | "fan" (static) | $=2^{>0}$ | $d_7$ | $c_7$ | "leg" | $=128$ | "fan" (static) | $=2^{>0}$ |

NOTE – We consider any "track"/"leg" or "pit"/"baton" accommodation bit a root bit.

TABLE IX
TRANSFER UNIT ENCODING—BINARY-PURE VARIANTS

| $N - N_{data} = N_{extra} = n_e \bullet N_{ctrl} = 2^7$ | $\xi$ | $t-\xi$ | $b$ | $r-\xi$ |
|---|---|---|---|---|
| 2•64; 4•32; 8•16; 16•8; 32•4; 64•2; 128•1 | 1 | $8 \bullet n_e$ | $n_e \bullet \log_2 N_{ctrl}$ | $n_e \bullet \log_2 n_e$ |

NOTE – That approach uses individual control displacement and control code bits.

TABLE X
ECHO BUDGET MEMO

| Pool | Samples Within | $m - n_e < h < m^{bis+\xi}$ | Signature |
|---|---|---|---|
| | | pattern | sum over 0-th bit row |
| Total | $pow(2, t) = pow(2, r) \times pow(2, b)$ | | |
| Native | $pow(2, t) - pow(2, 7 \bullet n_e)$ | 0/1 • • • • 0/1 | $\Sigma^{(h)}u_0 < n_e + \xi$ |
| Rest | $pow(2, 7 \bullet n_e)$ [i.e., excluding 0-th bit row] | 1 • • • • 1 | $\Sigma^{(h)}u_0 = n_e + \xi$ |

## RELAY RACE REPRESENTATION

We introduce the so called (because so working) relay race representation, our newbie, to be applied by us during an echo multiplexing round after now, instead of the so called (because so working, too) base change representation we applied afore, visualizing one pictorial of its possible functions, see Table V (example with $n_e = 4$, only roots are shown).

Unlike of its older predecessor, that is a more complicated, arithmetic-based procedure we plan to replace, the newbie is a completely (in general) or mostly (at least) logic-based, thus, simpler one. But also like of its predecessor, the newbie is still easily intelligible–perceptible and scalable–adaptable because both memory-less and pattern-ness, see Table VI.

Concerning the regular transfer unit capable to hold a single item—responding to either a data octet value or a control code observed during an octet time period—of the original duty, the application of the newbie necessitates for the relative position of the respective period to be described in the content of every instance of the regular transfer unit containing a control code, along with that code itself, see Table VII.

Differentiating, a completely logic-based option of the newbie results from the so called (because so selected) binary-pure approach on the transfer unit encoding, that defines only such situations when all the bits (if any) keeping a control code and all the bits keeping its position are separated explicitly and so treatable separately, see Tables VIII and IX.

When routinely involved during a round of the implemented linguistic multiplexing process, an option of the newbie gen-

TABLE XI
TRANSFER UNIT ENCODING—BINARY-CODED APPROACH

| Bit Role @ $n_e$ | | Bit Role if $N_{ctrl}$ | | D/C A-Bits | | Bit Role @ $n_e$ | | Bit Role if $N_{ctrl}$ | |
|---|---|---|---|---|---|---|---|---|---|
| "pit" (static) | any | "pit" (static) | any | $d_0$ | $c_0$ | "baton" | any | "baton" | any |
| "track" | any | "track" | any | $d_1$ | $c_1$ | "leg" | any | "leg" | any |
| "track" | any | "track" | any | $d_2$ | $c_2$ | "leg" | any | "leg" | any |
| "track" | ≥5 | "fan" | is divisible by $2^5$ | $d_3$ | $c_3$ | "leg" | ≥5 | "fan" | is divisible by $2^5$ |
| "track" | ≥9 | "fan" | $2^4$ | $d_4$ | $c_4$ | "leg" | ≥9 | "fan" | $2^4$ |
| "track" | ≥17 | "fan" | $2^3$ | $d_5$ | $c_5$ | "leg" | ≥17 | "fan" | $2^3$ |
| "track" | ≥33 | "fan" | $2^2$ | $d_6$ | $c_6$ | "leg" | ≥33 | "fan" | $2^2$ |
| "track" | ≥65 | "fan" | $2^1$ | $d_7$ | $c_7$ | "leg" | ≥65 | "fan" | $2^1$ |

NOTE – We consider any "fan" accommodation bit, only when present, an affix bit.

TABLE XII
TRANSFER UNIT ENCODING—BINARY-CODED VARIANTS

| Parameter→ | Multiplexing Round Duration, $n_e$ | In-unit Control Modulus, $N_{ctrl}$ |
|---|---|---|
| Constraints | $2^1 < n_e < 2^7$, $n_e$ is not a power of 2 | $2 < n_e \bullet N_{ctrl} < 128$ with $N_{ctrl} > 0$ |

NOTE – That approach reversibly merges displacement and code info in a control.

TABLE XIII
NESTED ECHO BUDGET—BINARY-CODED UNIT

| Pool | Samples Within | Applicability | Remark |
|---|---|---|---|
| Total | $N = N_{data} + N_{extra} = 2^8 + 2^{8-1} = 384$ | $^{(h)}$u : $m - n_e < h \leq m$ ($n_e$ transfer units) | only use of the native samples is in the scope of this paper |
| Native | $N_{data} + n_e \bullet N_{ctrl}$ [i.e., including data and ctrl] | | |
| Rest | $N - N_{data} - n_e \bullet N_{ctrl} = N_{extra} - n_e \bullet N_{ctrl}$ | | |

erates samples from its native echo pool. However, the total echo pool always provides an extra space called the rest echo pool, of distinguishable samples, besides of the native, which are free for other signals, e.g., when forced, see Table X.

Differentiating further, any mostly logic-based option of the newbie results from the so called (because so set, too) binary-coded approach on the transfer unit encoding, that defines the rest situations, i.e., when it is impossible to separate at least a part of the bits and, therefore, such the bits are treatable only together, see Tables XI, XII, and XIII.

To yet more understand how the newbie works, step by step, one could suppose a multi-leg relay race run, where each leg runner corresponds to a control-containing unit and the place the leg starts at corresponds to the relative position of the unit in the stream. Hence, the same one could suppose the function of the newbie as a selective overlay of the successively made photo shots, each fixing either a moment when two adjacent legs pass the baton, or that moment when the final-leg runner crosses the finish line and so ends the race.[3] By our turn, we try to help in this, visualizing a case, see Table XIV (example with $n_e = 9$ and three controls in some positions).

Closing the introduction and its accompanying comprehensive and (we believe) exhaustive description of our newbie, we also need to remember that any selected option of the newbie results in a function that is fully reversible, unambiguously as well as clearly and easily, see Table XV (example to turn back the representation done just above).

[3]Running, a leg pushes back the track under the leg's strong legs.

TABLE XIV
REPRESENTATION FLOW DEMONSTRATION

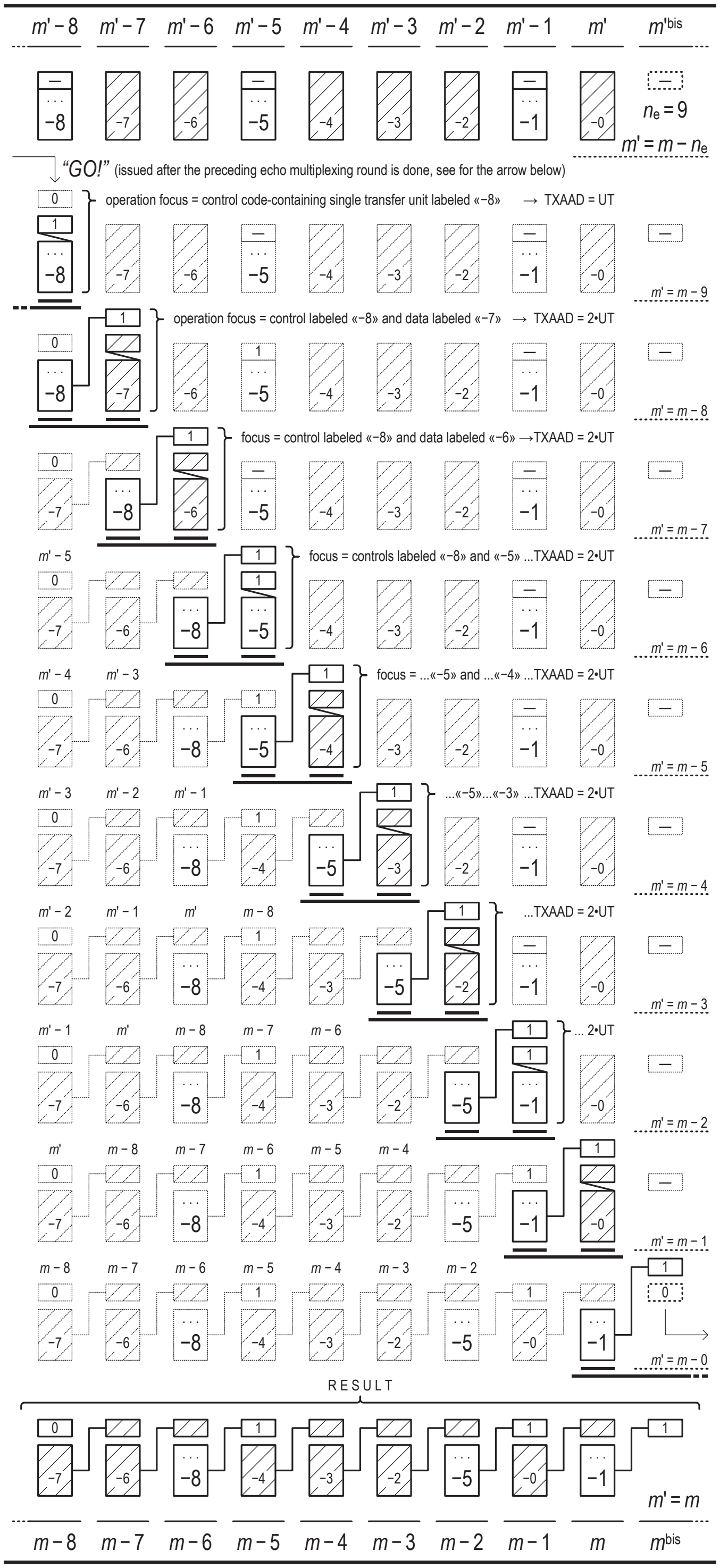

TABLE XV
DEMULTIPLEXING PATH RECONSTRUCTION

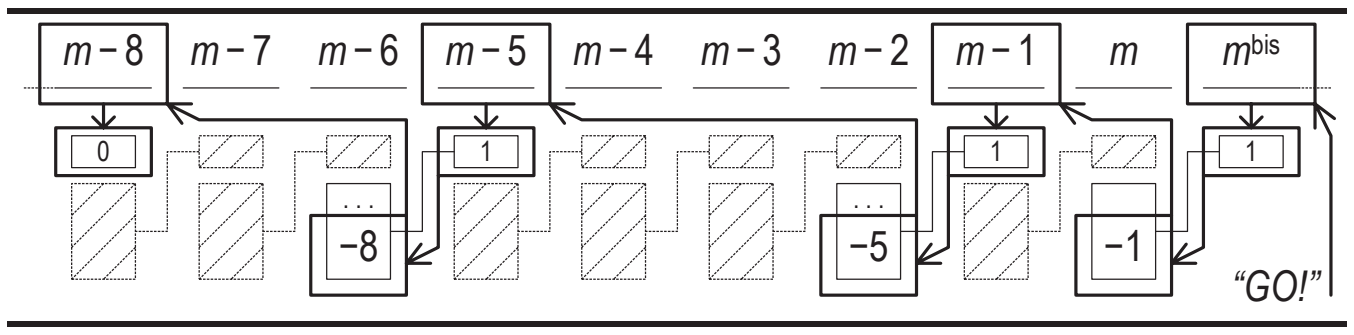

TABLE XVI
TRANSMIT ARGUMENT ACCUMULATION DELAY

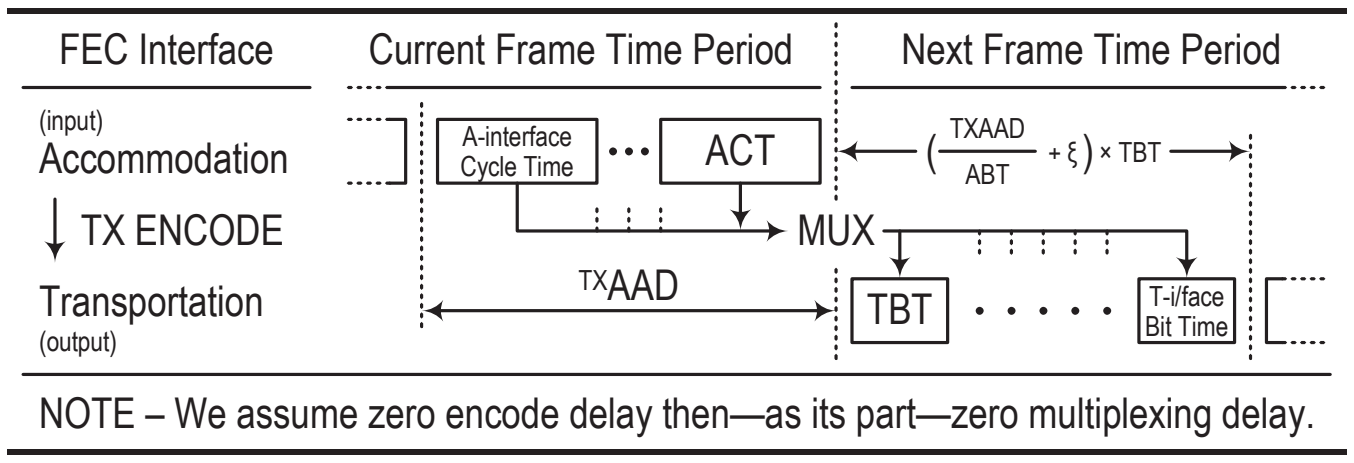

TABLE XVII
PAYLOAD MAPPING

| T-Bit | Usual Input, rate ≤ 4,095/4,096 | | Input with AMs, rate ≥ 1/4,096 | | Size |
|---|---|---|---|---|---|
| 0.0 | $^{(1)}u_0$ | 1st complete round | — | RAM/NAM bit # 0 | (in t-bits) |
| ⋮ | ⋮ | ⋮ | ⋮ | ⋮ | |
| 51.2 | $^{(64^{bis})}u_0$ | 1st complete round | — | RAM/NAM bit # 512 | |
| 51.3 | $^{(65)}u_0$ | 2nd complete round | — | RAM/NAM bit # 513 | |
| ⋮ | ⋮ | ⋮ | ⋮ | ⋮ | |
| 102.5 | $^{(128^{bis})}u_0$ | 2nd complete round | — | RAM/NAM bit # 1025 | 1,282 |
| 102.6 | $^{(129)}u_0$ | 3rd complete round | — | RAM/NAM bit # 1026 | |
| ⋮ | ⋮ | ⋮ | ⋮ | ⋮ | |
| 127.9 | $^{(160)}u_5$ | 3rd complete round | — | RAM/NAM bit # 1279 | |
| 128.0 | $^{(160)}u_6$ | 3rd complete round | — | AM rest padding # 1 | |
| 128.1 | $^{(160)}u_7$ | 3rd complete round | — | AM rest padding # 2 | Σ 1,282 |
| 128.2 | $^{(161)}u_0$ | 3rd complete round | $^{(161)}u_0$ | 3rd partial round | |
| ⋮ | ⋮ | ⋮ | ⋮ | ⋮ | 257 |
| 153.8 | $^{(192^{bis})}u_0$ | 3rd complete round | $^{(192^{bis})}u_0$ | 3rd partial round | Σ 1,539 |
| 153.9 | $^{(193)}u_0$ | 4th complete round | $^{(193)}u_0$ | 4th complete round | |
| ⋮ | ⋮ | ⋮ | ⋮ | ⋮ | 3,591 |
| 512.9 | $^{(640^{bis})}u_0$ | 10th complete round | $^{(640^{bis})}u_0$ | 10th complete round | Σ 513×10 |
| 513.0 | — | spare bit # 0 | — | spare bit # 0 | |
| ⋮ | ⋮ | ⋮ | ⋮ | ⋮ | $s = 10$ |
| 513.9 | — | spare bit # 9 | — | spare bit # 9 | Σ 514×10 |

TABLE XVIII
ACHIEVABLE RESOLUTION

| Scenario | "Enough" | "Octet" | "Limit" | $2^s - 1$ |
|---|---|---|---|---|
| $f$ @ M<br>1 / $f$ → | 10 GHz @ 512<br>100 ps @ ~50% of Mmax | 12,5 GHz @ 640<br>80 ps @ ~63% of Mmax | ~20 GHz @ 1,023<br>~50 ps @ 100% of Mmax | 1,023<br>(Mmax) |

TABLE XIX
ACCOMMODATION SERVICE IMPLEMENTATION

| Transfer Unit→ | Data (D/C=D) | Control (D/C=C) |
|---|---|---|
| Passable<br>(comment) | choice one of $N_{data} = 256$ items<br>(data octet values, 0x00 to 0xFF) | choice one of $N_{ctrl} = 2$ items<br>(control codes, data error + idle/IFG) |

TABLE XX
BACKWARD APPLICABILITY

| Protocol | $v$ | $^*n_p$ | = | $n_e$ | × | $k$ | + | $^*n_e$ | × | $^*k$ | $s$ | $N_{ctrl}$ |
|---|---|---|---|---|---|---|---|---|---|---|---|---|
| 2.5/5GBASE-T | 1,723 | 212 | = | 16 | × | 13 | + | 4 | × | 1 | 13 | 8 |
| 25/40GBASE-T | 3,211 | 400 | = | 64 | × | 6 | + | 16 | × | 1 | 4 | 2 |
| 10GBASE-T, MGBASE-T1 (M=2.5/5/10) | 3,250 | 400 | = | 16 | × | 25 | | — | | — | 25 | 8 |
| | | 400 | = | 32 | × | 12 | + | 16 | × | 1 | 27 | 4 |
| | | 400 | = | 64 | × | 6 | + | 16 | × | 1 | 43 | 2 |
| 10GBASE-KR | 2,080 | 256 | = | 64 | × | 4 | | — | | — | 28 | 2 |
| 1000BASE-T1 | 3,645 | 450 | = | 64 | × | 7 | + | 2 | × | 1 | 37 | 2 |

## CONCLUSION

Thus, we successfully establish the new representation procedure featuring, among others, a very small transmit argument accumulation delay, TXAAD, of just twice the accommodation interface cycle time (ACT), see Table XVI.[4]

Based on this new procedure, we implement the two initially planned linguistic multiplexing processes setting the input of the usual FEC frame or the input of the FEC frame with AMs (RAMs or NAMs), respectively, see Table XVII.

The process responsible for the input of the usual FEC frame sets $5{,}130$ transportation bits of that input in accordance with the content of $n_\mathrm{p} = 640$ instances of the regular transfer unit, leaving the rest $s = 514 \times 10 - 5{,}130 = 10$ t-bits spare.

The process responsible for the input of the FEC frame with AMs first preserves $1{,}280 + 2 = 1{,}282$ t-bits for the payload of AMs and its padding, after sets $257 + 3{,}591 = 3{,}848$ t-bits in accordance with $n_\mathrm{p} = 480$ instances, leaving the same rest of $s = 514 \times 10 - 1{,}282 - 3{,}848 = 10$ t-bits spare, too.

Applying these processes to the so modified 100GBASE-R FEC sublayer [1], we enable an extra, event translation ability in a form of those spare transportation bits, a ten ($s = 10$) per every frame issued by the sublayer, see Table XVIII.[5]

Moreover, we make sure that the so modified sublayer fulfills the duty of its original one, appropriately and sufficiently, highlighting the essential details of the service provided at its accommodation side, i.e., on its top, see Table XIX.

Finally, we find the new representation procedure applicable in similar Ethernet protocols with a very long binary transport word, like considered earlier in [2], [3], [4], [5], [6], and [7], varying in their speed and duty, see Table XX.[6]

[4] A binary-coded option of the proposed representation procedure is similar to an inversed (in time) replica of the 80B/81B block encoding introduced in 1000BASE-T1 we regarded in [7]. Such the replica will demonstrate the same (with every option of the proposed procedure) transmit argument accumulation delay of just 2 ACTs versus 10 ACTs as for the original encoding.

[5] Assuming the "enough" and "octet" variants, we could re-map the payload of NAMs into the most (latest) eight spare t-bits of every issued FEC frame, which is free of the need to propagate an event. With a single pulse per second event signaling, it would give a bit stream of about 39.062 Mbps for each per-lane payload of NAMs, that allows for the so re-modified 100GBASE-R-like sublayer to issue no dedicated FEC frame with NAMs at all.

[6] All designs proposed in this table assume no transfer unit gathering [4].

# Data Coding Means and Event Coding Means Multiplexed Over the 10G/25G-EPON FEC Inputs

Alexander Ivanov

*Abstract*—**Among existing Ethernet protocols featuring a very long binary transport word, there are some whose word we could call not very but extremely long. In this paper, we consider one of such the protocols, 25G-EPON. Also, we consider its lower-speed companion/predecessor, 10G-EPON. The objective of our search is the same as before: we try to extend the selected coding means with an event delivery function that is primarily intended to give a basic support for a precise synchronization task.**

*Index Terms*—**Ethernet, linguistic multiplexing, multiplexing, precise synchronization, synchronization, Ethernet Passive Optical Network, EPON, 10G-EPON, 25G-EPON.**

## Introduction

ETHERNET Passive Optical Network, or EPON in short, introduces a number of multi-gigabit physical layers and their corresponding FECs, two of those are in the focus of this paper, with speeds of 10 Gbps [1] and 25 Gbps [2].

Considering the 10 Gbps Forward Error Correction option, 10G-EPON FEC [1], we imply a tree-like, point-to-multipoint network whose trunk node runs a sole instance of the 10 Gbps optical line terminal, 10G-OLT, as well as whose branch nodes run instances of the 10 Gbps optical network unit, 10G-ONU, solely and jointly forming the 10 Gbps downstream, 10G-DS, and the 10 Gbps upstream, 10G-US, of information that travels from this trunk into all these branches and back, respectively, based on the unified framing rules, see Table I.

Considering the 25 Gbps Forward Error Correction option, 25G-EPON FEC [2], we imply a tree-like, point-to-multipoint network whose trunk node runs a sole instance of the 25 Gbps optical line terminal, 25G-OLT, as well as whose branch nodes run instances of the 25 Gbps optical network unit, 25G-ONU, solely and jointly forming the 25 Gbps downstream, 25G-DS, and the 25 Gbps upstream, 25G-US, of information that travels from this trunk into all these branches and back, respectively, based on the differentiated rules, see Tables II and III.

TABLE I
ORIGINAL 10G-EPON FRAMING

| Hierarchy and Elements | Contents, Comments | Size, t-bits |
|---|---|---|
| FEC [Forward Error Correction] frame | up to about 5.04 Mio frames/s | 31 × 66 = 2,046 |
| ├ **Input**, accessible by user | max 216 data octets→1,728 a-bits | 27 × 65 = **1,755** |
| ├ **65 bits** of $^{64B}/_{66B}$ block **#1** | ├ first 8 data octets or controls | ├ 8×8+1 = 65 |
| ├ **65 bits** of $^{64B}/_{66B}$ block **#2** | ├ next 8 data octets or controls | ├ 8×8+1 = 65 |
| . . . . . . . . . . . . . . . . . . | . . . . . . . . . . . . . . . . . . . . . | . . . . . . . . |
| ├ **65 bits** of $^{64B}/_{66B}$ block **#26** | ├ prev 8 data octets or controls | ├ 8×8+1 = 65 |
| └ **65 bits** of $^{64B}/_{66B}$ block **#27** | └ last 8 data octets or controls | └ 8×8+1 = 65 |
| ├ Parity, auto-generated | calculated over only the bits above | 32 × 8 = 256 |
| ├ body bits of 66B block #28 | ├ 1st eight GF($2^8$) parity symbols | ├ 8×8+0 = 64 |
| ├ body bits of 66B block #29 | ├ 2nd eight GF($2^8$) parity symbols | ├ 8×8+0 = 64 |
| ├ body bits of 66B block #30 | ├ 3rd eight GF($2^8$) parity symbols | ├ 8×8+0 = 64 |
| └ body bits of 66B block #31 | └ 4th eight GF($2^8$) parity symbols | └ 8×8+0 = 64 |
| └ Rest, auto-generated, too | intended to complete the blocks | 27×1 + 4×2 = 35 |
| ├ one bit of $^{64B}/_{66B}$ block #1 | ├ inverted sync of $^{64B}/_{66B}$ block #1 | ├ 0×0+1 = 1 |
| ├ one bit of $^{64B}/_{66B}$ block #2 | ├ inverted sync of $^{64B}/_{66B}$ block #2 | ├ 0×0+1 = 1 |
| . . . . . . . . . . . . . . . . . . | . . . . . . . . . . . . . . . . . . . . . . . . | . . . . . . . . |
| ├ one bit of $^{64B}/_{66B}$ block #26 | ├ inverted sync of $^{64B}/_{66B}$ block #26 | ├ 0×0+1 = 1 |
| ├ one bit of $^{64B}/_{66B}$ block #27 | ├ inverted sync of $^{64B}/_{66B}$ block #27 | ├ 0×0+1 = 1 |
| ├ sync bits of 66B block #28 | ├ always 00 [signals the block is not a data/ctrl] | ├ 0×0+2 = 2 |
| ├ sync bits of 66B block #29 | ├ always 11 [signals the block is not a data/ctrl] | ├ 0×0+2 = 2 |
| ├ sync bits of 66B block #30 | ├ always 11 [signals the block is not a data/ctrl] | ├ 0×0+2 = 2 |
| └ sync bits of 66B block #31 | └ always 00 [signals the block is not a data/ctrl] | └ 0×0+2 = 2 |

Regardless of a particular FEC option, the only its piece we are interested in, is the input, as a vector of transportation bits accessible by the user, with nothing else.

Thru the rest of this paper, we describe a way to enable an extra, event transfer ability inside the coding means based on the mentioned FEC options of EPON, also accounting for its asymmetric[1] and asynchronous[2] nature.

A manuscript of this work was submitted to IEEE Communications Letters December 26, 2022 and rejected for fair reasons during its peer review.

Please sorry for the author has no time to find this work a new home, peer reviewed or not, except of arXiv, and just hopes there it meets its reader, one or maybe various, whom the author beforehand thanks for their regard.

A. Ivanov is with JSC Continuum, Yaroslavl, the Russian Federation.

[1]The asymmetric core (essence, nature) of EPON may or may not manifest itself quantitatively, i.e., there may be an inequality between those DS and US speeds (bit rates), but anyway does the such qualitatively, i.e., there must be the expected difference between the DS and US behaviors (operations). In the EPON architecture, the OLT is the exclusive source of the DS frames and they are visible to all the ONUs present in that network. The same time, any ONU can issue its US frames but they will be visible to the OLT only. Performing their operations, the OLT issues the DS frames continuously, never turning off its laser, while an ONU issues its US frames in a variable-length burst, turning on and off its laser before and after each of its bursts, respectively. Moreover, an ONU in 25G-EPON shows a number of the sizes its US frames can be of, that does neither the OLT that uses one fixed size regarding of the DS speed, nor any ONU in 10G-EPON. Therefore, that difference makes the underlying coding means of EPON more complex than the same means of the protocols considered in [3], [4], [5], [6], [7], [8], and [9], but not untreatable, including thanks to the design recipes raised in the development of those originals.

[2]The asynchronous core of EPON manifests itself in the so called floating, ambiguous—as it seems to the user—relationship between the time interval a portion of the duty is factually sensed and the time interval it is actually sent: this happens due to insertion and deletion of the FEC parity information. This is like in 100GBASE-R considered in [3], where it happens due to the same manipulations, but with the (off-duty) payload related to alignment markers, however oppositely, this is unlike in the other protocols considered in [4], [5], [6], [7], [8], and [9], where every transfer quantity (unit or span) time period relates with only one, pre-known quantity period (absolute) in some quantity-conveying stream, as it is observed at the accommodation service interface, as well as only one, pre-defined quantity position (relative) in the FEC input, as it is observed at the transportation service interface, respectively. Therefore, we should clear that in the context of EPON, 100GBASE-R, and similar coding means, asynchronous in their work (behavior), any association made between a given quantity period and its time period, must always be read conditional, regardless of such was explicitly stated or just implicitly assumed.

TABLE II
ORIGINAL 25G-EPON DOWNSTREAM FRAMING

| Hierarchy and Elements | Contents, Comments | Size, t-bits |
|---|---|---|
| FEC frame issued by OLT | about 1.52 Mio frames/s | 66 × 257 = 16,926 |
| ├ **Input**, accessible by user | max 1,792 data octets→14,336 a-bits | 56 × 257 = **14,392** |
| ├ 256B/257B coded block #1 | ├ first 32 data octets or controls | ├ 32×8+1 = 257 |
| ├ 256B/257B coded block #2 | ├ next 32 data octets or controls | ├ 32×8+1 = 257 |
| . . . . . . . . . . . . . . . . . . . . | . . . . . . . . . . . . . . . . . . . . . . . . | . . . . . . . . . . |
| └ 256B/257B coded block #56 | └ last 32 data octets or controls | └ 32×8+1 = 257 |
| ├ Parity } auto-generated | calculated over only the bits above | 10×257= { 2,560 |
| └ Delimiter } auto-generated | ten-bit-long constant pattern | 10 |

TABLE III
ORIGINAL 25G-EPON UPSTREAM FRAMING

| Hierarchy and Elements | Contents, Comments | Size, t-bits |
|---|---|---|
| FEC frame issued by ONU | sent in a variable-length burst | 2,827÷16,926 |
| ├ **Input**, accessible by user | 32÷1,792 d.o.→256÷14,336 a-bits | (!)**257÷14,392** |
| ├ 256B/257B block #1 (mandatory) | ├ first 32 data octets or controls | ├ 32×8+1 = 257 |
| ├ 256B/257B block #2 <optional> | ├ *extra* 32 data octets or controls | ├ <if any> + 257 |
| . . . . . . . . . . . . . . . . . . . . | . . . . . . . . . . . . . . . . . . . . . . . . | . . . . . . . . . . |
| └ 256B/257B block #56 <optional> | └ *extra* 32 data octets or controls | └ <if any> + 257 |
| ├ Parity } auto-generated | calculated over only the bits above | (!)2570= { 2,560 |
| └ Delimiter } auto-generated | ten-bit-long constant pattern | 10 |

TABLE IV
PREPARATORY BINARY-PURE 25G-EPON-COMPATIBLE 64B/66B-TO-64B/65B TRANSCODING

| Sync Bits | Body Bits = Explicit 64B/66B Coded Block Content | Implicit | \|\|\| | D/C | Transfer Span ($g = 8$), *ready to further multiplexing* | X Bits |
|---|---|---|---|---|---|---|
| 01 | $D_0$ $D_1$ $D_2$ $D_3$ $D_4$ $D_5$ $D_6$ $D_7$ | — | → | "D" | $D_0$ $D_1$ $D_2$ $D_3$ $D_4$ $D_5$ $D_6$ $D_7$ | — |
| 10 | 0x1e $C_0$ $C_1$ $C_2$ $C_3$ $C_4$ $C_5$ $C_6$ $C_7$ | — | → | "C" | disp X $C_0$ $C_1$ $C_2$ $C_3$ $C_4$ $C_5$ $C_6$ $C_7$ | $CMD_1$ |
| 10 | 0x2d $C_0$ $C_1$ $C_2$ $C_3$ $O_4$ $D_5$ $D_6$ $D_7$ | — | deprecated in 25G-EPON | | | |
| 10 | 0x55 $D_1$ $D_2$ $D_3$ $O_0$ $O_4$ $D_5$ $D_6$ $D_7$ | — | deprecated in 25G-EPON | | | |
| 10 | 0x4b $D_1$ $D_2$ $D_3$ $O_0$ $C_4$ $C_5$ $C_6$ $C_7$ | — | deprecated in 25G-EPON | | | |
| 10 | 0x66 $D_1$ $D_2$ $D_3$ $O_0$ — $D_5$ $D_6$ $D_7$ | $S_4$ | deprecated in 25G-EPON | | | |
| 10 | 0x33 $C_0$ $C_1$ $C_2$ $C_3$ — $D_5$ $D_6$ $D_7$ | $S_4$ | deprecated in 25G-EPON | | | |
| 10 | 0x78 $D_1$ $D_2$ $D_3$ $D_4$ $D_5$ $D_6$ $D_7$ | $S_0$ | → | "C" | disp X $D_1$ $D_2$ $D_3$ $D_4$ $D_5$ $D_6$ $D_7$ | $CMD_2$ |
| 10 | 0xff $D_0$ $D_1$ $D_2$ $D_3$ $D_4$ $D_5$ $D_6$ | $T_7$ | → | "C" | disp X $D_0$ $D_1$ $D_2$ $D_3$ $D_4$ $D_5$ $D_6$ | $CMD_3$ |
| 10 | 0xe1 $D_0$ $D_1$ $D_2$ $D_3$ $D_4$ $D_5$ — $C_7$ | $T_6$ | → | "C" | disp X $D_0$ $D_1$ $D_2$ $D_3$ $D_4$ $D_5$ $C_7$ | $CMD_4$ |
| 10 | 0xd2 $D_0$ $D_1$ $D_2$ $D_3$ $D_4$ — $C_6$ $C_7$ | $T_5$ | → | "C" | disp X $D_0$ $D_1$ $D_2$ $D_3$ $D_4$ $C_6$ $C_7$ | $CMD_4$ |
| 10 | 0xcc $D_0$ $D_1$ $D_2$ $D_3$ — $C_5$ $C_6$ $C_7$ | $T_4$ | → | "C" | disp X $D_0$ $D_1$ $D_2$ $D_3$ $C_5$ $C_6$ $C_7$ | $CMD_4$ |
| 10 | 0xb4 $D_0$ $D_1$ $D_2$ — $C_4$ $C_5$ $C_6$ $C_7$ | $T_3$ | → | "C" | disp X $D_0$ $D_1$ $D_2$ $C_4$ $C_5$ $C_6$ $C_7$ | $CMD_4$ |
| 10 | 0xaa $D_0$ $D_1$ — $C_3$ $C_4$ $C_5$ $C_6$ $C_7$ | $T_2$ | → | "C" | disp X $D_0$ $D_1$ $C_3$ $C_4$ $C_5$ $C_6$ $C_7$ | $CMD_4$ |
| 10 | 0x99 $D_0$ — $C_2$ $C_3$ $C_4$ $C_5$ $C_6$ $C_7$ | $T_1$ | → | "C" | disp X $D_0$ $C_2$ $C_3$ $C_4$ $C_5$ $C_6$ $C_7$ | $CMD_4$ |
| 10 | 0x87 — $C_1$ $C_2$ $C_3$ $C_4$ $C_5$ $C_6$ $C_7$ | $T_0$ | → | "C" | disp X $C_1$ $C_2$ $C_3$ $C_4$ $C_5$ $C_6$ $C_7$ | $CMD_4$ |
| | (This is not defined in the original 64B/66B block encoding and also not used in the original 25G-EPON physical encoding.) "FILLER" | | { | "C" | disp X 49 extras into the spare bits | $CMD_4$ |
| Block Type | Rest of the Block, D/C/O = Data/Control/Ordered set | Start/Terminate | – | Root Bits | Affix Bits, *simply bypassed during multiplexing* | ( choice 1 of 4 ) |

DESIGN CONDITIONS TO BE MANDATORILY MET:

$0 \le \text{disp} \le 31$, $0 \le X \le 3$, $CMD_1 \ne CMD_2 \ne CMD_3 \ne CMD_4$

$0 \le$ ▖, ▘ $\le 1$, ▖ $\ne$ ▘ ( ⊣ is the future baton bit )

TABLE V
25G-EPON-AIMED UPSTREAM LINGUISTIC MULTIPLEXING PROCESS CONFIGURATION

| $^{US}n_p$ as a multiple of $g$ | + $n_{FILLER}$ @ $^{US}n_p$ | → $*n_p$ | = $n_e$ | × $k$ | + $*n_e$ | × $*k$ | $v = v(*n_p)$ | $v/257$ | $s = s(v, {}^{US}n_p)$ |
|---|---|---|---|---|---|---|---|---|---|
| — ; 8 ; 16 ; 24 ; 32 | 32 ; 24 ; 16 ; 8 ; — | 32 | — | — | 32 | × 1 | 257 | 1 | 196 ; 147 ; 98 ; 49 ; — |
| 32 ; 40 ; 48 ; 56 ; 64 | 32 ; 24 ; 16 ; 8 ; — | 64 | — | — | 64 | × 1 | 514 | 2 | 197 ; 148 ; 99 ; 50 ; 1 |
| . . . | . . . | . . . | . . . | . . . | . . . | . . . | . . . | . . . | . . . |
| 192 ; 200 ; 208 ; 216 ; 224 | 32 ; 24 ; 16 ; 8 ; — | 224 | — | — | 224 | × 1 | 1,799 | 7 | 202 ; 153 ; 104 ; 55 ; 6 |
| 224 ; 232 ; 240 ; 248 ; 256 | 32 ; 24 ; 16 ; 8 ; — | 256 | 256 | × 1 | — | — | 2,056 | 8 | 203 ; 154 ; 105 ; 56 ; 7 |
| 256 ; 264 ; 272 ; 280 ; 288 | 32 ; 24 ; 16 ; 8 ; — | 288 | 256 | × 1 | + 32 | × 1 | 2,313 | 9 | 203 ; 154 ; 105 ; 56 ; 7 |
| : : : | : : : | : : : | : : : | : : : | : : : | : : : | : : : | : : : | : : : |
| 1,472 ; 1,480 ; 1,488 ; 1,496 ; 1,504 | 32 ; 24 ; 16 ; 8 ; — | 1,504 | 256 | × 5 | + 224 | × 1 | 12,079 | 47 | 237 ; 188 ; 139 ; 90 ; 41 |
| 1,504 ; 1,512 ; 1,520 ; 1,528 ; 1,536 | 32 ; 24 ; 16 ; 8 ; — | 1,536 | 256 | × 6 | — | — | 12,336 | 48 | 238 ; 189 ; 140 ; 91 ; 42 |
| 1,536 ; 1,544 ; 1,552 ; 1,560 ; 1,568 | 32 ; 24 ; 16 ; 8 ; — | 1,568 | 256 | × 6 | + 32 | × 1 | 12,593 | 49 | 238 ; 189 ; 140 ; 91 ; 42 |
| . . . | . . . | . . . | . . . | . . . | . . . | . . . | . . . | . . . | . . . |
| 1,728 ; 1,736 ; 1,744 ; 1,752 ; 1,760 | 32 ; 24 ; 16 ; 8 ; — | 1,760 | 256 | × 6 | + 224 | × 1 | 14,135 | 55 | 244 ; 195 ; 146 ; 97 ; 48 |
| 1,760 ; 1,768 ; 1,776 ; 1,784 ; 1,792 | 32 ; 24 ; 16 ; 8 ; — | 1,792 | 256 | × 7 | — | — | 14,392 | 56 | 245 ; 196 ; 147 ; 98 ; 49 |

TABLE VI
25G-EPON-AIMED DOWNSTREAM LINGUISTIC MULTIPLEXING PROCESS CONFIGURATION

| $^{DS}n_p$ as a multiple of $g$ | + $n_{FILLER}$ @ $^{DS}n_p$ | → $*n_p$ | = $n_e$ | × $k$ | + $*n_e$ | × $*k$ | $v = v(*n_p)$ | $v/257$ | $s = s(v, {}^{DS}n_p)$ |
|---|---|---|---|---|---|---|---|---|---|
| — ; 8 ; 16 ; · · · · · · · · ; 1,792 | 1,792 ; · · · · · · ; 8 ; — | 1,792 | 256 | × 7 | — | — | 14,392 | 56 | 11,025 ; · · · · · ; 98 ; 49 |
| Comment→ 1•8 2•8 · · · · · · · · 224•8 | 224•8 · · · · · · 1•8 0•8 | fixed | seven complete rounds | | no partial round anyway | | fixed | fixed | (224+1)•49 · · · · · (1+1)•49 1•49 |

TABLE VII
COMPLETE ECHO MULTIPLEXING ROUND EXAMPLE

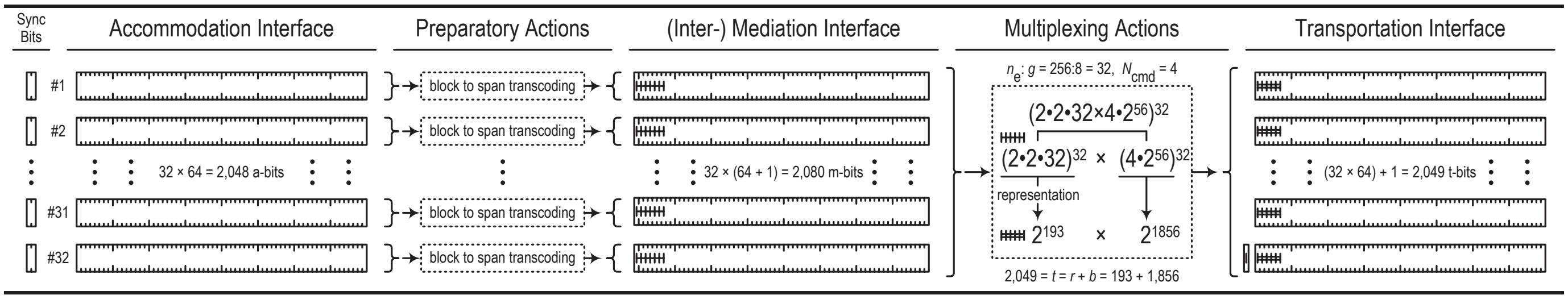

## GATHERING'N'RELAYING

We fuse an appropriately selected (bespoken) variant of the relay race representation approach, as it is developed form [3] from [4] from [5] from earlier, with an appropriately adapted (modified) version of the standard gathering approach, as it is developed from [6] from [7] from [8] from earlier, respectively, negotiating also various changes necessary to enable and then conduct such the fusion, i.e., assuming $g > 1$, $N_{\mathrm{ctrl}} \to N_{\mathrm{cmd}}$, $1 \to g$, $n_{\mathrm{e}} \to n_{\mathrm{e}} : g$, $m \to m/g$, $h \to h/g$, etc.[3]

Instead of performing on every separate transfer unit in the stream, how it was before this fusion, the resulting, fused procedure performs on pre-collected (and properly pre-processed) transfer spans. In that stream, every span covers over the eight ($g = 8$) successive transfer unit periods corresponded with its eight units, one of those is considered its head.[4]

Via and thanks to the measures mentioned above, we divide the only, transfer unit stream, as it was given initially, into the two streams, deliberately managed to be disproportional in the coverage, as so much as $g > 1$, the smaller one, stream of the head units, consisting of just one per every $g$ units in the initial stream, and the bigger one, stream of the rest units, consisting of all except one per every $g$ units, respectively.

While the gathering pass of that procedure may address the content of any transfer unit, its relaying pass will address only and only the content of the stream of the head units.[5]

So we establish the so called gathering'n'relaying approach, or GnR in short, expected to be useful in this work.

## SPARE'N'EXTRA

We complement a situationally available (given) amount of the spare bits, as it is developed from [9], with a purposefully controllable (managed) amount of the bits, each of those we consider an extra into (the favor of) the spare bits. We consider that these both—the spare bits, if any, and the extra bits, when necessary—convey some but off-duty information.[6]

In the FEC input and further in the FEC frame, these extra bits cause for they hit either a free-standing, span-long, multi-bit "filler" run conveying a fixed number of transportation bits accessible by the user, that is expectedly less than the number of transportation bits occupied by that whole span (due to the transcoding overhead), or a number of such the runs.[7]

Treating the extra bits so, i.e., on the block-based manner above, fits very well the EPON nature: because exactly block-based manipulations, including insertion and deletion, already happen in the nodes, thanks inside each to its first-in-first-out buffers, all involved in the standard, FIFO-based definition of the FEC sublayer operation, see [1] and [2].

The same time, treating the extras on that manner—already in itself and yet more when complemented along the spares—enables us to reach our objective as well as to give the coding background a simpler $L$-level description expressed in a form of some $L$-staged scheme, $L$-equation system, $L$-site pipeline, $L$-operation routine, $L$-step algorithm, etc.[8]

So we establish the so called spare'n'extra approach, or SnX in short, expected to be not less useful in the work, too.

[3]Unlike the rest, the change $N_{\mathrm{ctrl}} \to N_{\mathrm{cmd}}$ has a symbolic (visual) sense only: we introduce this renewed symbol, $N_{\mathrm{cmd}}$, to replace the previous one, $N_{\mathrm{ctrl}}$, not in the usage, but in the meaning in the new circumstances, where the head unit of a transfer span determines—only and only—how to interpret that span period in the whole, while neither the unit period the head takes its place in, nor any of the separate unit periods the span one consists of.

[4]Because now we need to account for just a single ("head") per $g$ transfer units gathered into a span, instead of every unit in the stream, as it was actual in [3], we refine the calculating and constraining expressions, $(m - h)$ and $n_{\mathrm{e}} \times N_{\mathrm{ctrl}}$, initially applied in [3], with $(m - h)/g$ and $(n_{\mathrm{e}} : g) \times N_{\mathrm{cmd}}$, respectively, that enables us to re-use the relay race representation procedure introduced in [3]. With $g = 1$ and $N_{\mathrm{cmd}} = N_{\mathrm{ctrl}}$, these refined expressions became the initial definitions given in [3]'s Tables VII and IX.

[5]Assuming the 64B/66B block encoding, that features 15 different control block types, and its possible reduction, that could be applicable in a particular case, we distinguish the following head transfer unit encoding variants, all are binary-pure, when considered in the view and terms of [3]:

"V" : $0 \le \mathrm{V} < N_{\mathrm{cmd}} = 2^0(=1)$ , $0 \le |\mathrm{disp}| < n_{\mathrm{e}} : g = 2^7(=128)$
"W" : $0 \le \mathrm{W} < N_{\mathrm{cmd}} = 2^1(=2)$ , $0 \le |\mathrm{disp}| < n_{\mathrm{e}} : g = 2^6(=64)$
"X" : $0 \le \mathrm{X} < N_{\mathrm{cmd}} = 2^2(=4)$ , $0 \le |\mathrm{disp}| < n_{\mathrm{e}} : g = 2^5(=32)$
"Y" : $0 \le \mathrm{Y} < N_{\mathrm{cmd}} = 2^3(=8)$ , $0 \le |\mathrm{disp}| < n_{\mathrm{e}} : g = 2^4(=16)$
"Z" : $0 \le \mathrm{Z} < N_{\mathrm{cmd}} = 2^4(=16)$ , $0 \le |\mathrm{disp}| < n_{\mathrm{e}} : g = 2^3(=8)$

It is clear to see that, for each of the variants, $(n_{\mathrm{e}} : g) \times N_{\mathrm{cmd}} = 2^7 = 128$.

[6]If such information should be exposed off the sublayer, the implementer can charge the last (16th) free room (0x00) in the space of 4-bit Hamming-distanced block type codes of the original 64B/66B block encoding.

[7]Each one of the spare bits occupies a maybe separate but certain position in the currently issued FEC input. Oppositely, all the extras into the spare bits appear in multi-bit patterns maybe anyway spread across that input.

[8]Describing a comprehensive function definable—in static, not dynamic—memory-less, we find vivid, clear, and easy to consider that function in a form of an abstract coding scheme exposing a number of bit signal slices drawn on the boundaries (i.e., arguments and results) of its stages, like the following:

$\mathrm{M}_0$ = A : slice # 0 mediation = accommodation // $\mathrm{m}_0$-bits = a-bits
$\mathrm{M}_1$ : slice # 1 intermediation // $\mathrm{m}_1$-bits
. . . . . . . . . . . . . . . . . . . . . . . . . . . . . . . . . . . . . . . . . . . . . . . . . . . . . . . . . . . .
$\mathrm{M}_{L-1}$ : slice # $L-1$ intermediation // $\mathrm{m}_{L-1}$-bits
$\mathrm{M}_L$ = T : slice # $L$ mediation = transportation // $\mathrm{m}_L$-bits = t-bits

where $L$ is the number of the stages in the scheme. Because of just two stages completing it—gathering (G), implemented via transcoding, then relaying (R), implemented via multiplexing—the scheme we proposed in this consideration has no need to expose more than these $L + 1$ bit signal slices, where $L = 2$:

$\mathrm{M}_0$ = A : accommodation service interface // $\mathrm{m}_0$-bits = a-bits
$\mathrm{M}_1$ = M : only ($L-1 = 1$) intermediation interface // $\mathrm{m}_1$-bits = m-bits
$\mathrm{M}_2$ = T : transportation service interface // $\mathrm{m}_2$-bits = t-bits

In practice, we used to see mediation and intermediation the same in mention.

TABLE VIII
PREPARATORY 10G-EPON-COMPATIBLE TRANSCODING

| D/C | Transfer Span, 0 ≤ disp ≤ 15, 0 ≤ Y ≤ 7, $CMD_1$ ≠ ... ≠ $CMD_8$ | | | | | | | | | | Y Bits |
|---|---|---|---|---|---|---|---|---|---|---|---|
| "D" | | | $D_0$ | $D_1$ | $D_2$ | $D_3$ | $D_4$ | $D_5$ | $D_6$ | $D_7$ | — |
| "C" | - | disp | Y | $C_0$ | $C_1$ | $C_2$ | $C_3$ | $C_4$ | $C_5$ | $C_6$ | $C_7$ | $CMD_1$ |
| "C" | - | disp | Y | $C_0$ | $C_1$ | $C_2$ | $C_3$ $O_4$ | $D_5$ | $D_6$ | $D_7$ | $CMD_2$ |
| "C" | - | disp | Y | $D_1$ | $D_2$ | $D_3$ | $O_0$ $O_4$ | $D_5$ | $D_6$ | $D_7$ | $CMD_3$ |
| "C" | - | disp | Y | $D_1$ | $D_2$ | $D_3$ | $O_0$ $C_4$ | $C_5$ | $C_6$ | $C_7$ | $CMD_4$ |
| "C" | - | disp | Y | $D_1$ | $D_2$ | $D_3$ | $O_0$ | $D_5$ | $D_6$ | $D_7$ | $CMD_5$ |
| "C" | - | disp | Y | $C_0$ | $C_1$ | $C_2$ | $C_3$ | $D_5$ | $D_6$ | $D_7$ | $CMD_5$ |
| "C" | - | disp | Y | $D_1$ | $D_2$ | $D_3$ | $D_4$ | $D_5$ | $D_6$ | $D_7$ | $CMD_6$ |
| "C" | - | disp | Y | $D_0$ | $D_1$ | $D_2$ | $D_3$ | $D_4$ | $D_5$ | $D_6$ | $CMD_7$ |
| "C" | - | disp | Y | $D_0$ | $D_1$ | $D_2$ | $D_3$ | $D_4$ | $D_5$ | $C_7$ | $CMD_8$ |
| "C" | - | disp | Y | $D_0$ | $D_1$ | $D_2$ | $D_3$ | $D_4$ | $C_6$ | $C_7$ | $CMD_8$ |
| "C" | - | disp | Y | $D_0$ | $D_1$ | $D_2$ | $D_3$ | $C_5$ | $C_6$ | $C_7$ | $CMD_8$ |
| "C" | - | disp | Y | $D_0$ | $D_1$ | $D_2$ | $C_4$ | $C_5$ | $C_6$ | $C_7$ | $CMD_8$ |
| "C" | - | disp | Y | $D_0$ | $D_1$ | $C_3$ | $C_4$ | $C_5$ | $C_6$ | $C_7$ | $CMD_8$ |
| "C" | - | disp | Y | $D_0$ | $C_2$ | $C_3$ | $C_4$ | $C_5$ | $C_6$ | $C_7$ | $CMD_8$ |
| "C" | - | disp | Y | $C_1$ | $C_2$ | $C_3$ | $C_4$ | $C_5$ | $C_6$ | $C_7$ | $CMD_8$ |
| "C" | - | disp | Y | "filler" | "filler" | "filler" | "filler" | "filler" | "filler" | "filler" | $CMD_8$ |
| Root Bits | Affix Bits, simply bypassed during multiplexing | | | | | | | | | | (choice 1 of 8) |

TABLE IX
10G-EPON-AIMED MULTIPLEXING CONFIGURATION

| $*n_p$ | = | $n_e$ | × | $k$ | + | $*n_e$ | × | $*k$ | $g$ | $\xi$ | $v$ | $v/65$ | $s_{min}$ |
|---|---|---|---|---|---|---|---|---|---|---|---|---|---|
| 216 | | 128 | × | 1 | + | 88 | × | 1 | 8 | 1 | 1,755 | 27 | 25 |
| Comment→ | | one complete round | | | | and one partial round | | | fixed | fixed | fixed | fixed | @ no fillers |

TABLE X
ECHO BUDGET MEMO

| Stream | $t$ | = | $b$ | + | $r$ | $n_e : g$ | Total | Rest (i.e., not Native) |
|---|---|---|---|---|---|---|---|---|
| 10G-DS | 1,025 | | 944 | | 81 | 16 | $2^{1025}$ | $2^{7\cdot(1)\cdot16+8\cdot(8-1)\cdot16}=2^{1008}$ |
| 10G-DS | 705 | | 649 | | 56 | 11 | $2^{705}$ | $2^{7\cdot(1)\cdot11+8\cdot(8-1)\cdot11}=2^{693}$ |
| 10G-US | 1,025 | | 944 | | 81 | 16 | $2^{1025}$ | $2^{7\cdot(1)\cdot16+8\cdot(8-1)\cdot16}=2^{1008}$ |
| 10G-US | 705 | | 649 | | 56 | 11 | $2^{705}$ | $2^{7\cdot(1)\cdot11+8\cdot(8-1)\cdot11}=2^{693}$ |
| 25G-DS | 2,049 | | 1,856 | | 193 | 32 | $2^{2049}$ | $2^{7\cdot(1)\cdot32+8\cdot(8-1)\cdot32}=2^{2016}$ |
| 25G-US | 2,049 | | 1,856 | | 193 | 32 | $2^{2049}$ | $2^{7\cdot(1)\cdot32+8\cdot(8-1)\cdot32}=2^{2016}$ |
| 25G-US | 1,793 | | 1,624 | | 169 | 28 | $2^{1793}$ | $2^{7\cdot(1)\cdot28+8\cdot(8-1)\cdot28}=2^{1764}$ |
| 25G-US | 1,537 | | 1,392 | | 145 | 24 | $2^{1537}$ | $2^{7\cdot(1)\cdot24+8\cdot(8-1)\cdot24}=2^{1512}$ |
| 25G-US | 1,281 | | 1,160 | | 121 | 20 | $2^{1281}$ | $2^{7\cdot(1)\cdot20+8\cdot(8-1)\cdot20}=2^{1260}$ |
| 25G-US | 1,025 | | 928 | | 97 | 16 | $2^{1025}$ | $2^{7\cdot(1)\cdot16+8\cdot(8-1)\cdot16}=2^{1008}$ |
| 25G-US | 769 | | 696 | | 73 | 12 | $2^{769}$ | $2^{7\cdot(1)\cdot12+8\cdot(8-1)\cdot12}=2^{756}$ |
| 25G-US | 513 | | 464 | | 49 | 8 | $2^{513}$ | $2^{7\cdot(1)\cdot8+8\cdot(8-1)\cdot8}=2^{504}$ |
| 25G-US | 257 | | 232 | | 25 | 4 | $2^{257}$ | $2^{7\cdot(1)\cdot4+8\cdot(8-1)\cdot4}=2^{252}$ |

TABLE XI
BACKWARD APPLICABILITY

| Protocol with $g = 8$ | $N_{cmd}$ | $*n_p$ | = | $n_e$ | × | $k$ | + | $*n_e$ | × | $*k$ | payload $v$ | $s_{min}$ |
|---|---|---|---|---|---|---|---|---|---|---|---|---|
| 100GBASE-R | 5 | 640 | = | 200 | × | 3 | + | 40 | × | 1 | 5,140 | 16 |
| 2.5/5GBASE-T | 8 | 200 | = | 128 | × | 1 | + | 72 | × | 1 | 1,625 | 23 |
| 40GBASE-T | 5 | 400 | = | 200 | × | 2 | | — | | — | 3,208 | 6 |
| 25GBASE-T | 8 | 400 | = | 128 | × | 3 | + | 16 | × | 1 | 3,208 | 4 |
| MultiGBASE-T1, 10GBASE-T | 8 | 400 | = | 128 | × | 3 | + | 16 | × | 1 | 3,250 | 46 |
| 10GBASE-KR | 8 | 256 | = | 128 | × | 2 | | — | | — | 2,080 | 30 |

## CONCLUSION

With the GnR and SnX approaches established, we are ready to prepare an appropriate coding means for each of the EPON FEC options, doing the following, step by step.

First, we define the preparatory transcoding, from 64B/66B coded blocks into eight-unit transfer spans, that is compatible with (equivalent for) the 25G-EPON original downstream and upstream duty organization rules (transmission structures), and also provides an over-duty means to transfer a variable number of user-defined transportation bits, see Table IV.

Second, we then propose the linguistic multiplexing process capable to treat on all the expectable FEC input options of the 25G-EPON upstream, that is the most complex among all the considered, as well as the process capable to treat on the FEC input option of the 25G-EPON downstream, that is much more simpler one, see Tables V and VI, respectively.

Considering each of those multiplexing processes, we imply that it involves the transcoding as many times as it is necessary to perform an echo multiplexing round, see Table VII.

Third, similarly to the done above, we define the preparatory a-block-to-a-span transcoding compatible with the 10G-EPON original duty organization, plus one extension, see Table VIII,[9] and propose the linguistic multiplexing process capable to treat on the single, unified 10G-EPON FEC input option, that is the simplest one among the considered, see Table IX.[10]

Finalizing the description, we conclude that the new design recipes, elaborated then introduced during this consideration, are useful, further extensible, see Table X, and easy applicable in another Ethernet coding means featuring a very long binary transport word, see Table XI.

[9] There is no statement made in [1] about a reduction of the number of the control block types, permissible in 10G-EPON, therefore here we use the "Y" encoding variant capable to reflect all the original types unreduced. Oppositely, it is explicitly stated in [2] that 25G-EPON permits just 10 out of 15 possible control block types specified in the original 64B/66B block encoding, therefore earlier, see Table IV, we used the "X" variant corresponding to that reduction allowed normatively. (Of course, a reduction simplifies the transcoding.)

[10] In Tables IX and XI, $s_{min}$ is the value for the case of no "filler" spans.

# Data Coding Means and Event Coding Means Multiplexed Over the 10GPASS-XR FEC Inputs

Alexander Ivanov

***Abstract*—EPoC is an extension of the EPON communication technology, targeting into the field and running over the infrastructure of conventional cable television networks. From the view of our research, it is an appropriate Ethernet protocol featuring an extremely long binary transport word plus some redundancy we could convert into an extra, event transfer ability.**

**Thru the paper, we consider the relatively young but actually known multi-gigabit EPoC protocol, 10GPASS-XR.**

## Introduction

EPON (Ethernet Passive Optical Networks) Protocol over Coax, abbreviated to EPoC, introduces the 10GPASS-XR family of physical layers, consisting of these paired members, 10GPASS-XR-D and 10GPASS-XR-U, for downstream (DS) and upstream (US), respectively [1].[1]

Considering the EPoC downstream, see Table I, we read we practice a single, fixed FEC (Forward Error Correction) input option continuously, once per every frame issued by the coax line terminal (CLT), because that node forms the downstream solely, being the trunk in the EPoC tree.

Considering the EPoC upstream, see Table II, we, each time, practice one reasonable of many allowable FEC input options, selecting that option once per each frame in a burst issued by a coax network unit (CNU), because such the nodes forms the upstream jointly, each being a branch in the tree.

In its original, EPoC provides no QoS (Quality of Service) support, but almost any option of its FEC input provides some redundancy accessible to the user. Based upon this, there could be found a way to bring the EPoC coding means up to a state providing a basic QoS support, leastwise.

## QoS Support Plan

Modifying the original EPoC coding means thru the rest of this paper, we follow the plan drafted below.

Regarding the downstream, we purport its behavior to be manageable continuously, see Table III once, hence refine the organization of its FEC input option to be preserving an extra room enough to send a periodic event per every issued frame, over the original duty, not delaying the duty.

Regarding the downstream, too, we imply it to provide a sort of seamy preemption, enabling two distinct classes of service (CoS) the user can clearly distinguish and freely associate with every (but namely) control-conveying transfer span.

Regarding the upstream, we purport its behavior just to be manageable intermittently, see Table III more, therefore refine

TABLE I
ORIGINAL DOWNSTREAM FRAMING

| Hierarchy and Elements | Contents, Comments | Size, t-bits |
|---|---|---|
| FEC frame issued by CLT | sent continuously | Σ = 16,140 |
| ├ Input, accessible by user | max 1,760 data octets→14,080 a-bits | 220 × 65 = 14,300 |
| ├ $^{64B}/_{65B}$ coded block #1 | ├ first 8 data octets or controls | ├ 8×8+1 = 65 |
| . . . . . . . . . . . . . . . . . . | . . . . . . . . . . . . . . . . . . . . | . . . . . . . |
| └ $^{64B}/_{65B}$ coded block #220 | └ last 8 data octets or controls | └ 8×8+1 = 65 |
| ├ CRC-40 } auto-generated | calculated over all the bits above | 1,840 = { 40 |
| └ Parity } auto-generated | calculated over all the bits above | 1,800 |

TABLE II
ORIGINAL UPSTREAM FRAMING

| Hierarchy and Elements | | Contents, Comments | | | | | | | | Size, t-bits |
|---|---|---|---|---|---|---|---|---|---|---|
| FEC frame issued by CNU | | sent in a variable-length burst | | | | | | | | Σ 385÷16,140 |
| ├ Input, accessible by user | | max 8÷1,760 d.o.→64÷14,080 a-bits | | | | | | | | (!)65÷14,300 |
| ├ $^{64B}/_{65B}$ block #1 | <mandatory> | ‹m› | ‹m› | ‹m› | ‹m› | ‹m› | ‹m› | | | ├ 8×8+1 = 65 |
| ├ $^{64B}/_{65B}$ block #2 | <optional> | ‹m› | ‹m› | ‹m› | ‹m› | ‹m› | ‹o› | | | ├ <if any> +65 |
| . . . . . . . . . . . . . . . . . . | | . . . | . . . | . . . | . . . | . . . | . . . | | | . . . . . . . |
| ├ $^{64B}/_{65B}$ block #12 | <optional> | ‹m› | ‹m› | ‹m› | ‹m› | ‹m› | ‹o› | | | ├ <if any> +65 |
| ├ $^{64B}/_{65B}$ block #13 | <optional> | ‹m› | ‹m› | ‹m› | ‹m› | — | — | | | ├ <if any> +65 |
| . . . . . . . . . . . . . . . . . . | | . . . | . . . | . . . | . . . | . . . | . . . | | | . . . . . . . |
| ├ $^{64B}/_{65B}$ block #25 | <optional> | ‹m› | ‹m› | ‹m› | ‹m› | — | — | | | ├ <if any> +65 |
| ├ $^{64B}/_{65B}$ block #26 | <optional> | ‹m› | ‹m› | ‹m› | ‹o› | — | — | | | ├ <if any> +65 |
| . . . . . . . . . . . . . . . . . . | | . . . | . . . | . . . | . . . | . . . | . . . | | | . . . . . . . |
| ├ $^{64B}/_{65B}$ block #76 | <optional> | ‹m› | ‹m› | ‹m› | ‹o› | — | — | | | ├ <if any> +65 |
| ├ $^{64B}/_{65B}$ block #77 | <optional> | ‹m› | ‹m› | — | — | — | — | | | ├ <if any> +65 |
| . . . . . . . . . . . . . . . . . . | | . . . | . . . | . . . | . . . | . . . | . . . | | | . . . . . . . |
| ├ $^{64B}/_{65B}$ block #101 | <optional> | ‹m› | ‹m› | — | — | — | — | | | ├ <if any> +65 |
| ├ $^{64B}/_{65B}$ block #102 | <optional> | ‹m› | ‹o› | — | — | — | — | | | ├ <if any> +65 |
| . . . . . . . . . . . . . . . . . . | | . . . | . . . | . . . | . . . | . . . | . . . | | | . . . . . . . |
| └ $^{64B}/_{65B}$ block #220 | <optional> | ‹m› | ‹o› | — | — | — | — | | | └ <if any> +65 |
| ├ CRC-40 | } auto-generated | ‹m› | ‹m› | ‹m› | ‹m› | ‹m› | ‹m› | 40 | 40 | 40 |
| └ Parity | } auto-generated | ‹m› | ‹m› | ‹m› | ‹m› | ‹m› | ‹m› | 280 | 900 | 1,800 |
| ($S^{hortened}$) $L_{ong}$ / $M_{edium}$ / $S_{hort}$ → | | L | SL | M | SM | S | SS | (S)S | (S)M | S(L) |

<m> = mandatory / <o> = optional 64B/65B coded block

TABLE III
POSSIBLE QOS MODELING

| Continuous Management Strategy | Intermittent Management Strategy |
|---|---|
| every period should be manageable<br>e.g., every span time period at the accommodation interface | some periods may be manageable<br>e.g., just a period per a multi-period run in the duty stream |

A manuscript of this work was submitted to IEEE Communications Letters December 26, 2022 and rejected for fair reasons during its peer review.

Please sorry for the author has no time to find this work a new home, peer reviewed or not, except of arXiv, and just hopes there it meets its reader, one or maybe various, whom the author beforehand thanks for their regard.

A. Ivanov is with JSC Continuum, Yaroslavl, the Russian Federation.

[1]Being a descendant of EPON, EPoC inherits a lot from that technology, including also its asymmetric and asynchronous nature, see [2], inalienably.

TABLE IV
MODIFIED REPRESENTATION ANALYSIS

| Scenario | $m-3g$ | $m-2g$ | $m-1g$ | $m$ | $m^{bis}$ |
|---|---|---|---|---|---|
| pit/baton row→ D D D D, rest of rows → | choice between 0÷1, same as before | choice between 0÷1, same as before | choice between 0÷1, same as before | choice between 0÷1, same as before | fixed to always 0; choice between |
| $C_q$D D D | always 0 | 0÷1 | 0÷1 | 0÷1 | 1÷$N_{CoS}$ |
| D $C_q$D D | 0÷1 | always 0 | 0÷1 | 0÷1 | 1÷$N_{CoS}$ |
| $C_qC_q$D D | always 0 | 1÷$N_{CoS}$ | 0÷1 | 0÷1 | 1÷$N_{CoS}$ |
| D D $C_q$D | 0÷1 | 0÷1 | always 0 | 0÷1 | 1÷$N_{CoS}$ |
| $C_q$D $C_q$D | always 0 | 0÷1 | 1÷$N_{CoS}$ | 0÷1 | 1÷$N_{CoS}$ |
| D $C_qC_q$D | 0÷1 | always 0 | 1÷$N_{CoS}$ | 0÷1 | 1÷$N_{CoS}$ |
| $C_qC_qC_q$D | always 0 | 1÷$N_{CoS}$ | 1÷$N_{CoS}$ | 0÷1 | 1÷$N_{CoS}$ |
| D D D $C_q$ | 0÷1 | 0÷1 | 0÷1 | always 0 | 1÷$N_{CoS}$ |
| $C_q$D D $C_q$ | always 0 | 0÷1 | 0÷1 | 1÷$N_{CoS}$ | 1÷$N_{CoS}$ |
| D $C_q$D $C_q$ | 0÷1 | always 0 | 0÷1 | 1÷$N_{CoS}$ | 1÷$N_{CoS}$ |
| $C_qC_q$D $C_q$ | always 0 | 1÷$N_{CoS}$ | 0÷1 | 1÷$N_{CoS}$ | 1÷$N_{CoS}$ |
| D D $C_qC_q$ | 0÷1 | 0÷1 | always 0 | 1÷$N_{CoS}$ | 1÷$N_{CoS}$ |
| $C_q$D $C_qC_q$ | always 0 | 0÷1 | 1÷$N_{CoS}$ | 1÷$N_{CoS}$ | 1÷$N_{CoS}$ |
| D $C_qC_qC_q$ | 0÷1 | always 0 | 1÷$N_{CoS}$ | 1÷$N_{CoS}$ | 1÷$N_{CoS}$ |
| $C_qC_qC_qC_q$ | always 0 | 1÷$N_{CoS}$ | 1÷$N_{CoS}$ | 1÷$N_{CoS}$ | 1÷$N_{CoS}$ |
| pit/baton row→ generalized for $n_e : g = 4$ | choice between 0÷1, rest is as before | choice between 0,1÷$N_{CoS}$, rest is as before | choice between 0,1÷$N_{CoS}$, rest is as before | choice between 0,1÷$N_{CoS}$, rest is as before | choice between 0,1÷$N_{CoS}$, time→ |

TABLE V
MODIFIED PBR ANALYSIS

| Scenario | $m-(H-1)g$ | $m-(H-2)g$ | · · · · · | $m-(H-H)g = m$ | $m^{bis}$ |
|---|---|---|---|---|---|
| $\{D/C_q\}^H$ | choice one of $2^1=2$ distinct items × | choice one of $1+N_{CoS}$ distinct items × | · · · · · × | choice one of $1+N_{CoS}$ distinct items × | choice one of $1+N_{CoS}$ distinct items |
| Repetitions per a round → | only once | either $H = n_e : g$ (complete round) or $H = {}^*n_e : g$ (partial round) times | | | |

TABLE VI
BLOCK TRANSCODING RULES

| Implicit CTRL | D/$C_q$ Transfer Span, 0 ≤ @ ≤ 24, 0 ≤ # ≤ 4, $CMD_1 \neq \ldots \neq CMD_5$ | | | | | | | | | # Value |
|---|---|---|---|---|---|---|---|---|---|---|
| — | "D" | | $D_0$ | $D_1$ | $D_2$ | $D_3$ | $D_4$ | $D_5$ | $D_6$ $D_7$ | — |
| — | "$C_q$" | @ with # | $C_0$ | $C_1$ | $C_2$ | $C_3$ | $C_4$ | $C_5$ | $C_6$ $C_7$ | $CMD_1$ |
| $S_4$ | "$C_q$" | @ with # | $C_0$ | $C_1$ | $C_2$ | $C_3$ | — | $D_5$ | $D_6$ $D_7$ | $CMD_2$ |
| $S_0$ | "$C_q$" | @ with # | $D_1$ | $D_2$ | $D_3$ | $D_4$ | $D_5$ | $D_6$ | $D_7$ | $CMD_3$ |
| $T_7$ | "$C_q$" | @ with # | $D_0$ | $D_1$ | $D_2$ | $D_3$ | $D_4$ | $D_5$ | $D_6$ | $CMD_4$ |
| $T_6$ | "$C_q$" | @ with # | $D_0$ | $D_1$ | $D_2$ | $D_3$ | $D_4$ | $D_5$ | $C_7$ | $CMD_5$ |
| $T_5$ | "$C_q$" | @ with # | $D_0$ | $D_1$ | $D_2$ | $D_3$ | $D_4$ | $C_6$ | $C_7$ | |
| $T_4$ | "$C_q$" | @ with # | $D_0$ | $D_1$ | $D_2$ | $D_3$ | $C_5$ | $C_6$ | $C_7$ | |
| $T_3$ | "$C_q$" | @ with # | $D_0$ | $D_1$ | $D_2$ | $C_4$ | $C_5$ | $C_6$ | $C_7$ | |
| $T_2$ | "$C_q$" | @ with # | $D_0$ | $D_1$ | $C_3$ | $C_4$ | $C_5$ | $C_6$ | $C_7$ | |
| $T_1$ | "$C_q$" | @ with # | $D_0$ | $C_2$ | $C_3$ | $C_4$ | $C_5$ | $C_6$ | $C_7$ | |
| $T_0$ | "$C_q$" | @ with # | $C_1$ | $C_2$ | $C_3$ | $C_4$ | $C_5$ | $C_6$ | $C_7$ | |
| ev | "$C_q$" | @ with # | user-defined event information | | | | | | | |
| (not sent) | Root Bits | | Affix Bits, simply bypassed during multiplexing | | | | | | | (choice 1 of 5) |

TABLE VII
EVENT INSERTION RULE

| Gathered [into spans] Transfer Unit Time Periods→ ↓Scenario | | | | | |
|---|---|---|---|---|---|
| | $(m-g)$, $(m-g)-1$, $(m-g)-2$, ⋮, $(m-g)-(g-1)$ | $(m)$, $(m)-1$, $(m)-2$, ⋮, $(m)-(g-1)$ | $(m-g)$, $(m-g)-1$, $(m-g)-2$, ⋮, $(m-g)-(g-1)$ | $(m)$, $(m)-1$, $(m)-2$, ⋮, $(m)-(g-1)$ | $(m+g)$, $(m+g)-1$, $(m+g)-2$, ⋮, $(m+g)-(g-1)$ |
| $^{(m)}$event ↓ $^{(m)}$span | D/$C_q$ Transfer Span of Head at $(m-g)$ | D/$C_q$ Transfer Span of Head at $(m)$ | D/$C_q$ Transfer Span of Head at $(m-g)$ | EVENT associated to Span of Head at $(m)$ | D/$C_q$ Transfer Span of Head at $(m)$ |
| Stream→ | Before Insertion | | After (prior split and then) Insertion | | |

the organization of its allowable FEC input options to be also preserving an over-duty capacity, however, as when and much as it is possible with a particular option.

Regardless of a given stream, we moreover purport that its behavior should also reflect for there may be a sporadic event that expedites (accompanies) a duty-related transfer span, but, expectedly, at the cost of delaying the rest of the original duty going next that so expedited (accompanied) span.

## TWIST'N'MERGE

For dealing with the new circumstances, we try the so called twist'n'merge approach (TnM) resulting from the next.

Let us consider (the content of) every transfer unit—in the argument stream—separate (when $g = 1$) or designated head (when $g > 1$), regular and consisting of the so called pit/baton, track/leg, and fan/fan rows (or PBR, TLR, and FFR in short) characterized by their moduli (positive integers) of $N_{\rm PB} \geq 1$, $N_{\rm TL} \geq 1$, and $N_{\rm FF} \geq 1$, respectively, whatever any given unit conveys a data value (D), or a control code (Cq):

D : $N_{\rm FF} \times N_{\rm TL} \times N_{\rm pit} \leq N$ // . . . . . . data //
Cq : $N_{\rm FF} \times N_{\rm TL} \times N_{\rm btn} \leq N$ // . . . control //
D/Cq : $N_{\rm FF} \times N_{\rm TL} \times N_{\rm PB} = N$ // data or not //

which implies that $N_{\rm PB} = \max\{N_{\rm pit}, N_{\rm btn}\}$, where $N_{\rm pit} \geq 1$ and $N_{\rm btn} = 1 + N_{\rm CoS}$ are the PBR limits for all such data- and control-conveying units, respectively, while $N_{\rm CoS} \geq 1$ is the number of the classes of service we need to support.

Hence, every treating on one given series of $H$ such transfer units, $H \leq N_{\rm TL}$, with the latest unit occupying the period $m$ in the argument stream, generates one corresponding series of some renewed units—in the function stream—that is one unit (exactly, one baton) longer than the argument series because it integrates all the D/Cq information in itself now:

earliest : $N_{\rm FF} \times N_{\rm TL} \times N_{\rm pit}$ @ $(1-H)g + m < m$
next : $N_{\rm FF} \times N_{\rm TL} \times N_{\rm PB}$ @ $(2-H)g + m < m$
. . . . . . . . . . . . . . . . . . . . . . . . . . . . . . . . . . . . . . . . . . . . . . .
latest : $N_{\rm FF} \times N_{\rm TL} \times N_{\rm PB}$ @ $(H-H)g + m = m$
latest$^{\rm bis}$ : $N_{\rm btn}$ @ $m^{\rm bis}$

as well as (the content of) the PBRs and TLRs of whose units may be *twisted* during the work of an appropriate logic-based representation option—developed from [2], [3], and earlier—intended to convert from the argument into the function.

Finally, (the content of) any rows of any transfer units in the function stream can be *merged* via the work of an appropriate arithmetic-based representation option—developed from [4], [5], [6], [7], [8], and earlier—intended to inscribe (the content of) the rows into the desired space, anyway usefully, including somehow homo- and/or heterogeneously.

TABLE VIII
DOWNSTREAM FEC INPUT PREPARATIONS

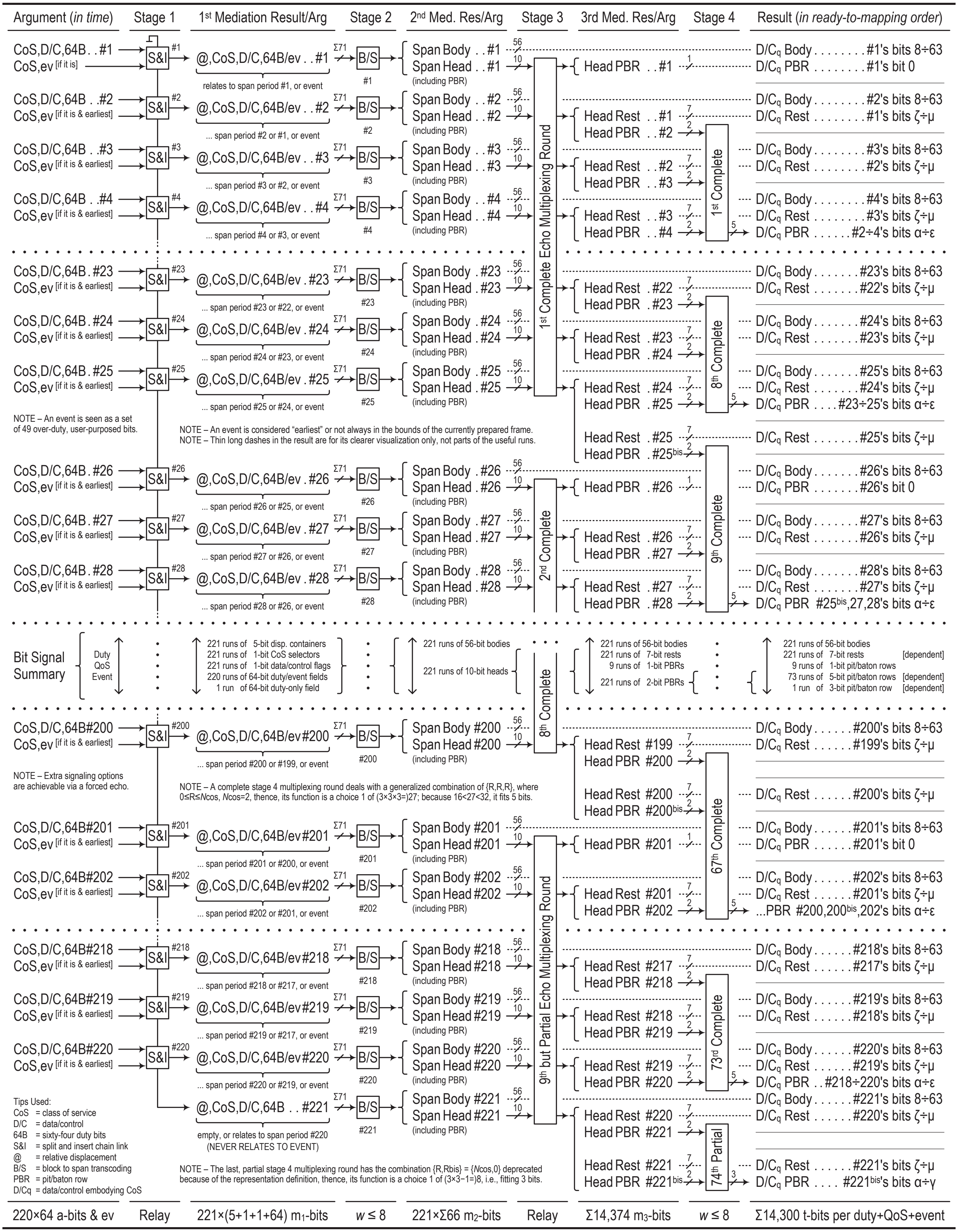

TABLE IX
DOWNSTREAM LINGUISTIC MULTIPLEXING DESCRIPTION

| Scope | $n_p$ | → | $*n_p$ | : | $g$ | = | $*n_p:g$ | = | $n_e:g$ | × | $k$ | + | $*n_e:g$ | × | $*k$ | $N_{CoS}$ | $N_r$ | $t$ | = | $b$ | + | $r$ | $*t$ | = | $*b$ | + | $*r$ | $w$ | $v$ | $s$ |
|---|---|---|---|---|---|---|---|---|---|---|---|---|---|---|---|---|---|---|---|---|---|---|---|---|---|---|---|---|---|---|
| Stage 1 | 1,760 | | 1,768 | : | 8 | = | 221 | = | 1 | × | 221 | | — | | — | 1+1 | — | 71 | = | 71 | | — | — | | — | | — | — | 15,691 | <0 |
| Stage 2 | — | | 1,768 | : | 8 | = | 221 | = | 1 | × | 221 | | — | | — | 1+1 | $5^3$ | 66 | = | 59 | + | 7 | — | | — | | — | ≤8 | 14,586 | <0 |
| Stage 3 | — | | 1,768 | : | 8 | = | 221 | = | 25 | × | 8 | + | 21 | × | 1 | 1+1 | — | 1,626 | = | 1,400 | + | 226 | 1,366 | = | 1,176 | + | 190 | — | 14,374 | <0 |
| Stage 4 | — | | 1,768 | : | 8 | = | 221 | = | 3 | × | 73 | + | 2 | × | 1 | 1+1 | $3^1$ | $194\frac{72}{73}$ | = | $189\frac{72}{73}$ | + | 5 | 66 | = | 63 | + | 3 | ≤8 | 14,300 | — |

TABLE X
UPSTREAM LINGUISTIC MULTIPLEXING VARIANTS

| $n_{p,max} : g = *n_p : g$ | $k$ | $*n_e : g$ with $n_e : g = 25$ | $v : 65$ with $N_{CoS} = 1$ and $N_r = N_@ \times N_\# = 25 \times 5 = 5^3$ | $s$ |
|---|---|---|---|---|
| — ; 1 ; ⋯ ; 24 | — | — ; 1 ; ⋯ ; 24 | — ; 1 ; ⋯ ; 24 | — ; — ; ⋯ ; 23 |
| 25 ; ⋯ ; 49 | 1 | — ; ⋯ ; 24 | 25 ; ⋯ ; 49 | 24 ; ⋯ ; 47 |
| 50 ; ⋯ ; 66 $\xrightarrow{\text{break}}$ 68 ; ⋯ ; 74 | 2 | — ; ⋯ ; 16 $\xrightarrow{\text{break}}$ 18 ; ⋯ ; 24 | 50 ; ⋯ ; 66 $\xrightarrow{\text{skip}}$ 67 ; ⋯ ; 73 | 48 ; ⋯ ; 63 $\xrightarrow{\text{wrap}}$ — ; ⋯ ; 6 |
| 75 ; ⋯ ; 99 | 3 | — ; ⋯ ; 24 | 74 ; ⋯ ; 98 | 7 ; ⋯ ; 30 |
| 100 ; ⋯ ; 124 | 4 | — ; ⋯ ; 24 | 99 ; ⋯ ; 123 | 31 ; ⋯ ; 54 |
| 125 ; ⋯ ; 134 $\xrightarrow{\text{break}}$ 136 ; ⋯ ; 149 | 5 | — ; ⋯ ; 9 $\xrightarrow{\text{break}}$ 11 ; ⋯ ; 24 | 123 ; ⋯ ; 133 $\xrightarrow{\text{skip}}$ 134 ; ⋯ ; 147 | 55 ; ⋯ ; 63 $\xrightarrow{\text{wrap}}$ — ; ⋯ ; 13 |
| 150 ; ⋯ ; 174 | 6 | — ; ⋯ ; 24 | 148 ; ⋯ ; 172 | 14 ; ⋯ ; 37 |
| 175 ; ⋯ ; 199 | 7 | — ; ⋯ ; 24 | 173 ; ⋯ ; 197 | 38 ; ⋯ ; 61 |
| 200 ; ⋯ ; 202 $\xrightarrow{\text{break}}$ 204 ; ⋯ ; 223 | 8 | — ; ⋯ ; 2 $\xrightarrow{\text{break}}$ 4 ; ⋯ ; 23 | 198 ; ⋯ ; 200 $\xrightarrow{\text{skip}}$ 201 ; ⋯ ; 220 | 62 ; ⋯ ; 63 $\xrightarrow{\text{wrap}}$ — ; ⋯ ; 19 |

## MODIFIED EPOC DOWNSTREAM

Following the QoS support plan regarding the downstream, we assume $g = 8$ and $N_{\text{CoS}} = 2$, as well as $N_{\text{pit}} = 2$, thence $N_{\text{btn}} = 1 + N_{\text{CoS}} = 3$, $N_{\text{PB}} = \max\{N_{\text{pit}}, N_{\text{btn}}\} = 3 < 4 = 2^2$, while $N_@ = 25 = 5^2$ with $N_\# = 5 = 5^1$, thence $N_{\text{TL}} = N_@ \times N_\# = 5^{2+1} = 125 < 128 = 2^7$, and $N_{\text{FF}} = 1 = 2^0$, and so construct the modified EPoC downstream coding means, based upon the TnM approach, see Tables IV and V (model), then VI and VII, and then VIII alone, successively.

The modified downstream input reflects nothing more than exact $220 + 1$ regular transfer spans, passing up to one event per every frame, with no delay in the original duty.

Describing this renovated means in the sense and terms of the underlying linguistic multiplexing process, see Table IX, we—once again but now in another details—respond for how the means deals with the single, fixed FEC input option of the original EPoC downstream, stage by stage.

## MODIFIED EPOC UPSTREAM

Following the QoS support plan regarding the upstream, we assume $g = 8$ but $N_{\text{CoS}} = 1$, thence $N_{\text{PB}} = N_{\text{pit}} = N_{\text{btn}} = 1 + N_{\text{CoS}} = 2 = 2^1$, $N_{\text{TL}} = N_@ \times N_\# = 5^3 = 125 < 2^7$, and $N_{\text{FF}} = 1 = 2^0$, and so construct the modified EPoC upstream coding means, based upon the TnM approach, too. Alongside up to 223 regular transfer spans, the modified upstream input may also reflect a variable run of the spare bits [9].

Describing this renovated means in the sense and terms of the underlying linguistic multiplexing processes, see Table X, we explicitly respond that the means is linear with no one and with a few exceptions—necessary for decoding the upstream unambiguously—in how it deals with the volume, $v$, and the target scope, $*n_{\text{p}}$, over all the allowable FEC input options of the original EPoC upstream, respectively.

## CONCLUSION

We described a way the EPoC coding means could achieve a more complicated state, where it becomes capable to provide an event-reflecting service as well as a basic support for QoS, both within the physical layer it operates on.

We also find the design principles contributing into the way applicable in similar Ethernet protocols featuring an extremely or just very long binary transport word.[2]

[2] Among the considered before EPoC, there are a gitabit and various multigigabit protocols, featuring no and eight-unit native gathering, respectively.